\documentclass[11pt]{article}

\usepackage{booktabs}
\usepackage[margin=15pt,font=small,labelfont={bf,sf},justification=justified]{caption}
\usepackage[superscript,biblabel]{cite}
\usepackage{color}
\usepackage{float}
\usepackage{graphicx}
\usepackage{lscape}
\usepackage[bbgreekl]{mathbbol}
\usepackage{mathtools}
\usepackage{multibib}
\usepackage{multirow}
\usepackage{tabularx}
\usepackage{titleref}
\usepackage{url}

\usepackage{tikz}
\usepackage{amsmath}

\usetikzlibrary{fadings}
\usetikzlibrary{patterns}
\usetikzlibrary{shadows.blur}
\usetikzlibrary{shapes}

\newif\ifbembo
\newif\ifcharter
\newif\iferewhon
\newif\iflibertine
\newif\iflibertinealt
\newif\ifpalantino
\newif\iftimesnewroman

\bembofalse
\chartertrue
\erewhonfalse
\libertinefalse
\libertinealtfalse
\palantinofalse
\timesnewromanfalse

\ifbembo
  \usepackage[p,osf]{ETbb} 
  \usepackage[scaled=.95,type1]{cabin} 
  \usepackage[varqu,varl]{zi4}
  \usepackage[T1]{fontenc}
  \usepackage[libertine,vvarbb]{newtxmath}
  \usepackage[cal=boondoxo,bb=boondox]{mathalfa}
\fi

\ifcharter
  \usepackage[scaled=.98,p]{XCharter}
  \usepackage[scaled=1.04,varqu,varl]{inconsolata}
  \usepackage[type1]{cabin}
  \usepackage[uprightscript,xcharter,vvarbb,scaled=1.05]{newtxmath}
  \usepackage[cal=euler,scr=boondoxo,bb=boondox]{mathalpha}  
\fi

\iferewhon
  \usepackage[p,osf,scaled=.98,space]{erewhon} 
  \usepackage[varqu,varl]{inconsolata} 
  \usepackage[type1,scaled=.95]{cabin} 
  \usepackage[utopia,vvarbb]{newtxmath}
\fi

\iflibertine
  \usepackage[sb]{libertinus}
  \usepackage[T1]{fontenc}
  \usepackage{textcomp}
  \usepackage[varqu,varl]{zi4}
  \usepackage{libertinust1math} 
  \usepackage[scr=boondoxo,bb=boondox]{mathalpha} 
\fi

\iflibertinealt
  \usepackage[sb]{libertinus}
  \usepackage[T1]{fontenc}
  \usepackage{textcomp}
  \usepackage{libertinust1math} 
  \usepackage[scr=boondoxo,bb=boondox]{mathalpha} 
\fi

\ifpalantino
  \usepackage[largesc]{newpxtext}
  \usepackage[vvarbb]{newpxmath}
\fi

\iftimesnewroman
  \usepackage{newtxtext} 
  \usepackage[scaled=0.95]{inconsolata} 
  \usepackage[vvarbb]{newtxmath} 
\fi

\usepackage[T1]{fontenc}
\usepackage[final,protrusion=true,expansion=true]{microtype}

\usepackage{titling}
\usepackage{authblk} 
\pretitle{\begin{center}\bfseries\sffamily\LARGE}
\posttitle{\end{center}}

\predate{\begin{center}\normalsize}
\postdate{\end{center}}

\usepackage{abstract}

\usepackage{sectsty} 
\allsectionsfont{\sffamily}

\usepackage[left=1in,right=1in,top=1in,bottom=1in,includefoot,heightrounded]{geometry}

\newcites{supp}{SUPPLEMENTAL REFERENCES}

\makeatletter
\patchcmd{\LS@rot}{90}{-90}{}{}
\patchcmd{\endlandscape}{90}{-90}{}{}
\makeatother

\newcommand*{\tran}{^{\mkern-1.5mu\mathsf{T}}}
\newcommand{\cc}{{\text{c}}}

\newcommand{\C}{\mathbf{C}}
\newcommand{\U}{\mathbf{U}}
\newcommand{\bu}{\mathbf{u}}
\newcommand{\x}{\mathbf{x}}
\newcommand{\jumpinverse}[1]{\left\llbracket#1\right\rrbracket}

\begin{document}

\title{An Immersed Interface Method for Incompressible Flows and Near Contact}

\author[1]{Michael J. Facci}
\author[1]{Qi Sun}
\author[1--5]{Boyce E.~Griffith\thanks{Corresponding author: boyceg@email.unc.edu}}
\affil[1]{Department of Mathematics, University of North Carolina, Chapel Hill, NC, USA}
\affil[2]{Department of Biomedical Engineering, University of North Carolina, Chapel Hill, NC, USA}
\affil[3]{Carolina Center for Interdisciplinary Applied Mathematics, University of North Carolina, Chapel Hill, NC, USA}
\affil[4]{Computational Medicine Program, University of North Carolina School of Medicine, Chapel Hill, NC, USA}
\affil[5]{McAllister Heart Institute, University of North Carolina School of Medicine, Chapel Hill, NC, USA}

\vspace{\baselineskip}

\maketitle
\begin{abstract}

We present an enhanced immersed interface method for simulating incompressible fluid flows in thin gaps between closely spaced immersed boundaries. This regime, common in engineered structures such as tribological interfaces and bearing assemblies, poses significant computational challenges because of limitations in grid resolution and the prohibitive cost of mesh refinement near contact. The immersed interface method imposes jump conditions that capture stress discontinuities generated by forces that are concentrated along immersed boundaries.
Our approach introduces a bilinear velocity interpolation operator in two spatial dimensions that incorporates jump conditions from multiple nearby interfaces when they occupy the same interpolation stencil. 
Furthermore, we demonstrate that this methodology for resolving distinct interfaces that are in near contact can also be applied to achieve improved accuracy for single-interface geometries involving intrinsically sharp features, in which multiple interface segments intersect a single grid cell.
Numerical results demonstrate substantial improvements in both interface and Eulerian velocity accuracy compared to lubrication-based approaches employed with both the immersed boundary and immersed interface methods, even for interfaces whose separation is one-fiftieth of a computational grid cell. Additionally, we demonstrate that this formulation effectively resolves flows near sharp geometric features. The proposed method improves upon previous interpolation schemes and eliminates the need for prior knowledge of interface orientation or geometry. This makes it broadly applicable to a wide range of fluid--structure interaction problems involving near-contact dynamics.
\end{abstract}


\section{Introduction}

Simulating the dynamics of interactions between fluid flows and immersed boundaries has a broad range of applications in science and engineering. These include heart valves \cite{griffith_immersed_2012,peskin_flow_1972}, bio-locomotion for swimmers \cite{zeng_immersed_2023}, inferior vena cava filters\cite{kolahdouz_sharp_2023}, red blood cell aggregation\cite{XU20131810, rbcs}, and prosthetic joint lubrication\cite{LIU2006905}. The immersed boundary (IB) method and its extensions are widely used for such applications. Peskin and Printz\cite{PESKIN199333} showed that the delta function formulation for singular forces reduces to jump conditions associated with interfacial forces concentrated along fluid–structure boundaries.
The immersed interface method (IIM) is a particular approach to such models that enforces these jump conditions to couple a fluid to an immersed boundary.

The IIM originated in work by Mayo{\cite{mayo_fast_1984}} and by LeVeque and Li{\cite{leveque_immersed_1994}} for the solution of
elliptic partial differential equations with singular forces. The IIM was
extended to the two-dimensional solution of the incompressible Navier-Stokes
equations by Lee and LeVeque\cite{lee_immersed_2003} and Lai and Li \cite{li_immersed_2001}. 
Jump conditions for velocity, pressure, and their derivatives for fluid flows in three spatial dimensions were derived by Xu and Wang\cite{xu_systematic_2006}.
Immersed interface methods developed by Thekkethil and Sharma \cite{thekkethil_level_2019} and Xu et al. \cite{xu_level-set_2020} use level set representations of the
geometry. In contrast,
the IIM described by Kolahdouz \cite{kolahdouz_immersed_2020,kolahdouz_sharp_2021} deals with
interfaces represented as triangulated surfaces. Such surface representations can be readily used with finite element structural models. 

A challenging regime arises in fluid–structure interaction (FSI) problems involving thin layers of fluid between closely spaced interfaces, particularly where the separation distance is much less than the background grid spacing. 
These scenarios are common in applications involving mechanical bearings, in which a thin fluid film lubricates relative motion between adjacent components. In principle, it is possible to resolve these narrow gaps, but doing so requires computational grids that are multiple orders of magnitude more refined than those used in this study. 
Classical notions of convergence and accuracy are challenging to apply for these inherently under-resolved configurations, because the flow field in the narrow gap remains under-resolved. 
Studies in tribology have addressed these difficulties by deriving asymptotic limits for fluid properties in the narrow gaps. In some configurations, as the gap spacing becomes arbitrarily small, the velocity field becomes linear. These analytical results provide a rich set of benchmarks that we use to test our current methods.

Several studies have incorporated lubrication theory \cite{Acheson} to improve accuracy in these near touching regimes. Fai and Rycroft \cite{FAI2018319} coupled lubrication models with the IB method, while Yacoubi et al. \cite{YACOUBI2019108919} integrated lubrication theory with an IIM by introducing analytic singular forces within the lubrication layer. These methods, however, have notable limitations. The IB-based approach is constrained by its smoothed force and velocity spreading operators, which generally reduce accuracy near immersed boundaries. 
Furthermore, the prior IIM-based approach with lubrication forces uses cubic spline–based interface representations, which restricts its versatility compared to models that use piecewise linear elements, the standard in many engineering analyses. 
In addition, because these methods directly use lubrication theory, their implementations are complex and require prior knowledge of the immersed geometry's position, compared to a model which does not require an explicit lubrication model.

This paper presents an enhanced immersed interface method designed for simulating near-contact interactions between two discrete $C^0$ interfaces by imposing jump conditions at a neighboring interface in the spreading and interpolation operators. 
We choose to represent our surface geometries using a piecewise linear discretization to match the engineering standard for meshing complex geometries. This choice of geometry inherently limits the order of jump conditions, but as we show herein, we achieve substantial improvements in the Eulerian velocity field using only the lowest order jump conditions. 
Our novel bilinear velocity interpolation operator serves to maintain excellent accuracy in the Eulerian velocity field in which the separation between two interfaces is less than the background grid spacing. 
When two distinct interfaces lie within an interpolation stencil, the operator incorporates weighted jump conditions from both interfaces to capture discontinuities in the velocity gradient. This operator implicitly assumes that the fluid has a linear profile in gap, which is reasonable in light of the velocity fields' asymptotic limits. 
Under grid refinement, this interpolation operator converges to the modified velocity interpolation operator by Kolahdouz et al\cite{kolahdouz_immersed_2020}, which achieved global second order and local almost second order convergence in the Eulerian velocity field. Numerical experiments demonstrate notable improvements in velocity accuracy compared to prior lubrication theory approaches and show strong agreement with analytical solutions.

The numerical challenges of near contact naturally extend to geometries with sharp features, where two discrete elements on the same interface may lie within an interpolation or spreading stencil. At the discrete level, the problem of handling near contact and sharp corners is equivalent.
Therefore, the methodology developed here for resolving distinct proximal interfaces naturally extends to these sharp geometries.
We impose jump conditions on the proximal interfacial element in both the interpolation and spreading operators. We demonstrate that this approach yields substantial accuracy improvements in these circumstances, serving as a viable alternative to the discontinuous Galerkin (DG) representation of jump conditions employed by Facci et al. \cite{faccisharp}.

The primary contribution of this work is to establish that the enhanced IIM provides a robust and accurate framework in two spatial dimensions for simulating fluid flows in thin fluid layers between multiple immersed interfaces, and flows around geometries with sharp corners. It captures interface and grid velocities accurately for separations as small as one-fiftieth of a grid cell without requiring prior knowledge of interface orientation or separation distance.

\section{Continuous Equations of Motion}

This section outlines the equations of motion for a viscous incompressible fluid with interfacial forces concentrated on an immersed boundary. This boundary both applies force to the fluid and moves with the local fluid velocity. The fluid domain, $\Omega $, is partitioned
into two subregions, $\Omega_t^+$ (exterior) and $\Omega_t^-$ (interior), by an
interface $\Gamma_t = \overline{\Omega_t^+} \bigcap \overline{\Omega_t^-}$, as in Figure 1.  We use the initial coordinates $\mathbf{X}$ of the boundary as Lagrangian coordinates, so that $\mathbf{X} \in \Gamma_0$, with corresponding current coordinates $\boldsymbol{\chi}(\mathbf{X},t) \in \Gamma_t$. 
The jump conditions express the discontinuity in fluid traction generated by the singular force along the interface. 
For a general function $\phi(\mathbf{x},t)$ and position $\mathbf{x}= \mathbf{\boldsymbol{\chi}}
  (\mathbf{X}, t)$ along the interface, the jump condition is
\begin{equation}
  \llbracket \phi (\mathbf{x}, t) \rrbracket \equiv \phi^+ (\mathbf{x}, t) - \phi^- (\mathbf{x}, t) \equiv \lim_{\epsilon
  \rightarrow 0^+} \phi (\mathbf{x} + \epsilon \mathbf{n}
  (\mathbf{X}, t), t) - \lim_{\varepsilon \rightarrow 0^-} \phi
  (\mathbf{x} - \varepsilon \mathbf{n} (\mathbf{X}, t), t).
\end{equation}
The equations of motion are
\begin{align}
  \rho \frac{\mathrm{D} \mathbf{u}(\mathbf{x},t)}{Dt}& = - \nabla p (\mathbf{x}, t) + \mu
  \nabla^2 \mathbf{u} (\mathbf{x}, t), & \mathbf{x} \in \Omega, \\
  \nabla \cdot \mathbf{u} (\mathbf{x}, t) & = 0, & \mathbf{x} \in
  \Omega, \\
  \llbracket \mathbf{u} (\mathbf{\boldsymbol{\chi}} (\mathbf{X}, t), t) \rrbracket &
  = 0,  &\mathbf{X} \in \Gamma_0,  \\
  \llbracket p (\mathbf{\boldsymbol{\chi}} (\mathbf{X}, t), t) \rrbracket & = -\jmath^{-
  1} (\mathbf{X}, t)\, \mathbf{F} (\mathbf{X}, t) \cdot \mathbf{n} (\mathbf{X}, t), & \mathbf{X} \in \Gamma_0, \\
  \mu\left\llbracket  \frac{\partial \mathbf{u} (\mathbf{\boldsymbol{\chi}}
  (\mathbf{X}, t), t)}{\partial \mathbf{x}_i} \right\rrbracket &
  =   (\mathbb{I}- \mathbf{n} (\mathbf{X}, t)
  \mathbf{n} (\mathbf{X}, t)\tran) \jmath^{- 1} (\mathbf{X}, t)\, \mathbf{F} (\mathbf{X}, t) n_i  (\mathbf{X}, t), &
  \mathbf{X} \in \Gamma_0, 
  \\  \mathbf{U} (\mathbf{X}, t)  &
  = \mathbf{u} (\boldsymbol{\chi}(\mathbf{X}, t), t),  &\mathbf{X} \in \Gamma_0, 
  \\ \frac{\partial \boldsymbol{\chi(\mathbf{X},t)}}{\partial t}     &
  = \mathbf{U}(\mathbf{X},t),  &\mathbf{X} \in \Gamma_0, 
\end{align}
in which $\rho$, $\mu$, $\mathbf{u} (\mathbf{x}, t)$, and $p
(\mathbf{x}, t)$, are the fluid's mass
density, dynamic viscosity, velocity, and pressure, and $t\in[0,t_\mathrm{end}]$. 
$\mathbf{F} (\mathbf{X}, t)$ and $\mathbf{U} (\mathbf{X}, t)$ are the interfacial force and velocity, respectively, in Lagrangian material coordinates, and $\mathbf{n} (\mathbf{X}, t)$ is the surface normal vector in the interface's current configuration.
 $\jmath^{- 1} =
\frac{{\mathrm{d}a}}{{\mathrm{d}A}}$ relates surface area along the interface in the current $(\mathrm{d}a)$ and reference $(\mathrm{d}A)$ configurations. For rigid body motions, $\jmath^{- 1} = 1$. 

Equations (4)--(6) were derived in previous work by Peskin and Printz\cite{PESKIN199333}, Lai and Li\cite{li_immersed_2001}, and  Xu and Wang\cite{xu_systematic_2006}. Peskin and Printz showed that the delta function formulation for singular forces reduces to these jump conditions across the interface. Thus, our continuous equations of motion do not explicitly contain the Eulerian forcing. The effects of the jump conditions, however, arise in the discrete equations of motion through correction terms that take the form of a discretized body force concentrated in the vicinity of the immersed boundary.
Our numerical tests consider cases with both stationary and moving rigid boundaries. Both cases involve forces that are computed using a penalty method. In either case, we view the boundary as having a \textit{prescribed} configuration $\boldsymbol{\xi} (\mathbf{X},t)$ and an \textit{actual} configuration $\mathbf{\boldsymbol{\chi}} (\mathbf{X}, t)$.
These two configurations are connected through penalty forces that are of the form

\begin{equation}
    \mathbf{F} (\mathbf{X}, t)  =  \kappa (\boldsymbol{\xi} (\mathbf{X},
    t) - \mathbf{\boldsymbol{\chi}} (\mathbf{X}, t)) + \eta \left( \mathbf{V}(\mathbf{X},
    t) - \mathbf{U}
    (\mathbf{X}, t) \right).
\end{equation}
in which $\kappa$ is the penalty spring stiffness and $\eta$ is the penalty spring damping. As $\kappa \rightarrow \infty$, this constraint exactly
imposes the motion\cite{IMFSI-griffith}. 
This study considers some stationary interfaces, so that $\boldsymbol{\chi}(\mathbf{X},t) = \mathbf{X}$.
The simplified force model becomes $\mathbf{F} (\mathbf{X}, t) = \kappa (\mathbf{X} - \mathbf{\boldsymbol{\chi}} (\mathbf{X}, t)) - \eta  \mathbf{U}
    (\mathbf{X}, t) $.
    Because the total interface displacement in our numerical experiments is between two to three orders of magnitude smaller than the domain length, we use $\mathbf{X}$ to mediate the coupling between the interface and the fluid without loss of accuracy. In a prior study, Hua and Peskin~\cite{HUA2022111435} used the reference configuration to mediate the fluid-interface coupling for an immersed boundary tethered by a spring force to its target position and demonstrated that it provides sufficient accuracy for their stability analyses.
\section{Discrete Equations of Motion}

This section describes the spatial and temporal discretizations of the equations of motion. It also details our modified velocity interpolation which accounts for multiple interfaces in one grid cell. We also describe the projection of the discrete jump conditions a continuous basis, and the discrete force spreading operator.
\subsection{Interface Representation}

We use a finite element representation of the immersed interface with
triangulation $\mathcal{T}_h$ of $\Gamma_0$, the reference configuration.
Consider elements $U_i$ such that $\mathcal{T}_h = \bigcup_i U_i$, with $i$ indexing
the mesh elements. The nodes of the mesh elements are $\{
\mathbf{X}_j \}_{j = 1}^M$ and have corresponding nodal (Lagrangian) basis functions $\{ \psi_j (\mathbf{X})
\}_{j = 1}^M$. Herein, we consider piecewise linear interface representations, so that $\psi_j$ is a piecewise linear Lagrange polynomial. The nodal basis functions $\psi_j(\mathbf{X})$ are continuous across elements but are not differentiable at the nodes. The current location of the interfacial
nodes at time $t$ are $\{ \mathbf{\boldsymbol{\chi}}_j (t) \}_{j = 1}^M$. In the finite
element space dictated by the subspace $S_h = \text{span} \{ \psi_j
(\mathbf{X}) \}_{j = 1}^M$, the configuration of the interface is
\begin{equation}
\mathbf{\boldsymbol{\chi}}_h (\mathbf{X}, t) = \sum_{j = 1}^M \mathbf{\boldsymbol{\chi}}_j (t)
  \psi_j (\mathbf{X}).
\end{equation}

\begin{figure}[t]
\center{}

\tikzset{every picture/.style={line width=0.75pt}} 

\begin{tikzpicture}[x=0.75pt,y=0.75pt,yscale=-1,xscale=1]

\draw  [draw opacity=0] (146,47) -- (385.82,47) -- (385.82,263.82) -- (146,263.82) -- cycle ; \draw  [color={rgb, 255:red, 155; green, 155; blue, 155 }  ,draw opacity=1 ] (146,47) -- (146,263.82)(166,47) -- (166,263.82)(186,47) -- (186,263.82)(206,47) -- (206,263.82)(226,47) -- (226,263.82)(246,47) -- (246,263.82)(266,47) -- (266,263.82)(286,47) -- (286,263.82)(306,47) -- (306,263.82)(326,47) -- (326,263.82)(346,47) -- (346,263.82)(366,47) -- (366,263.82) ; \draw  [color={rgb, 255:red, 155; green, 155; blue, 155 }  ,draw opacity=1 ] (146,47) -- (385.82,47)(146,67) -- (385.82,67)(146,87) -- (385.82,87)(146,107) -- (385.82,107)(146,127) -- (385.82,127)(146,147) -- (385.82,147)(146,167) -- (385.82,167)(146,187) -- (385.82,187)(146,207) -- (385.82,207)(146,227) -- (385.82,227)(146,247) -- (385.82,247) ; \draw  [color={rgb, 255:red, 155; green, 155; blue, 155 }  ,draw opacity=1 ]  ;
\draw  [line width=1.5]  (146,47) -- (385.82,47) -- (385.82,263.82) -- (146,263.82) -- cycle ;
\draw  [line width=1.5]  (193.57,162.82) .. controls (201.97,128.58) and (234.65,100.82) .. (266.57,100.82) .. controls (298.49,100.82) and (317.56,128.58) .. (309.16,162.82) .. controls (300.76,197.06) and (268.07,224.82) .. (236.16,224.82) .. controls (204.24,224.82) and (185.17,197.06) .. (193.57,162.82) -- cycle ;
\draw  [fill={rgb, 255:red, 0; green, 0; blue, 0 }  ,fill opacity=1 ] (199,141) .. controls (199,138.79) and (200.79,137) .. (203,137) .. controls (205.21,137) and (207,138.79) .. (207,141) .. controls (207,143.21) and (205.21,145) .. (203,145) .. controls (200.79,145) and (199,143.21) .. (199,141) -- cycle ;
\draw  [fill={rgb, 255:red, 0; green, 0; blue, 0 }  ,fill opacity=1 ] (211,125) .. controls (211,122.79) and (212.79,121) .. (215,121) .. controls (217.21,121) and (219,122.79) .. (219,125) .. controls (219,127.21) and (217.21,129) .. (215,129) .. controls (212.79,129) and (211,127.21) .. (211,125) -- cycle ;
\draw  [fill={rgb, 255:red, 0; green, 0; blue, 0 }  ,fill opacity=1 ] (227,113) .. controls (227,110.79) and (228.79,109) .. (231,109) .. controls (233.21,109) and (235,110.79) .. (235,113) .. controls (235,115.21) and (233.21,117) .. (231,117) .. controls (228.79,117) and (227,115.21) .. (227,113) -- cycle ;
\draw  [fill={rgb, 255:red, 0; green, 0; blue, 0 }  ,fill opacity=1 ] (190.57,160.82) .. controls (190.57,158.61) and (192.36,156.82) .. (194.57,156.82) .. controls (196.78,156.82) and (198.57,158.61) .. (198.57,160.82) .. controls (198.57,163.03) and (196.78,164.82) .. (194.57,164.82) .. controls (192.36,164.82) and (190.57,163.03) .. (190.57,160.82) -- cycle ;
\draw  [fill={rgb, 255:red, 0; green, 0; blue, 0 }  ,fill opacity=1 ] (187.57,181.82) .. controls (187.57,179.61) and (189.36,177.82) .. (191.57,177.82) .. controls (193.78,177.82) and (195.57,179.61) .. (195.57,181.82) .. controls (195.57,184.03) and (193.78,185.82) .. (191.57,185.82) .. controls (189.36,185.82) and (187.57,184.03) .. (187.57,181.82) -- cycle ;
\draw  [fill={rgb, 255:red, 0; green, 0; blue, 0 }  ,fill opacity=1 ] (193.57,202.82) .. controls (193.57,200.61) and (195.36,198.82) .. (197.57,198.82) .. controls (199.78,198.82) and (201.57,200.61) .. (201.57,202.82) .. controls (201.57,205.03) and (199.78,206.82) .. (197.57,206.82) .. controls (195.36,206.82) and (193.57,205.03) .. (193.57,202.82) -- cycle ;
\draw  [fill={rgb, 255:red, 0; green, 0; blue, 0 }  ,fill opacity=1 ] (207.57,217.82) .. controls (207.57,215.61) and (209.36,213.82) .. (211.57,213.82) .. controls (213.78,213.82) and (215.57,215.61) .. (215.57,217.82) .. controls (215.57,220.03) and (213.78,221.82) .. (211.57,221.82) .. controls (209.36,221.82) and (207.57,220.03) .. (207.57,217.82) -- cycle ;
\draw  [fill={rgb, 255:red, 0; green, 0; blue, 0 }  ,fill opacity=1 ] (227.16,223.82) .. controls (227.16,221.61) and (228.95,219.82) .. (231.16,219.82) .. controls (233.37,219.82) and (235.16,221.61) .. (235.16,223.82) .. controls (235.16,226.03) and (233.37,227.82) .. (231.16,227.82) .. controls (228.95,227.82) and (227.16,226.03) .. (227.16,223.82) -- cycle ;
\draw  [fill={rgb, 255:red, 0; green, 0; blue, 0 }  ,fill opacity=1 ] (248.57,221.82) .. controls (248.57,219.61) and (250.36,217.82) .. (252.57,217.82) .. controls (254.78,217.82) and (256.57,219.61) .. (256.57,221.82) .. controls (256.57,224.03) and (254.78,225.82) .. (252.57,225.82) .. controls (250.36,225.82) and (248.57,224.03) .. (248.57,221.82) -- cycle ;
\draw  [fill={rgb, 255:red, 0; green, 0; blue, 0 }  ,fill opacity=1 ] (268.57,213.82) .. controls (268.57,211.61) and (270.36,209.82) .. (272.57,209.82) .. controls (274.78,209.82) and (276.57,211.61) .. (276.57,213.82) .. controls (276.57,216.03) and (274.78,217.82) .. (272.57,217.82) .. controls (270.36,217.82) and (268.57,216.03) .. (268.57,213.82) -- cycle ;
\draw  [fill={rgb, 255:red, 0; green, 0; blue, 0 }  ,fill opacity=1 ] (284.57,199.82) .. controls (284.57,197.61) and (286.36,195.82) .. (288.57,195.82) .. controls (290.78,195.82) and (292.57,197.61) .. (292.57,199.82) .. controls (292.57,202.03) and (290.78,203.82) .. (288.57,203.82) .. controls (286.36,203.82) and (284.57,202.03) .. (284.57,199.82) -- cycle ;
\draw  [fill={rgb, 255:red, 0; green, 0; blue, 0 }  ,fill opacity=1 ] (297.57,182.82) .. controls (297.57,180.61) and (299.36,178.82) .. (301.57,178.82) .. controls (303.78,178.82) and (305.57,180.61) .. (305.57,182.82) .. controls (305.57,185.03) and (303.78,186.82) .. (301.57,186.82) .. controls (299.36,186.82) and (297.57,185.03) .. (297.57,182.82) -- cycle ;
\draw  [fill={rgb, 255:red, 0; green, 0; blue, 0 }  ,fill opacity=1 ] (306,161) .. controls (306,158.79) and (307.79,157) .. (310,157) .. controls (312.21,157) and (314,158.79) .. (314,161) .. controls (314,163.21) and (312.21,165) .. (310,165) .. controls (307.79,165) and (306,163.21) .. (306,161) -- cycle ;
\draw  [fill={rgb, 255:red, 0; green, 0; blue, 0 }  ,fill opacity=1 ] (247,103) .. controls (247,100.79) and (248.79,99) .. (251,99) .. controls (253.21,99) and (255,100.79) .. (255,103) .. controls (255,105.21) and (253.21,107) .. (251,107) .. controls (248.79,107) and (247,105.21) .. (247,103) -- cycle ;
\draw  [fill={rgb, 255:red, 0; green, 0; blue, 0 }  ,fill opacity=1 ] (268.57,101.82) .. controls (268.57,99.61) and (270.36,97.82) .. (272.57,97.82) .. controls (274.78,97.82) and (276.57,99.61) .. (276.57,101.82) .. controls (276.57,104.03) and (274.78,105.82) .. (272.57,105.82) .. controls (270.36,105.82) and (268.57,104.03) .. (268.57,101.82) -- cycle ;
\draw  [fill={rgb, 255:red, 0; green, 0; blue, 0 }  ,fill opacity=1 ] (287,108) .. controls (287,105.79) and (288.79,104) .. (291,104) .. controls (293.21,104) and (295,105.79) .. (295,108) .. controls (295,110.21) and (293.21,112) .. (291,112) .. controls (288.79,112) and (287,110.21) .. (287,108) -- cycle ;
\draw  [fill={rgb, 255:red, 0; green, 0; blue, 0 }  ,fill opacity=1 ] (301,121) .. controls (301,118.79) and (302.79,117) .. (305,117) .. controls (307.21,117) and (309,118.79) .. (309,121) .. controls (309,123.21) and (307.21,125) .. (305,125) .. controls (302.79,125) and (301,123.21) .. (301,121) -- cycle ;
\draw  [fill={rgb, 255:red, 0; green, 0; blue, 0 }  ,fill opacity=1 ] (306.57,139.82) .. controls (306.57,137.61) and (308.36,135.82) .. (310.57,135.82) .. controls (312.78,135.82) and (314.57,137.61) .. (314.57,139.82) .. controls (314.57,142.03) and (312.78,143.82) .. (310.57,143.82) .. controls (308.36,143.82) and (306.57,142.03) .. (306.57,139.82) -- cycle ;
\draw    (337.82,205.82) -- (309.1,187.44) ;
\draw [shift={(306.57,185.82)}, rotate = 32.62] [fill={rgb, 255:red, 0; green, 0; blue, 0 }  ][line width=0.08]  [draw opacity=0] (8.93,-4.29) -- (0,0) -- (8.93,4.29) -- cycle    ;
\draw    (229,72) .. controls (219.38,74.66) and (225.1,89.99) .. (235.89,102.64) ;
\draw [shift={(237.82,104.82)}, rotate = 227.29] [fill={rgb, 255:red, 0; green, 0; blue, 0 }  ][line width=0.08]  [draw opacity=0] (8.93,-4.29) -- (0,0) -- (8.93,4.29) -- cycle    ;

\draw (155.36,58.4) node [anchor=north west][inner sep=0.75pt]  [font=\large]  {$\Omega _{t}^{+}$};
\draw (236.36,144.4) node [anchor=north west][inner sep=0.75pt]  [font=\large]  {$\Omega _{t}^{-}$};
\draw (312,207.4) node [anchor=north west][inner sep=0.75pt]  [font=\large]  {$\chi (\mathbf{X} ,t)$};
\draw (231,54.4) node [anchor=north west][inner sep=0.75pt]  [font=\normalsize]  {$ \begin{array}{l}
\Gamma _{t} =\ \overline{\Omega _{t}^{+}}\bigcap \overline{\Omega _{t}^{-}}\\
\end{array}$};

\end{tikzpicture}

  \caption{A Lagrangian coordinate system represents represents the interface, $\Gamma_t$. At time $t$, the position of the interface in Eulerian coordinates is $\boldsymbol{\chi}(\mathbf{X},t)$.}
  \label{10}
\end{figure}

\subsection{Modified Velocity Interpolation}
The configuration of the Lagrangian mesh is updated by applying the no slip condition $\mathbf{U}_h = \mathcal{I}[{\boldsymbol{X},\mathbf{F}} ](\mathbf{u})$, in which $\mathcal{I}$ is a modified bilinear interpolation in two spatial dimensions and the subsequent
$L^2 $ projection of $\mathbf{U}_h$ into the subspace $S_h$. For simplicity of notation, the subscript $h$ is omitted in this section.

\begin{figure}[H]
\center{}
  \resizebox{230pt}{!}{
\tikzset{every picture/.style={line width=0.75pt}} 

\begin{tikzpicture}[x=0.75pt,y=0.75pt,yscale=-1,xscale=1]

\draw   (169,129.74) -- (347.76,129.74) -- (347.76,295) -- (169,295) -- cycle ;
\draw  [color={rgb, 255:red, 0; green, 0; blue, 0 }  ,draw opacity=1 ][fill={rgb, 255:red, 74; green, 144; blue, 226 }  ,fill opacity=1 ] (163.87,129.74) .. controls (163.87,126.91) and (166.17,124.61) .. (169,124.61) .. controls (171.83,124.61) and (174.13,126.91) .. (174.13,129.74) .. controls (174.13,132.58) and (171.83,134.87) .. (169,134.87) .. controls (166.17,134.87) and (163.87,132.58) .. (163.87,129.74) -- cycle ;
\draw  [color={rgb, 255:red, 0; green, 0; blue, 0 }  ,draw opacity=1 ][fill={rgb, 255:red, 74; green, 144; blue, 226 }  ,fill opacity=1 ] (342.63,129.74) .. controls (342.63,126.91) and (344.93,124.61) .. (347.76,124.61) .. controls (350.59,124.61) and (352.89,126.91) .. (352.89,129.74) .. controls (352.89,132.58) and (350.59,134.87) .. (347.76,134.87) .. controls (344.93,134.87) and (342.63,132.58) .. (342.63,129.74) -- cycle ;
\draw  [color={rgb, 255:red, 0; green, 0; blue, 0 }  ,draw opacity=1 ][fill={rgb, 255:red, 74; green, 144; blue, 226 }  ,fill opacity=1 ] (342.63,295) .. controls (342.63,292.17) and (344.93,289.87) .. (347.76,289.87) .. controls (350.59,289.87) and (352.89,292.17) .. (352.89,295) .. controls (352.89,297.83) and (350.59,300.13) .. (347.76,300.13) .. controls (344.93,300.13) and (342.63,297.83) .. (342.63,295) -- cycle ;
\draw  [color={rgb, 255:red, 0; green, 0; blue, 0 }  ,draw opacity=1 ][fill={rgb, 255:red, 74; green, 144; blue, 226 }  ,fill opacity=1 ] (163.87,295) .. controls (163.87,292.17) and (166.17,289.87) .. (169,289.87) .. controls (171.83,289.87) and (174.13,292.17) .. (174.13,295) .. controls (174.13,297.83) and (171.83,300.13) .. (169,300.13) .. controls (166.17,300.13) and (163.87,297.83) .. (163.87,295) -- cycle ;
\draw  [dash pattern={on 0.84pt off 2.51pt}]  (168.76,191.08) -- (240.76,191.08) ;
\draw  [dash pattern={on 0.84pt off 2.51pt}]  (240.76,295.34) -- (240.76,191.08) ;
\draw    (170.76,321.34) -- (238.76,321.34) ;
\draw [shift={(241.76,321.34)}, rotate = 180] [fill={rgb, 255:red, 0; green, 0; blue, 0 }  ][line width=0.08]  [draw opacity=0] (8.93,-4.29) -- (0,0) -- (8.93,4.29) -- cycle    ;
\draw [shift={(167.76,321.34)}, rotate = 0] [fill={rgb, 255:red, 0; green, 0; blue, 0 }  ][line width=0.08]  [draw opacity=0] (8.93,-4.29) -- (0,0) -- (8.93,4.29) -- cycle    ;
\draw    (157,194) -- (157,293) ;
\draw [shift={(157,296)}, rotate = 270] [fill={rgb, 255:red, 0; green, 0; blue, 0 }  ][line width=0.08]  [draw opacity=0] (8.93,-4.29) -- (0,0) -- (8.93,4.29) -- cycle    ;
\draw [shift={(157,191)}, rotate = 90] [fill={rgb, 255:red, 0; green, 0; blue, 0 }  ][line width=0.08]  [draw opacity=0] (8.93,-4.29) -- (0,0) -- (8.93,4.29) -- cycle    ;
\draw    (244.76,321.34) -- (345.76,321.34) ;
\draw [shift={(348.76,321.34)}, rotate = 180] [fill={rgb, 255:red, 0; green, 0; blue, 0 }  ][line width=0.08]  [draw opacity=0] (8.93,-4.29) -- (0,0) -- (8.93,4.29) -- cycle    ;
\draw [shift={(241.76,321.34)}, rotate = 0] [fill={rgb, 255:red, 0; green, 0; blue, 0 }  ][line width=0.08]  [draw opacity=0] (8.93,-4.29) -- (0,0) -- (8.93,4.29) -- cycle    ;
\draw    (157,133.34) -- (157,188) ;
\draw [shift={(157,191)}, rotate = 270] [fill={rgb, 255:red, 0; green, 0; blue, 0 }  ][line width=0.08]  [draw opacity=0] (8.93,-4.29) -- (0,0) -- (8.93,4.29) -- cycle    ;
\draw [shift={(157,130.34)}, rotate = 90] [fill={rgb, 255:red, 0; green, 0; blue, 0 }  ][line width=0.08]  [draw opacity=0] (8.93,-4.29) -- (0,0) -- (8.93,4.29) -- cycle    ;
\draw [color={rgb, 255:red, 245; green, 166; blue, 35 }  ,draw opacity=1 ][line width=1.5]    (339.76,101.58) -- (141.76,280.58) ;
\draw  [color={rgb, 255:red, 0; green, 0; blue, 0 }  ,draw opacity=1 ][fill={rgb, 255:red, 208; green, 2; blue, 27 }  ,fill opacity=1 ] (235.63,191.08) .. controls (235.63,188.25) and (237.93,185.95) .. (240.76,185.95) .. controls (243.59,185.95) and (245.89,188.25) .. (245.89,191.08) .. controls (245.89,193.91) and (243.59,196.21) .. (240.76,196.21) .. controls (237.93,196.21) and (235.63,193.91) .. (235.63,191.08) -- cycle ;

\draw (171,298.4) node [anchor=north west][inner sep=0.75pt]    {$x_{1}$};
\draw (349.76,298.4) node [anchor=north west][inner sep=0.75pt]    {$x_{2}$};
\draw (349.76,133.14) node [anchor=north west][inner sep=0.75pt]    {$x_{3}$};
\draw (171,133.14) node [anchor=north west][inner sep=0.75pt]    {$x_{4}$};
\draw (295,173.4) node [anchor=north west][inner sep=0.75pt]  [font=\LARGE]  {$ \begin{array}{l}
\Omega _{t}^{+}\\
\end{array}$};
\draw (211,132.4) node [anchor=north west][inner sep=0.75pt]  [font=\LARGE]  {$ \begin{array}{l}
\Omega _{t}^{-}\\
\end{array}$};
\draw (200,328.4) node [anchor=north west][inner sep=0.75pt]    {$ph$};
\draw (267,327.4) node [anchor=north west][inner sep=0.75pt]    {$( 1-p) h$};
\draw (132,226.4) node [anchor=north west][inner sep=0.75pt]    {$qh$};
\draw (98.84,154.82) node [anchor=north west][inner sep=0.75pt]    {$( 1-q) h$};
\draw (242.76,194.48) node [anchor=north west][inner sep=0.75pt]    {$\boldsymbol{\alpha} _{k}$};
\draw (306,89.4) node [anchor=north west][inner sep=0.75pt]  [font=\large]  {$\Gamma _{t}^{1}$};
\end{tikzpicture}}

  \caption{Velocity interpolation from Cartesian grid velocities (blue) to quadrature point $\boldsymbol{\alpha}_k$ (red) on $\Gamma_t^1$. Velocities at $x_1$, $x_2$, and $x_3$ require correction terms to account for the discontinuity in the velocity gradient generated by the interface.}
\end{figure}
This section explores the modified interpolation step, as the $L^2$ projection is the same as used in Kolahdouz et al\cite{kolahdouz_immersed_2020}. 
Consider the bounding box of Eulerian nodes on which Cartesian velocities are evaluated. 
First, we identify all interfacial quadrature points, $\{\boldsymbol{\alpha}_k\}_{k=1}^{N}$, on $\Gamma_t^1$ that are located within the bounding box. 
We compute a modified bilinear interpolation of the corner velocities to $\boldsymbol{\alpha}_k$ using correction terms, as in Tan et al\cite{tan_immersed_2009} with the no slip condition applied.
For an isotropic grid spacing $h$, the interface velocity $\U(\boldsymbol{\alpha}_k)$ is
\begin{align}
\mathbf{U}(\boldsymbol{\alpha}_k) & =  (1-p)(1-q) (\mathbf{u}_1+ \mathbf{C}_1)  + p(1-q) (\mathbf{u}_2+ \mathbf{C}_2) + p q (\mathbf{u}_3+ \mathbf{C}_3) +  (1-p) q (\mathbf{u}_4 + \mathbf{C}_4) ,
\end{align}
in which 
\begin{align}
\C_1 &= 
\begin{cases}
    -h p\left\llbracket \dfrac{\partial \mathbf{u} (\boldsymbol{\alpha}_k, t)}{\partial x} \right\rrbracket 
    - h q \left\llbracket \dfrac{\partial \mathbf{u} (\boldsymbol{\alpha}_k, t)}{\partial y} \right\rrbracket, 
    & \boldsymbol{x} \in \Omega_t^+, \\
    0, & \boldsymbol{x} \in \Omega_t^-,
\end{cases} \\
\C_2 &= 
\begin{cases}
    h (1-p)\left\llbracket \dfrac{\partial \mathbf{u} (\boldsymbol{\alpha}_k, t)}{\partial x} \right\rrbracket 
    - h q \left\llbracket \dfrac{\partial \mathbf{u} (\boldsymbol{\alpha}_k, t)}{\partial y} \right\rrbracket, 
    & \boldsymbol{x} \in \Omega_t^+, \\
    0, & \boldsymbol{x} \in \Omega_t^-,
\end{cases} \\
\C_3 &= 
\begin{cases}
    h (1-p)\left\llbracket \dfrac{\partial \mathbf{u} (\boldsymbol{\alpha}_k, t)}{\partial x} \right\rrbracket 
    + h (1-q) \left\llbracket \dfrac{\partial \mathbf{u} (\boldsymbol{\alpha}_k, t)}{\partial y} \right\rrbracket, 
    & \boldsymbol{x} \in \Omega_t^+, \\
    0, & \boldsymbol{x} \in \Omega_t^-,
\end{cases} \\
\C_4 &= 
\begin{cases}
    -h p\left\llbracket \dfrac{\partial \mathbf{u} (\boldsymbol{\alpha}_k, t)}{\partial x} \right\rrbracket 
    + h (1-q) \left\llbracket \dfrac{\partial \mathbf{u} (\boldsymbol{\alpha}_k, t)}{\partial y} \right\rrbracket, 
    & \boldsymbol{x} \in \Omega_t^+, \\
    0, & \boldsymbol{x} \in \Omega_t^-,
\end{cases}
\end{align}
are correction terms account for the discontinuities in the fluid's shear stress generated by the interface. Figure 2 provides a schematic in which $x_1$, $x_2$, and $x_3$ are Cartesian velocity nodes that contribute to the correction terms. 
To account for an additional discontinuity in the shear stress generated by a second interface within the velocity interpolation stencil or a sharp corner on a single interface,
we cast rays, $\{\mathbf{r}_i\}_{i=1}^4$, from $\boldsymbol{\alpha}_k$ to each corner of the bounding box and locate any intersection points, $\tilde{\x}_i$, on the intersected element. 
Next, we evaluate $\left\llbracket  \dfrac{\partial \mathbf{u} (\tilde{\x}_i, t)}{\partial \mathbf{x}} \right\rrbracket$ by interpolating the jump condition from neighboring Lagrangian nodes on the element where $\tilde{\x}_i$ is located. Then, we add a second correction term to the modified bilinear interpolation:

\begin{equation}
\tilde{\C}_i= \begin{cases}
s(n^x_{1}, n^x_{2}, r_i^x)\, \xi_i \left\llbracket  \dfrac{\partial \mathbf{u} (\tilde{\x}_i, t)}{\partial x} \right\rrbracket + s(n^y_{1}, n^y_{2}, r_i^y)\, \zeta_i \left\llbracket  \dfrac{\partial \mathbf{u} (\tilde{\x}_i, t)}{\partial y} \right\rrbracket, & \text{if} \exists\, \tilde{\x}_i \in \Gamma_t^1 \cup \Gamma_t^2 ,\\
0, &\text{otherwise},
\end{cases}
\end{equation}
in which weights $\xi_i$ and $\zeta_i$ are the components of the unsigned distances from $\tilde{\x}_i$ to the corner of the bounding box, $r^x_i$ and $r^y_i$ are the components of $\mathbf{r}_i$, and 
\begin{equation}
s(n^j_{1}, n^j_{2}, r^j_i) = \begin{cases}
\mathrm{sign}(-n^j_{2}), &\text{if}\,\ n^j_{1}\cdot n^j_{2}>0,\\
\mathrm{sign}(r^j_i), &\text{if}\,\ n^j_{1}\cdot n^j_{2}<0 \land r^j_i\cdot n^j_{2}<0,\\
\mathrm{sign}(-r^j_i), &\text{if}\,\ n^j_{1}\cdot n^j_{2}<0 \land r^j_i\cdot n^j_{2}>0,\\
0, &\mathrm{otherwise,}
\end{cases}
\end{equation}
in which $n^j_1$ is the discrete normal vector at $\boldsymbol{\alpha}_k$, and $n^j_2$ is the discrete normal vector at $\tilde{\x}_i$ for the $j$th spatial dimension.
Figure 3 shows a schematic of the ray casting and jump condition weights $\xi_i$ and $\zeta_i$.

\begin{figure}[H]
\center{}
\resizebox{230pt}{!}{

\tikzset{every picture/.style={line width=0.75pt}} 

\begin{tikzpicture}[x=0.75pt,y=0.75pt,yscale=-1,xscale=1]

\draw  [dash pattern={on 4.5pt off 4.5pt}]  (536.18,116.69) -- (708.06,312.12) ;
\draw   (440.88,65.44) -- (708.06,65.44) -- (708.06,312.12) -- (440.88,312.12) -- cycle ;
\draw  [color={rgb, 255:red, 0; green, 0; blue, 0 }  ,draw opacity=1 ][fill={rgb, 255:red, 74; green, 144; blue, 226 }  ,fill opacity=1 ] (433.21,65.44) .. controls (433.21,61.21) and (436.65,57.78) .. (440.88,57.78) .. controls (445.11,57.78) and (448.55,61.21) .. (448.55,65.44) .. controls (448.55,69.67) and (445.11,73.1) .. (440.88,73.1) .. controls (436.65,73.1) and (433.21,69.67) .. (433.21,65.44) -- cycle ;
\draw  [color={rgb, 255:red, 0; green, 0; blue, 0 }  ,draw opacity=1 ][fill={rgb, 255:red, 74; green, 144; blue, 226 }  ,fill opacity=1 ] (700.39,65.44) .. controls (700.39,61.21) and (703.82,57.78) .. (708.06,57.78) .. controls (712.29,57.78) and (715.72,61.21) .. (715.72,65.44) .. controls (715.72,69.67) and (712.29,73.1) .. (708.06,73.1) .. controls (703.82,73.1) and (700.39,69.67) .. (700.39,65.44) -- cycle ;
\draw  [color={rgb, 255:red, 0; green, 0; blue, 0 }  ,draw opacity=1 ][fill={rgb, 255:red, 74; green, 144; blue, 226 }  ,fill opacity=1 ] (700.39,312.12) .. controls (700.39,307.89) and (703.82,304.46) .. (708.06,304.46) .. controls (712.29,304.46) and (715.72,307.89) .. (715.72,312.12) .. controls (715.72,316.35) and (712.29,319.77) .. (708.06,319.77) .. controls (703.82,319.77) and (700.39,316.35) .. (700.39,312.12) -- cycle ;
\draw  [color={rgb, 255:red, 0; green, 0; blue, 0 }  ,draw opacity=1 ][fill={rgb, 255:red, 74; green, 144; blue, 226 }  ,fill opacity=1 ] (433.21,312.12) .. controls (433.21,307.89) and (436.65,304.46) .. (440.88,304.46) .. controls (445.11,304.46) and (448.55,307.89) .. (448.55,312.12) .. controls (448.55,316.35) and (445.11,319.77) .. (440.88,319.77) .. controls (436.65,319.77) and (433.21,316.35) .. (433.21,312.12) -- cycle ;
\draw [color={rgb, 255:red, 245; green, 166; blue, 35 }  ,draw opacity=1 ][line width=1.5]    (602.53,42.55) -- (383.73,278.65) ;
\draw [color={rgb, 255:red, 144; green, 19; blue, 254 }  ,draw opacity=1 ][line width=1.5]    (759,92.42) -- (504,335.42) ;
\draw  [color={rgb, 255:red, 0; green, 0; blue, 0 }  ,draw opacity=1 ][fill={rgb, 255:red, 208; green, 2; blue, 27 }  ,fill opacity=1 ] (528.51,116.69) .. controls (528.51,112.47) and (531.94,109.04) .. (536.18,109.04) .. controls (540.41,109.04) and (543.84,112.47) .. (543.84,116.69) .. controls (543.84,120.92) and (540.41,124.35) .. (536.18,124.35) .. controls (531.94,124.35) and (528.51,120.92) .. (528.51,116.69) -- cycle ;
\draw  [color={rgb, 255:red, 0; green, 0; blue, 0 }  ,draw opacity=1 ][fill={rgb, 255:red, 144; green, 19; blue, 254 }  ,fill opacity=1 ][line width=1.5]  (619.83,218.58) .. controls (619.83,214.35) and (623.27,210.92) .. (627.5,210.92) .. controls (631.73,210.92) and (635.17,214.35) .. (635.17,218.58) .. controls (635.17,222.81) and (631.73,226.24) .. (627.5,226.24) .. controls (623.27,226.24) and (619.83,222.81) .. (619.83,218.58) -- cycle ;
\draw    (729.5,219.95) -- (729.5,308.95) ;
\draw [shift={(729.5,311.95)}, rotate = 270] [fill={rgb, 255:red, 0; green, 0; blue, 0 }  ][line width=0.08]  [draw opacity=0] (8.93,-4.29) -- (0,0) -- (8.93,4.29) -- cycle    ;
\draw [shift={(729.5,216.95)}, rotate = 90] [fill={rgb, 255:red, 0; green, 0; blue, 0 }  ][line width=0.08]  [draw opacity=0] (8.93,-4.29) -- (0,0) -- (8.93,4.29) -- cycle    ;
\draw    (700.5,329.58) -- (622.27,329.58) ;
\draw [shift={(619.27,329.58)}, rotate = 360] [fill={rgb, 255:red, 0; green, 0; blue, 0 }  ][line width=0.08]  [draw opacity=0] (8.93,-4.29) -- (0,0) -- (8.93,4.29) -- cycle    ;
\draw [shift={(703.5,329.58)}, rotate = 180] [fill={rgb, 255:red, 0; green, 0; blue, 0 }  ][line width=0.08]  [draw opacity=0] (8.93,-4.29) -- (0,0) -- (8.93,4.29) -- cycle    ;

\draw (450.77,324.12) node [anchor=north west][inner sep=0.75pt]  [font=\large]  {$x_{1}$};
\draw (717.94,324.12) node [anchor=north west][inner sep=0.75pt]  [font=\large]  {$x_{2}$};
\draw (717.94,77.45) node [anchor=north west][inner sep=0.75pt]  [font=\large]  {$x_{3}$};
\draw (450.77,77.45) node [anchor=north west][inner sep=0.75pt]  [font=\large]  {$x_{4}$};
\draw (556.73,106.38) node [anchor=north west][inner sep=0.75pt]  [font=\large]  {$\mathbf{\alpha }_{k}$};
\draw (457.1,204.77) node [anchor=north west][inner sep=0.75pt]  [font=\Large]  {$\Gamma _{t}^{1}$};
\draw (515.02,275.94) node [anchor=north west][inner sep=0.75pt]  [font=\Large]  {$\Gamma _{t}^{2}$};
\draw (600.83,155.46) node [anchor=north west][inner sep=0.75pt]  [font=\large]  {$\mathbf{r}_{2}$};
\draw (643.08,204.37) node [anchor=north west][inner sep=0.75pt]  [font=\large]  {$\tilde{x}_{2}$};
\draw (735,254.4) node [anchor=north west][inner sep=0.75pt]  [font=\large]  {$\zeta _{i}$};
\draw (655,337.4) node [anchor=north west][inner sep=0.75pt]  [font=\large]  {$\xi _{i}$};

\end{tikzpicture}}
\resizebox{210pt}{!}{

\tikzset{every picture/.style={line width=0.75pt}} 

\begin{tikzpicture}[x=0.75pt,y=0.75pt,yscale=-1,xscale=1]

\draw  [dash pattern={on 4.5pt off 4.5pt}]  (311.84,135.43) -- (189.81,356.15) ;
\draw   (189.81,81.33) -- (487.09,81.33) -- (487.09,356.15) -- (189.81,356.15) -- cycle ;
\draw  [color={rgb, 255:red, 0; green, 0; blue, 0 }  ,draw opacity=1 ][fill={rgb, 255:red, 74; green, 144; blue, 226 }  ,fill opacity=1 ] (181.28,81.33) .. controls (181.28,76.62) and (185.1,72.8) .. (189.81,72.8) .. controls (194.52,72.8) and (198.34,76.62) .. (198.34,81.33) .. controls (198.34,86.04) and (194.52,89.86) .. (189.81,89.86) .. controls (185.1,89.86) and (181.28,86.04) .. (181.28,81.33) -- cycle ;
\draw  [color={rgb, 255:red, 0; green, 0; blue, 0 }  ,draw opacity=1 ][fill={rgb, 255:red, 74; green, 144; blue, 226 }  ,fill opacity=1 ] (478.56,81.33) .. controls (478.56,76.62) and (482.38,72.8) .. (487.09,72.8) .. controls (491.8,72.8) and (495.62,76.62) .. (495.62,81.33) .. controls (495.62,86.04) and (491.8,89.86) .. (487.09,89.86) .. controls (482.38,89.86) and (478.56,86.04) .. (478.56,81.33) -- cycle ;
\draw  [color={rgb, 255:red, 0; green, 0; blue, 0 }  ,draw opacity=1 ][fill={rgb, 255:red, 74; green, 144; blue, 226 }  ,fill opacity=1 ] (478.56,356.15) .. controls (478.56,351.44) and (482.38,347.62) .. (487.09,347.62) .. controls (491.8,347.62) and (495.62,351.44) .. (495.62,356.15) .. controls (495.62,360.86) and (491.8,364.68) .. (487.09,364.68) .. controls (482.38,364.68) and (478.56,360.86) .. (478.56,356.15) -- cycle ;
\draw  [color={rgb, 255:red, 0; green, 0; blue, 0 }  ,draw opacity=1 ][fill={rgb, 255:red, 74; green, 144; blue, 226 }  ,fill opacity=1 ] (181.28,356.15) .. controls (181.28,351.44) and (185.1,347.62) .. (189.81,347.62) .. controls (194.52,347.62) and (198.34,351.44) .. (198.34,356.15) .. controls (198.34,360.86) and (194.52,364.68) .. (189.81,364.68) .. controls (185.1,364.68) and (181.28,360.86) .. (181.28,356.15) -- cycle ;
\draw [color={rgb, 255:red, 245; green, 166; blue, 35 }  ,draw opacity=1 ][line width=1.5]    (279.09,37.25) -- (331.09,199.25) ;
\draw  [color={rgb, 255:red, 0; green, 0; blue, 0 }  ,draw opacity=1 ][fill={rgb, 255:red, 208; green, 2; blue, 27 }  ,fill opacity=1 ] (303.31,135.43) .. controls (303.31,130.72) and (307.13,126.9) .. (311.84,126.9) .. controls (316.56,126.9) and (320.38,130.72) .. (320.38,135.43) .. controls (320.38,140.14) and (316.56,143.96) .. (311.84,143.96) .. controls (307.13,143.96) and (303.31,140.14) .. (303.31,135.43) -- cycle ;
\draw [color={rgb, 255:red, 245; green, 166; blue, 35 }  ,draw opacity=1 ][line width=1.5]    (331.09,199.25) -- (150.09,151.25) ;
\draw  [color={rgb, 255:red, 0; green, 0; blue, 0 }  ,draw opacity=1 ][fill={rgb, 255:red, 144; green, 19; blue, 254 }  ,fill opacity=1 ] (276.31,188.43) .. controls (276.31,183.72) and (280.13,179.9) .. (284.84,179.9) .. controls (289.56,179.9) and (293.38,183.72) .. (293.38,188.43) .. controls (293.38,193.14) and (289.56,196.96) .. (284.84,196.96) .. controls (280.13,196.96) and (276.31,193.14) .. (276.31,188.43) -- cycle ;
\draw [line width=0.75]    (169.09,189.25) -- (169.09,351.25) ;
\draw [shift={(169.09,354.25)}, rotate = 270] [fill={rgb, 255:red, 0; green, 0; blue, 0 }  ][line width=0.08]  [draw opacity=0] (8.93,-4.29) -- (0,0) -- (8.93,4.29) -- cycle    ;
\draw [shift={(169.09,186.25)}, rotate = 90] [fill={rgb, 255:red, 0; green, 0; blue, 0 }  ][line width=0.08]  [draw opacity=0] (8.93,-4.29) -- (0,0) -- (8.93,4.29) -- cycle    ;
\draw [line width=0.75]    (193.09,368.25) -- (287.09,368.25) ;
\draw [shift={(290.09,368.25)}, rotate = 180] [fill={rgb, 255:red, 0; green, 0; blue, 0 }  ][line width=0.08]  [draw opacity=0] (8.93,-4.29) -- (0,0) -- (8.93,4.29) -- cycle    ;
\draw [shift={(190.09,368.25)}, rotate = 0] [fill={rgb, 255:red, 0; green, 0; blue, 0 }  ][line width=0.08]  [draw opacity=0] (8.93,-4.29) -- (0,0) -- (8.93,4.29) -- cycle    ;

\draw (196.77,370) node [anchor=north west][inner sep=0.75pt]    {$x_{1}$};
\draw (494.05,364.11) node [anchor=north west][inner sep=0.75pt]    {$x_{2}$};
\draw (494.05,89.29) node [anchor=north west][inner sep=0.75pt]    {$x_{3}$};
\draw (196.77,89.29) node [anchor=north west][inner sep=0.75pt]    {$x_{4}$};
\draw (369.02,215.05) node [anchor=north west][inner sep=0.75pt]  [font=\Large,color={rgb, 255:red, 74; green, 74; blue, 74 }  ,opacity=1 ]  {$ \begin{array}{l}
\Omega _{t}^{+}\\
\end{array}$};
\draw (242.56,97.35) node [anchor=north west][inner sep=0.75pt]  [font=\Large,color={rgb, 255:red, 74; green, 74; blue, 74 }  ,opacity=1 ]  {$ \begin{array}{l}
\Omega _{t}^{-}\\
\end{array}$};
\draw (317.84,110.43) node [anchor=north west][inner sep=0.75pt]    {$\mathbf{\alpha }_{k}$};
\draw (297.6,46.27) node [anchor=north west][inner sep=0.75pt]  [font=\large]  {$\Gamma _{t}^{1}$};
\draw (256,249) node [anchor=north west][inner sep=0.75pt]    {$\mathbf{r}_{1}$};
\draw (138,262) node [anchor=north west][inner sep=0.75pt]  [font=\Large]  {$\zeta _{i}$};
\draw (240.09,368.25) node [anchor=north west][inner sep=0.75pt]  [font=\Large]  {$\xi _{i}$};
\draw (284.84,196.96) node [anchor=north west][inner sep=0.75pt]  [font=\large]  {$\tilde{x}_{1}$};

\end{tikzpicture}}

\caption{Schematics for velocity interpolation when ray $\mathbf{r}_i$ intersects a second element on a nearby interface within an interpolation stencil, or a neighboring element on the same interface forming a sharp corner. Weights $\xi$ and $\zeta$ are the $x$ and $y$ components of the unsigned distance from $\tilde{\x}_2$ to $\x_2$ (left) or $x_1$ (right). }
\end{figure}

Last, we compute the modified velocity interpolation at each $\boldsymbol{\alpha}_k$. The two correction modified interpolation scheme is

\begin{align}
\begin{split}
\mathbf{U}(\boldsymbol{\alpha}_k) & =  (1-p)(1-q) (\mathbf{u}_1 + \mathbf{C}_1 + \tilde{\mathbf{C}_1})  + p(1-q) (\mathbf{u}_2+ \mathbf{C}_2 + \tilde{\mathbf{C}_2})\\ &+ p q (\mathbf{u}_3+ \mathbf{C}_3 + \tilde{\mathbf{C}_3}) +  (1-p) q (\mathbf{u}_4 + \mathbf{C}_4 + \tilde{\mathbf{C}_4}). 
\end{split}
\end{align}
Figure 4 shows all of the elementary orientations in which two interfaces can arise in one bounding box. Other orientations can be achieved by $\frac{\pi}{2}$ rotations and swapping the positions of $\Gamma_t^1$ and $\Gamma_t^2$. The implemented algorithm handles each of these cases implicitly without prior information about the two interfaces' orientation or relative position.
It does not rely on a hard-coded list of configurations, but instead determines the appropriate intersection positions for an arbitrary pair of interfaces automatically.

\begin{figure}[H]
\center{}
\resizebox{450pt}{!}{

\tikzset{every picture/.style={line width=0.75pt}} 

\begin{tikzpicture}[x=0.75pt,y=0.75pt,yscale=-1,xscale=1]

\draw  [dash pattern={on 4.5pt off 4.5pt}]  (246.64,85.98) -- (346,56.32) ;
\draw   (191.56,56.32) -- (346,56.32) -- (346,199.09) -- (191.56,199.09) -- cycle ;
\draw  [color={rgb, 255:red, 0; green, 0; blue, 0 }  ,draw opacity=1 ][fill={rgb, 255:red, 74; green, 144; blue, 226 }  ,fill opacity=1 ] (187.12,56.32) .. controls (187.12,53.87) and (189.11,51.88) .. (191.56,51.88) .. controls (194,51.88) and (195.99,53.87) .. (195.99,56.32) .. controls (195.99,58.76) and (194,60.75) .. (191.56,60.75) .. controls (189.11,60.75) and (187.12,58.76) .. (187.12,56.32) -- cycle ;
\draw  [color={rgb, 255:red, 0; green, 0; blue, 0 }  ,draw opacity=1 ][fill={rgb, 255:red, 74; green, 144; blue, 226 }  ,fill opacity=1 ] (341.57,56.32) .. controls (341.57,53.87) and (343.55,51.88) .. (346,51.88) .. controls (348.45,51.88) and (350.43,53.87) .. (350.43,56.32) .. controls (350.43,58.76) and (348.45,60.75) .. (346,60.75) .. controls (343.55,60.75) and (341.57,58.76) .. (341.57,56.32) -- cycle ;
\draw  [color={rgb, 255:red, 0; green, 0; blue, 0 }  ,draw opacity=1 ][fill={rgb, 255:red, 74; green, 144; blue, 226 }  ,fill opacity=1 ] (341.57,199.09) .. controls (341.57,196.64) and (343.55,194.66) .. (346,194.66) .. controls (348.45,194.66) and (350.43,196.64) .. (350.43,199.09) .. controls (350.43,201.54) and (348.45,203.52) .. (346,203.52) .. controls (343.55,203.52) and (341.57,201.54) .. (341.57,199.09) -- cycle ;
\draw  [color={rgb, 255:red, 0; green, 0; blue, 0 }  ,draw opacity=1 ][fill={rgb, 255:red, 74; green, 144; blue, 226 }  ,fill opacity=1 ] (187.12,199.09) .. controls (187.12,196.64) and (189.11,194.66) .. (191.56,194.66) .. controls (194,194.66) and (195.99,196.64) .. (195.99,199.09) .. controls (195.99,201.54) and (194,203.52) .. (191.56,203.52) .. controls (189.11,203.52) and (187.12,201.54) .. (187.12,199.09) -- cycle ;
\draw [color={rgb, 255:red, 245; green, 166; blue, 35 }  ,draw opacity=1 ][line width=1.5]    (285,43.07) -- (158.52,179.72) ;
\draw [color={rgb, 255:red, 144; green, 19; blue, 254 }  ,draw opacity=1 ][line width=1.5]    (303.78,39.31) -- (364.53,177.71) ;
\draw  [color={rgb, 255:red, 0; green, 0; blue, 0 }  ,draw opacity=1 ][fill={rgb, 255:red, 208; green, 2; blue, 27 }  ,fill opacity=1 ] (242.21,85.98) .. controls (242.21,83.53) and (244.19,81.55) .. (246.64,81.55) .. controls (249.09,81.55) and (251.07,83.53) .. (251.07,85.98) .. controls (251.07,88.43) and (249.09,90.41) .. (246.64,90.41) .. controls (244.19,90.41) and (242.21,88.43) .. (242.21,85.98) -- cycle ;
\draw  [color={rgb, 255:red, 0; green, 0; blue, 0 }  ,draw opacity=1 ][fill={rgb, 255:red, 144; green, 19; blue, 254 }  ,fill opacity=1 ][line width=1.5]  (311.83,65.72) .. controls (311.83,63.27) and (313.81,61.29) .. (316.26,61.29) .. controls (318.71,61.29) and (320.69,63.27) .. (320.69,65.72) .. controls (320.69,68.17) and (318.71,70.15) .. (316.26,70.15) .. controls (313.81,70.15) and (311.83,68.17) .. (311.83,65.72) -- cycle ;
\draw  [dash pattern={on 4.5pt off 4.5pt}]  (472.51,85.73) -- (572.61,199.54) ;
\draw   (417.01,55.88) -- (572.61,55.88) -- (572.61,199.54) -- (417.01,199.54) -- cycle ;
\draw  [color={rgb, 255:red, 0; green, 0; blue, 0 }  ,draw opacity=1 ][fill={rgb, 255:red, 74; green, 144; blue, 226 }  ,fill opacity=1 ] (412.55,55.88) .. controls (412.55,53.42) and (414.55,51.42) .. (417.01,51.42) .. controls (419.48,51.42) and (421.48,53.42) .. (421.48,55.88) .. controls (421.48,58.35) and (419.48,60.34) .. (417.01,60.34) .. controls (414.55,60.34) and (412.55,58.35) .. (412.55,55.88) -- cycle ;
\draw  [color={rgb, 255:red, 0; green, 0; blue, 0 }  ,draw opacity=1 ][fill={rgb, 255:red, 74; green, 144; blue, 226 }  ,fill opacity=1 ] (568.14,55.88) .. controls (568.14,53.42) and (570.14,51.42) .. (572.61,51.42) .. controls (575.07,51.42) and (577.07,53.42) .. (577.07,55.88) .. controls (577.07,58.35) and (575.07,60.34) .. (572.61,60.34) .. controls (570.14,60.34) and (568.14,58.35) .. (568.14,55.88) -- cycle ;
\draw  [color={rgb, 255:red, 0; green, 0; blue, 0 }  ,draw opacity=1 ][fill={rgb, 255:red, 74; green, 144; blue, 226 }  ,fill opacity=1 ] (568.14,199.54) .. controls (568.14,197.08) and (570.14,195.08) .. (572.61,195.08) .. controls (575.07,195.08) and (577.07,197.08) .. (577.07,199.54) .. controls (577.07,202) and (575.07,204) .. (572.61,204) .. controls (570.14,204) and (568.14,202) .. (568.14,199.54) -- cycle ;
\draw  [color={rgb, 255:red, 0; green, 0; blue, 0 }  ,draw opacity=1 ][fill={rgb, 255:red, 74; green, 144; blue, 226 }  ,fill opacity=1 ] (412.55,199.54) .. controls (412.55,197.08) and (414.55,195.08) .. (417.01,195.08) .. controls (419.48,195.08) and (421.48,197.08) .. (421.48,199.54) .. controls (421.48,202) and (419.48,204) .. (417.01,204) .. controls (414.55,204) and (412.55,202) .. (412.55,199.54) -- cycle ;
\draw [color={rgb, 255:red, 245; green, 166; blue, 35 }  ,draw opacity=1 ][line width=1.5]    (511.15,42.55) -- (383.73,180.05) ;
\draw [color={rgb, 255:red, 144; green, 19; blue, 254 }  ,draw opacity=1 ][line width=1.5]    (597.04,138.93) -- (480.84,206.88) ;
\draw  [color={rgb, 255:red, 0; green, 0; blue, 0 }  ,draw opacity=1 ][fill={rgb, 255:red, 208; green, 2; blue, 27 }  ,fill opacity=1 ] (468.04,85.73) .. controls (468.04,83.27) and (470.04,81.27) .. (472.51,81.27) .. controls (474.98,81.27) and (476.98,83.27) .. (476.98,85.73) .. controls (476.98,88.19) and (474.98,90.19) .. (472.51,90.19) .. controls (470.04,90.19) and (468.04,88.19) .. (468.04,85.73) -- cycle ;
\draw  [color={rgb, 255:red, 0; green, 0; blue, 0 }  ,draw opacity=1 ][fill={rgb, 255:red, 144; green, 19; blue, 254 }  ,fill opacity=1 ][line width=1.5]  (541.32,168.85) .. controls (541.32,166.38) and (543.32,164.39) .. (545.79,164.39) .. controls (548.25,164.39) and (550.25,166.38) .. (550.25,168.85) .. controls (550.25,171.31) and (548.25,173.31) .. (545.79,173.31) .. controls (543.32,173.31) and (541.32,171.31) .. (541.32,168.85) -- cycle ;
\draw  [dash pattern={on 4.5pt off 4.5pt}]  (704.05,85.45) -- (648.32,199.88) ;
\draw   (648.32,55.44) -- (804.57,55.44) -- (804.57,199.88) -- (648.32,199.88) -- cycle ;
\draw  [color={rgb, 255:red, 0; green, 0; blue, 0 }  ,draw opacity=1 ][fill={rgb, 255:red, 74; green, 144; blue, 226 }  ,fill opacity=1 ] (643.84,55.44) .. controls (643.84,52.97) and (645.85,50.96) .. (648.32,50.96) .. controls (650.8,50.96) and (652.81,52.97) .. (652.81,55.44) .. controls (652.81,57.92) and (650.8,59.93) .. (648.32,59.93) .. controls (645.85,59.93) and (643.84,57.92) .. (643.84,55.44) -- cycle ;
\draw  [color={rgb, 255:red, 0; green, 0; blue, 0 }  ,draw opacity=1 ][fill={rgb, 255:red, 74; green, 144; blue, 226 }  ,fill opacity=1 ] (800.08,55.44) .. controls (800.08,52.97) and (802.09,50.96) .. (804.57,50.96) .. controls (807.04,50.96) and (809.05,52.97) .. (809.05,55.44) .. controls (809.05,57.92) and (807.04,59.93) .. (804.57,59.93) .. controls (802.09,59.93) and (800.08,57.92) .. (800.08,55.44) -- cycle ;
\draw  [color={rgb, 255:red, 0; green, 0; blue, 0 }  ,draw opacity=1 ][fill={rgb, 255:red, 74; green, 144; blue, 226 }  ,fill opacity=1 ] (800.08,199.88) .. controls (800.08,197.41) and (802.09,195.4) .. (804.57,195.4) .. controls (807.04,195.4) and (809.05,197.41) .. (809.05,199.88) .. controls (809.05,202.36) and (807.04,204.37) .. (804.57,204.37) .. controls (802.09,204.37) and (800.08,202.36) .. (800.08,199.88) -- cycle ;
\draw  [color={rgb, 255:red, 0; green, 0; blue, 0 }  ,draw opacity=1 ][fill={rgb, 255:red, 74; green, 144; blue, 226 }  ,fill opacity=1 ] (643.84,199.88) .. controls (643.84,197.41) and (645.85,195.4) .. (648.32,195.4) .. controls (650.8,195.4) and (652.81,197.41) .. (652.81,199.88) .. controls (652.81,202.36) and (650.8,204.37) .. (648.32,204.37) .. controls (645.85,204.37) and (643.84,202.36) .. (643.84,199.88) -- cycle ;
\draw [color={rgb, 255:red, 245; green, 166; blue, 35 }  ,draw opacity=1 ][line width=1.5]    (758.14,42.17) -- (615.71,154.64) ;
\draw [color={rgb, 255:red, 144; green, 19; blue, 254 }  ,draw opacity=1 ][line width=1.5]    (773.38,204.57) -- (614.66,165.81) ;
\draw  [color={rgb, 255:red, 0; green, 0; blue, 0 }  ,draw opacity=1 ][fill={rgb, 255:red, 208; green, 2; blue, 27 }  ,fill opacity=1 ] (699.57,85.45) .. controls (699.57,82.98) and (701.58,80.97) .. (704.05,80.97) .. controls (706.53,80.97) and (708.54,82.98) .. (708.54,85.45) .. controls (708.54,87.93) and (706.53,89.94) .. (704.05,89.94) .. controls (701.58,89.94) and (699.57,87.93) .. (699.57,85.45) -- cycle ;
\draw  [color={rgb, 255:red, 0; green, 0; blue, 0 }  ,draw opacity=1 ][fill={rgb, 255:red, 144; green, 19; blue, 254 }  ,fill opacity=1 ][line width=1.5]  (655.42,176.91) .. controls (655.42,174.43) and (657.43,172.42) .. (659.9,172.42) .. controls (662.38,172.42) and (664.39,174.43) .. (664.39,176.91) .. controls (664.39,179.38) and (662.38,181.39) .. (659.9,181.39) .. controls (657.43,181.39) and (655.42,179.38) .. (655.42,176.91) -- cycle ;
\draw  [dash pattern={on 4.5pt off 4.5pt}]  (145.64,300.9) -- (245,414.01) ;
\draw  [dash pattern={on 4.5pt off 4.5pt}]  (145.64,300.9) -- (90.56,414.01) ;
\draw   (90.56,271.24) -- (245,271.24) -- (245,414.01) -- (90.56,414.01) -- cycle ;
\draw  [color={rgb, 255:red, 0; green, 0; blue, 0 }  ,draw opacity=1 ][fill={rgb, 255:red, 74; green, 144; blue, 226 }  ,fill opacity=1 ] (86.12,271.24) .. controls (86.12,268.79) and (88.11,266.8) .. (90.56,266.8) .. controls (93,266.8) and (94.99,268.79) .. (94.99,271.24) .. controls (94.99,273.68) and (93,275.67) .. (90.56,275.67) .. controls (88.11,275.67) and (86.12,273.68) .. (86.12,271.24) -- cycle ;
\draw  [color={rgb, 255:red, 0; green, 0; blue, 0 }  ,draw opacity=1 ][fill={rgb, 255:red, 74; green, 144; blue, 226 }  ,fill opacity=1 ] (240.57,271.24) .. controls (240.57,268.79) and (242.55,266.8) .. (245,266.8) .. controls (247.45,266.8) and (249.43,268.79) .. (249.43,271.24) .. controls (249.43,273.68) and (247.45,275.67) .. (245,275.67) .. controls (242.55,275.67) and (240.57,273.68) .. (240.57,271.24) -- cycle ;
\draw  [color={rgb, 255:red, 0; green, 0; blue, 0 }  ,draw opacity=1 ][fill={rgb, 255:red, 74; green, 144; blue, 226 }  ,fill opacity=1 ] (240.57,414.01) .. controls (240.57,411.56) and (242.55,409.58) .. (245,409.58) .. controls (247.45,409.58) and (249.43,411.56) .. (249.43,414.01) .. controls (249.43,416.46) and (247.45,418.44) .. (245,418.44) .. controls (242.55,418.44) and (240.57,416.46) .. (240.57,414.01) -- cycle ;
\draw  [color={rgb, 255:red, 0; green, 0; blue, 0 }  ,draw opacity=1 ][fill={rgb, 255:red, 74; green, 144; blue, 226 }  ,fill opacity=1 ] (86.12,414.01) .. controls (86.12,411.56) and (88.11,409.58) .. (90.56,409.58) .. controls (93,409.58) and (94.99,411.56) .. (94.99,414.01) .. controls (94.99,416.46) and (93,418.44) .. (90.56,418.44) .. controls (88.11,418.44) and (86.12,416.46) .. (86.12,414.01) -- cycle ;
\draw [color={rgb, 255:red, 245; green, 166; blue, 35 }  ,draw opacity=1 ][line width=1.5]    (199.11,258.12) -- (58.32,369.29) ;
\draw [color={rgb, 255:red, 144; green, 19; blue, 254 }  ,draw opacity=1 ][line width=1.5]    (267.81,354.85) -- (56.76,388.64) ;
\draw  [color={rgb, 255:red, 0; green, 0; blue, 0 }  ,draw opacity=1 ][fill={rgb, 255:red, 208; green, 2; blue, 27 }  ,fill opacity=1 ] (141.21,300.9) .. controls (141.21,298.45) and (143.19,296.47) .. (145.64,296.47) .. controls (148.09,296.47) and (150.07,298.45) .. (150.07,300.9) .. controls (150.07,303.35) and (148.09,305.33) .. (145.64,305.33) .. controls (143.19,305.33) and (141.21,303.35) .. (141.21,300.9) -- cycle ;
\draw  [color={rgb, 255:red, 0; green, 0; blue, 0 }  ,draw opacity=1 ][fill={rgb, 255:red, 144; green, 19; blue, 254 }  ,fill opacity=1 ][line width=1.5]  (102.77,380.39) .. controls (102.77,377.94) and (104.75,375.96) .. (107.2,375.96) .. controls (109.65,375.96) and (111.63,377.94) .. (111.63,380.39) .. controls (111.63,382.84) and (109.65,384.82) .. (107.2,384.82) .. controls (104.75,384.82) and (102.77,382.84) .. (102.77,380.39) -- cycle ;
\draw  [color={rgb, 255:red, 0; green, 0; blue, 0 }  ,draw opacity=1 ][fill={rgb, 255:red, 144; green, 19; blue, 254 }  ,fill opacity=1 ][line width=1.5]  (198.36,365.32) .. controls (198.36,362.88) and (200.34,360.89) .. (202.79,360.89) .. controls (205.24,360.89) and (207.22,362.88) .. (207.22,365.32) .. controls (207.22,367.77) and (205.24,369.75) .. (202.79,369.75) .. controls (200.34,369.75) and (198.36,367.77) .. (198.36,365.32) -- cycle ;
\draw  [dash pattern={on 4.5pt off 4.5pt}]  (373.71,300.9) -- (473.07,271.24) ;
\draw  [dash pattern={on 4.5pt off 4.5pt}]  (373.71,300.9) -- (473.07,414.01) ;
\draw   (318.63,271.24) -- (473.07,271.24) -- (473.07,414.01) -- (318.63,414.01) -- cycle ;
\draw  [color={rgb, 255:red, 0; green, 0; blue, 0 }  ,draw opacity=1 ][fill={rgb, 255:red, 74; green, 144; blue, 226 }  ,fill opacity=1 ] (314.19,271.24) .. controls (314.19,268.79) and (316.18,266.8) .. (318.63,266.8) .. controls (321.07,266.8) and (323.06,268.79) .. (323.06,271.24) .. controls (323.06,273.68) and (321.07,275.67) .. (318.63,275.67) .. controls (316.18,275.67) and (314.19,273.68) .. (314.19,271.24) -- cycle ;
\draw  [color={rgb, 255:red, 0; green, 0; blue, 0 }  ,draw opacity=1 ][fill={rgb, 255:red, 74; green, 144; blue, 226 }  ,fill opacity=1 ] (468.64,271.24) .. controls (468.64,268.79) and (470.62,266.8) .. (473.07,266.8) .. controls (475.52,266.8) and (477.5,268.79) .. (477.5,271.24) .. controls (477.5,273.68) and (475.52,275.67) .. (473.07,275.67) .. controls (470.62,275.67) and (468.64,273.68) .. (468.64,271.24) -- cycle ;
\draw  [color={rgb, 255:red, 0; green, 0; blue, 0 }  ,draw opacity=1 ][fill={rgb, 255:red, 74; green, 144; blue, 226 }  ,fill opacity=1 ] (468.64,414.01) .. controls (468.64,411.56) and (470.62,409.58) .. (473.07,409.58) .. controls (475.52,409.58) and (477.5,411.56) .. (477.5,414.01) .. controls (477.5,416.46) and (475.52,418.44) .. (473.07,418.44) .. controls (470.62,418.44) and (468.64,416.46) .. (468.64,414.01) -- cycle ;
\draw  [color={rgb, 255:red, 0; green, 0; blue, 0 }  ,draw opacity=1 ][fill={rgb, 255:red, 74; green, 144; blue, 226 }  ,fill opacity=1 ] (314.19,414.01) .. controls (314.19,411.56) and (316.18,409.58) .. (318.63,409.58) .. controls (321.07,409.58) and (323.06,411.56) .. (323.06,414.01) .. controls (323.06,416.46) and (321.07,418.44) .. (318.63,418.44) .. controls (316.18,418.44) and (314.19,416.46) .. (314.19,414.01) -- cycle ;
\draw [color={rgb, 255:red, 245; green, 166; blue, 35 }  ,draw opacity=1 ][line width=1.5]    (427.18,258.12) -- (286.39,369.29) ;
\draw [color={rgb, 255:red, 144; green, 19; blue, 254 }  ,draw opacity=1 ][line width=1.5]    (458.87,255.79) -- (400.16,426.71) ;
\draw  [color={rgb, 255:red, 0; green, 0; blue, 0 }  ,draw opacity=1 ][fill={rgb, 255:red, 208; green, 2; blue, 27 }  ,fill opacity=1 ] (369.28,300.9) .. controls (369.28,298.45) and (371.26,296.47) .. (373.71,296.47) .. controls (376.16,296.47) and (378.14,298.45) .. (378.14,300.9) .. controls (378.14,303.35) and (376.16,305.33) .. (373.71,305.33) .. controls (371.26,305.33) and (369.28,303.35) .. (369.28,300.9) -- cycle ;
\draw  [color={rgb, 255:red, 0; green, 0; blue, 0 }  ,draw opacity=1 ][fill={rgb, 255:red, 144; green, 19; blue, 254 }  ,fill opacity=1 ][line width=1.5]  (419.48,357.98) .. controls (419.48,355.53) and (421.46,353.54) .. (423.91,353.54) .. controls (426.36,353.54) and (428.34,355.53) .. (428.34,357.98) .. controls (428.34,360.42) and (426.36,362.41) .. (423.91,362.41) .. controls (421.46,362.41) and (419.48,360.42) .. (419.48,357.98) -- cycle ;
\draw  [color={rgb, 255:red, 0; green, 0; blue, 0 }  ,draw opacity=1 ][fill={rgb, 255:red, 144; green, 19; blue, 254 }  ,fill opacity=1 ][line width=1.5]  (448.25,278.04) .. controls (448.25,275.6) and (450.23,273.61) .. (452.68,273.61) .. controls (455.13,273.61) and (457.11,275.6) .. (457.11,278.04) .. controls (457.11,280.49) and (455.13,282.48) .. (452.68,282.48) .. controls (450.23,282.48) and (448.25,280.49) .. (448.25,278.04) -- cycle ;
\draw  [dash pattern={on 4.5pt off 4.5pt}]  (817.21,301.42) -- (916.57,414.53) ;
\draw  [dash pattern={on 4.5pt off 4.5pt}]  (817.21,301.42) -- (916.57,271.76) ;
\draw  [dash pattern={on 4.5pt off 4.5pt}]  (817.21,301.42) -- (762.12,414.53) ;
\draw   (762.12,271.76) -- (916.57,271.76) -- (916.57,414.53) -- (762.12,414.53) -- cycle ;
\draw  [color={rgb, 255:red, 0; green, 0; blue, 0 }  ,draw opacity=1 ][fill={rgb, 255:red, 74; green, 144; blue, 226 }  ,fill opacity=1 ] (757.69,271.76) .. controls (757.69,269.31) and (759.68,267.32) .. (762.12,267.32) .. controls (764.57,267.32) and (766.56,269.31) .. (766.56,271.76) .. controls (766.56,274.2) and (764.57,276.19) .. (762.12,276.19) .. controls (759.68,276.19) and (757.69,274.2) .. (757.69,271.76) -- cycle ;
\draw  [color={rgb, 255:red, 0; green, 0; blue, 0 }  ,draw opacity=1 ][fill={rgb, 255:red, 74; green, 144; blue, 226 }  ,fill opacity=1 ] (912.13,271.76) .. controls (912.13,269.31) and (914.12,267.32) .. (916.57,267.32) .. controls (919.01,267.32) and (921,269.31) .. (921,271.76) .. controls (921,274.2) and (919.01,276.19) .. (916.57,276.19) .. controls (914.12,276.19) and (912.13,274.2) .. (912.13,271.76) -- cycle ;
\draw  [color={rgb, 255:red, 0; green, 0; blue, 0 }  ,draw opacity=1 ][fill={rgb, 255:red, 74; green, 144; blue, 226 }  ,fill opacity=1 ] (912.13,414.53) .. controls (912.13,412.08) and (914.12,410.1) .. (916.57,410.1) .. controls (919.01,410.1) and (921,412.08) .. (921,414.53) .. controls (921,416.98) and (919.01,418.96) .. (916.57,418.96) .. controls (914.12,418.96) and (912.13,416.98) .. (912.13,414.53) -- cycle ;
\draw  [color={rgb, 255:red, 0; green, 0; blue, 0 }  ,draw opacity=1 ][fill={rgb, 255:red, 74; green, 144; blue, 226 }  ,fill opacity=1 ] (757.69,414.53) .. controls (757.69,412.08) and (759.68,410.1) .. (762.12,410.1) .. controls (764.57,410.1) and (766.56,412.08) .. (766.56,414.53) .. controls (766.56,416.98) and (764.57,418.96) .. (762.12,418.96) .. controls (759.68,418.96) and (757.69,416.98) .. (757.69,414.53) -- cycle ;
\draw [color={rgb, 255:red, 245; green, 166; blue, 35 }  ,draw opacity=1 ][line width=1.5]    (870.68,258.64) -- (729.89,369.81) ;
\draw [color={rgb, 255:red, 144; green, 19; blue, 254 }  ,draw opacity=1 ][line width=1.5]    (912.33,252.54) -- (736.12,416.78) ;
\draw  [color={rgb, 255:red, 0; green, 0; blue, 0 }  ,draw opacity=1 ][fill={rgb, 255:red, 208; green, 2; blue, 27 }  ,fill opacity=1 ] (812.78,301.42) .. controls (812.78,298.97) and (814.76,296.99) .. (817.21,296.99) .. controls (819.66,296.99) and (821.64,298.97) .. (821.64,301.42) .. controls (821.64,303.87) and (819.66,305.85) .. (817.21,305.85) .. controls (814.76,305.85) and (812.78,303.87) .. (812.78,301.42) -- cycle ;
\draw  [color={rgb, 255:red, 0; green, 0; blue, 0 }  ,draw opacity=1 ][fill={rgb, 255:red, 144; green, 19; blue, 254 }  ,fill opacity=1 ][line width=1.5]  (832.65,323.45) .. controls (832.65,321.01) and (834.63,319.02) .. (837.08,319.02) .. controls (839.53,319.02) and (841.51,321.01) .. (841.51,323.45) .. controls (841.51,325.9) and (839.53,327.89) .. (837.08,327.89) .. controls (834.63,327.89) and (832.65,325.9) .. (832.65,323.45) -- cycle ;
\draw  [color={rgb, 255:red, 0; green, 0; blue, 0 }  ,draw opacity=1 ][fill={rgb, 255:red, 144; green, 19; blue, 254 }  ,fill opacity=1 ][line width=1.5]  (875.86,283.47) .. controls (875.86,281.02) and (877.84,279.04) .. (880.29,279.04) .. controls (882.74,279.04) and (884.72,281.02) .. (884.72,283.47) .. controls (884.72,285.92) and (882.74,287.9) .. (880.29,287.9) .. controls (877.84,287.9) and (875.86,285.92) .. (875.86,283.47) -- cycle ;
\draw  [color={rgb, 255:red, 0; green, 0; blue, 0 }  ,draw opacity=1 ][fill={rgb, 255:red, 144; green, 19; blue, 254 }  ,fill opacity=1 ][line width=1.5]  (777.41,374.73) .. controls (777.41,372.28) and (779.4,370.3) .. (781.84,370.3) .. controls (784.29,370.3) and (786.28,372.28) .. (786.28,374.73) .. controls (786.28,377.18) and (784.29,379.16) .. (781.84,379.16) .. controls (779.4,379.16) and (777.41,377.18) .. (777.41,374.73) -- cycle ;
\draw  [dash pattern={on 4.5pt off 4.5pt}]  (596.64,328.9) -- (692,414.01) ;
\draw  [dash pattern={on 4.5pt off 4.5pt}]  (596.64,328.9) -- (537.56,414.01) ;
\draw   (537.56,271.24) -- (692,271.24) -- (692,414.01) -- (537.56,414.01) -- cycle ;
\draw  [color={rgb, 255:red, 0; green, 0; blue, 0 }  ,draw opacity=1 ][fill={rgb, 255:red, 74; green, 144; blue, 226 }  ,fill opacity=1 ] (533.12,271.24) .. controls (533.12,268.79) and (535.11,266.8) .. (537.56,266.8) .. controls (540,266.8) and (541.99,268.79) .. (541.99,271.24) .. controls (541.99,273.68) and (540,275.67) .. (537.56,275.67) .. controls (535.11,275.67) and (533.12,273.68) .. (533.12,271.24) -- cycle ;
\draw  [color={rgb, 255:red, 0; green, 0; blue, 0 }  ,draw opacity=1 ][fill={rgb, 255:red, 74; green, 144; blue, 226 }  ,fill opacity=1 ] (687.57,271.24) .. controls (687.57,268.79) and (689.55,266.8) .. (692,266.8) .. controls (694.45,266.8) and (696.43,268.79) .. (696.43,271.24) .. controls (696.43,273.68) and (694.45,275.67) .. (692,275.67) .. controls (689.55,275.67) and (687.57,273.68) .. (687.57,271.24) -- cycle ;
\draw  [color={rgb, 255:red, 0; green, 0; blue, 0 }  ,draw opacity=1 ][fill={rgb, 255:red, 74; green, 144; blue, 226 }  ,fill opacity=1 ] (687.57,414.01) .. controls (687.57,411.56) and (689.55,409.58) .. (692,409.58) .. controls (694.45,409.58) and (696.43,411.56) .. (696.43,414.01) .. controls (696.43,416.46) and (694.45,418.44) .. (692,418.44) .. controls (689.55,418.44) and (687.57,416.46) .. (687.57,414.01) -- cycle ;
\draw  [color={rgb, 255:red, 0; green, 0; blue, 0 }  ,draw opacity=1 ][fill={rgb, 255:red, 74; green, 144; blue, 226 }  ,fill opacity=1 ] (533.12,414.01) .. controls (533.12,411.56) and (535.11,409.58) .. (537.56,409.58) .. controls (540,409.58) and (541.99,411.56) .. (541.99,414.01) .. controls (541.99,416.46) and (540,418.44) .. (537.56,418.44) .. controls (535.11,418.44) and (533.12,416.46) .. (533.12,414.01) -- cycle ;
\draw [color={rgb, 255:red, 245; green, 166; blue, 35 }  ,draw opacity=1 ][line width=1.5]    (720.27,296.31) -- (505.27,354.31) ;
\draw [color={rgb, 255:red, 144; green, 19; blue, 254 }  ,draw opacity=1 ][line width=1.5]    (714.81,354.85) -- (503.76,388.64) ;
\draw  [color={rgb, 255:red, 0; green, 0; blue, 0 }  ,draw opacity=1 ][fill={rgb, 255:red, 208; green, 2; blue, 27 }  ,fill opacity=1 ] (592.21,328.9) .. controls (592.21,326.45) and (594.19,324.47) .. (596.64,324.47) .. controls (599.09,324.47) and (601.07,326.45) .. (601.07,328.9) .. controls (601.07,331.35) and (599.09,333.33) .. (596.64,333.33) .. controls (594.19,333.33) and (592.21,331.35) .. (592.21,328.9) -- cycle ;
\draw  [color={rgb, 255:red, 0; green, 0; blue, 0 }  ,draw opacity=1 ][fill={rgb, 255:red, 144; green, 19; blue, 254 }  ,fill opacity=1 ][line width=1.5]  (554.77,380.39) .. controls (554.77,377.94) and (556.75,375.96) .. (559.2,375.96) .. controls (561.65,375.96) and (563.63,377.94) .. (563.63,380.39) .. controls (563.63,382.84) and (561.65,384.82) .. (559.2,384.82) .. controls (556.75,384.82) and (554.77,382.84) .. (554.77,380.39) -- cycle ;
\draw  [color={rgb, 255:red, 0; green, 0; blue, 0 }  ,draw opacity=1 ][fill={rgb, 255:red, 144; green, 19; blue, 254 }  ,fill opacity=1 ][line width=1.5]  (637.36,366.32) .. controls (637.36,363.88) and (639.34,361.89) .. (641.79,361.89) .. controls (644.24,361.89) and (646.22,363.88) .. (646.22,366.32) .. controls (646.22,368.77) and (644.24,370.75) .. (641.79,370.75) .. controls (639.34,370.75) and (637.36,368.77) .. (637.36,366.32) -- cycle ;

\draw (193.56,202.49) node [anchor=north west][inner sep=0.75pt]    {$x_{1}$};
\draw (348,202.49) node [anchor=north west][inner sep=0.75pt]    {$x_{2}$};
\draw (348,59.72) node [anchor=north west][inner sep=0.75pt]    {$x_{3}$};
\draw (193.56,59.72) node [anchor=north west][inner sep=0.75pt]    {$x_{4}$};
\draw (248.64,89.38) node [anchor=north west][inner sep=0.75pt]    {$\mathbf{\alpha }_{k}$};
\draw (199.34,93.13) node [anchor=north west][inner sep=0.75pt]  [font=\large]  {$\Gamma _{t}^{1}$};
\draw (317.61,122.02) node [anchor=north west][inner sep=0.75pt]  [font=\large]  {$\Gamma _{t}^{2}$};
\draw (278.25,75.4) node [anchor=north west][inner sep=0.75pt]    {$\mathbf{r}$};
\draw (419.01,202.94) node [anchor=north west][inner sep=0.75pt]    {$x_{1}$};
\draw (574.61,202.94) node [anchor=north west][inner sep=0.75pt]    {$x_{2}$};
\draw (574.61,59.28) node [anchor=north west][inner sep=0.75pt]    {$x_{3}$};
\draw (419.01,59.28) node [anchor=north west][inner sep=0.75pt]    {$x_{4}$};
\draw (480.51,76.13) node [anchor=north west][inner sep=0.75pt]    {$\mathbf{\alpha }_{k}$};
\draw (421.65,140.12) node [anchor=north west][inner sep=0.75pt]  [font=\large]  {$\Gamma _{t}^{1}$};
\draw (483.34,170.51) node [anchor=north west][inner sep=0.75pt]  [font=\large]  {$\Gamma _{t}^{2}$};
\draw (507.66,105.13) node [anchor=north west][inner sep=0.75pt]    {$\mathbf{r}$};
\draw (650.32,203.28) node [anchor=north west][inner sep=0.75pt]    {$x_{1}$};
\draw (806.57,203.28) node [anchor=north west][inner sep=0.75pt]    {$x_{2}$};
\draw (806.57,58.84) node [anchor=north west][inner sep=0.75pt]    {$x_{3}$};
\draw (650.32,58.84) node [anchor=north west][inner sep=0.75pt]    {$x_{4}$};
\draw (706.05,88.85) node [anchor=north west][inner sep=0.75pt]    {$\mathbf{\alpha }_{k}$};
\draw (651.61,89.63) node [anchor=north west][inner sep=0.75pt]  [font=\large]  {$\Gamma _{t}^{1}$};
\draw (744.58,172.79) node [anchor=north west][inner sep=0.75pt]  [font=\large]  {$\Gamma _{t}^{2}$};
\draw (684.9,127.32) node [anchor=north west][inner sep=0.75pt]    {$\mathbf{r}$};
\draw (92.56,417.41) node [anchor=north west][inner sep=0.75pt]    {$x_{1}$};
\draw (247,417.41) node [anchor=north west][inner sep=0.75pt]    {$x_{2}$};
\draw (247,274.64) node [anchor=north west][inner sep=0.75pt]    {$x_{3}$};
\draw (92.56,274.64) node [anchor=north west][inner sep=0.75pt]    {$x_{4}$};
\draw (155.63,292.3) node [anchor=north west][inner sep=0.75pt]    {$\mathbf{\alpha }_{k}$};
\draw (95.69,301.53) node [anchor=north west][inner sep=0.75pt]  [font=\large]  {$\Gamma _{t}^{1}$};
\draw (216.06,336.85) node [anchor=north west][inner sep=0.75pt]  [font=\large]  {$\Gamma _{t}^{2}$};
\draw (127.67,340.39) node [anchor=north west][inner sep=0.75pt]    {$\mathbf{r}$};
\draw (181.65,324.85) node [anchor=north west][inner sep=0.75pt]    {$\mathbf{r}$};
\draw (320.63,417.41) node [anchor=north west][inner sep=0.75pt]    {$x_{1}$};
\draw (475.07,417.41) node [anchor=north west][inner sep=0.75pt]    {$x_{2}$};
\draw (475.07,274.64) node [anchor=north west][inner sep=0.75pt]    {$x_{3}$};
\draw (320.63,274.64) node [anchor=north west][inner sep=0.75pt]    {$x_{4}$};
\draw (363.18,305.9) node [anchor=north west][inner sep=0.75pt]    {$\mathbf{\alpha }_{k}$};
\draw (321.76,346.53) node [anchor=north west][inner sep=0.75pt]  [font=\large]  {$\Gamma _{t}^{1}$};
\draw (412.55,390.99) node [anchor=north west][inner sep=0.75pt]  [font=\large]  {$\Gamma _{t}^{2}$};
\draw (411.83,273.4) node [anchor=north west][inner sep=0.75pt]    {$\mathbf{r}$};
\draw (406.48,319.65) node [anchor=north west][inner sep=0.75pt]    {$\mathbf{r}$};
\draw (764.12,417.93) node [anchor=north west][inner sep=0.75pt]    {$x_{1}$};
\draw (918.57,417.93) node [anchor=north west][inner sep=0.75pt]    {$x_{2}$};
\draw (918.57,275.16) node [anchor=north west][inner sep=0.75pt]    {$x_{3}$};
\draw (764.12,275.16) node [anchor=north west][inner sep=0.75pt]    {$x_{4}$};
\draw (824.68,296.42) node [anchor=north west][inner sep=0.75pt]    {$\mathbf{\alpha }_{k}$};
\draw (764.23,300.05) node [anchor=north west][inner sep=0.75pt]  [font=\large]  {$\Gamma _{t}^{1}$};
\draw (818.93,341.13) node [anchor=north west][inner sep=0.75pt]  [font=\large]  {$\Gamma _{t}^{2}$};
\draw (855.12,271.13) node [anchor=north west][inner sep=0.75pt]    {$\mathbf{r}$};
\draw (874.26,350.44) node [anchor=north west][inner sep=0.75pt]    {$\mathbf{r}$};
\draw (788.15,328.25) node [anchor=north west][inner sep=0.75pt]    {$\mathbf{r}$};
\draw (168.32,24.4) node [anchor=north west][inner sep=0.75pt]    {$\mathbf{( a)}$};
\draw (391.32,26.4) node [anchor=north west][inner sep=0.75pt]    {$\mathbf{( b)}$};
\draw (619.32,26.4) node [anchor=north west][inner sep=0.75pt]    {$\mathbf{( c)}$};
\draw (63.32,244.16) node [anchor=north west][inner sep=0.75pt]    {$\mathbf{( d)}$};
\draw (297.32,244.16) node [anchor=north west][inner sep=0.75pt]    {$\mathbf{( e)}$};
\draw (741.32,245.16) node [anchor=north west][inner sep=0.75pt]    {$\mathbf{( g)}$};
\draw (539.56,417.41) node [anchor=north west][inner sep=0.75pt]    {$x_{1}$};
\draw (694,417.41) node [anchor=north west][inner sep=0.75pt]    {$x_{2}$};
\draw (694,274.64) node [anchor=north west][inner sep=0.75pt]    {$x_{3}$};
\draw (539.56,274.64) node [anchor=north west][inner sep=0.75pt]    {$x_{4}$};
\draw (584.63,302.3) node [anchor=north west][inner sep=0.75pt]    {$\mathbf{\alpha }_{k}$};
\draw (541.69,317.53) node [anchor=north west][inner sep=0.75pt]  [font=\large]  {$\Gamma _{t}^{1}$};
\draw (663.06,335.85) node [anchor=north west][inner sep=0.75pt]  [font=\large]  {$\Gamma _{t}^{2}$};
\draw (583.67,350.39) node [anchor=north west][inner sep=0.75pt]    {$\mathbf{r}$};
\draw (624.65,338.85) node [anchor=north west][inner sep=0.75pt]    {$\mathbf{r}$};
\draw (510.32,244.16) node [anchor=north west][inner sep=0.75pt]    {$\mathbf{( f)}$};
\draw (308,70.72) node [anchor=north west][inner sep=0.75pt]    {$\tilde{x}_{3}$};
\draw (538,171.72) node [anchor=north west][inner sep=0.75pt]    {$\tilde{x}_{2}$};
\draw (669.9,156.31) node [anchor=north west][inner sep=0.75pt]    {$\tilde{x}_{1}$};
\draw (198.36,370.72) node [anchor=north west][inner sep=0.75pt]    {$\tilde{x}_{2}$};
\draw (109.2,383.79) node [anchor=north west][inner sep=0.75pt]    {$\tilde{x}_{1}$};
\draw (433.52,342.65) node [anchor=north west][inner sep=0.75pt]    {$\tilde{x}_{2}$};
\draw (450.25,281.44) node [anchor=north west][inner sep=0.75pt]    {$\tilde{x}_{3}$};
\draw (561.2,383.79) node [anchor=north west][inner sep=0.75pt]    {$\tilde{x}_{1}$};
\draw (631.79,370.72) node [anchor=north west][inner sep=0.75pt]    {$\tilde{x}_{2}$};
\draw (783.84,378.13) node [anchor=north west][inner sep=0.75pt]    {$\tilde{x}_{1}$};
\draw (849.79,313.72) node [anchor=north west][inner sep=0.75pt]    {$\tilde{x}_{2}$};
\draw (882.29,286.87) node [anchor=north west][inner sep=0.75pt]    {$\tilde{x}_{3}$};

\end{tikzpicture}}

\caption{Velocity interpolation schematics are shown for various orientations of $\Gamma_t^1$ and $\Gamma_t^2$ when both interfaces lie within one grid cell. The intersection of ray $\mathbf{r}$ with $\Gamma_t^2$ is denoted by $\tilde{x}$. Panels (a)-(c) show the possible orientations when $\Gamma_t^1$ and $\Gamma_t^2$ each isolate one node. Panels (d) and (e) show the possible orientations when $\Gamma_t^1$ isolates one node and $\Gamma_t^2$ isolates two nodes. Panel (f) shows the orientation in which both $\Gamma_t^1$ and $\Gamma_t^2$ isolate two nodes. Panel (g) shows the orientation in which $\Gamma_t^1$ isolates one node and $\Gamma_t^2$ isolates three nodes. All other possible orientations can be achieved by rotating both interfaces and swapping $\Gamma_t^1$ with $\Gamma_t^2$. Each orientation is implicitly handled in our implemented algorithm.}
\end{figure}

Finally, these interpolated velocities are then used to compute the $L^2$ projection of the velocity at mesh nodes, as in Kolahdouz et al. These nodal velocities are represented in the finite element basis as $\mathbf{U}_h (\mathbf{X}, t)  = \sum_{j = 1}^M  \mathbf{U}_j(t) \,\psi_j (\mathbf{X})$.

For piecewise linear velocity fields with discontinuities in the velocity gradient along the interface, such as those found in linear shear flow, the interpolation operator exactly reconstructs the Lagrangian velocity. 
Consider a continuous piecewise linear velocity field with discontinuities in its derivative along an immersed interface. 
Let $\boldsymbol{\alpha}_k\in \Omega_0$ and $\mathbf{L}_0(\x) = \mathbf{A}_0\x +\mathbf{b}_0$ for $\x \in \Omega_0$, and denote the jump condition between domains $\Omega_n$ and $\Omega_{n-1}$ as $\left\llbracket \frac{\partial \mathbf{u}}{\partial \x}\right\rrbracket_n = \left.\frac{\partial \mathbf{u}}{\partial \x} \right|_{\Omega_n} - \left.\frac{\partial \mathbf{u}}{\partial \x} \right|_{\Omega_{n-1}}$. 
In the simplest case, where node $\x_i\in\Omega_0$, one may readily compute the nodal contributions to the velocity interpolation without correction terms:
\begin{equation}
\bu(\x_i) = \bu(\boldsymbol{\alpha}_k) + \left.\frac{\partial \mathbf{u}}{\partial \x} \right|_{\Omega_0}\cdot(\x_i-\boldsymbol{\alpha}_k).
\end{equation}
Then, $\bu(\x_i) = \mathbf{L}_0(\x)$ and the linear interpolation is exact.
However, when node $\x_i$ is separated from $\boldsymbol{\alpha}_k$ by two discontinuities in the velocity gradient at $\Gamma_t^1$ and $\Gamma_t^2$, ray $\mathbf{r}_i$ intersects $\Gamma_t^1$ at $\boldsymbol{\gamma}_1$ and intersects $\Gamma_t^2$ at $\boldsymbol{\gamma}_2$. The nodal contributions become
\begin{equation}
\bu(\x_i) = \bu(\boldsymbol{\alpha}_k) +\left.\frac{\partial \mathbf{u}}{\partial \x} \right|_{\Omega_0}\cdot(\boldsymbol{\gamma}_1 - \boldsymbol{\alpha}_k)
+ \left.\frac{\partial \mathbf{u}}{\partial \x} \right|_{\Omega_1}\cdot(\boldsymbol{\gamma}_2 - \boldsymbol{\gamma}_1)
+\left.\frac{\partial \mathbf{u}}{\partial \x} \right|_{\Omega_2}\cdot(\x_i - \boldsymbol{\gamma}_2).
\end{equation}
Substituting the jump conditions and simplifying terms yields
\begin{equation}
\bu(\x_i) = \mathbf{L}_0(\x_i) +\left\llbracket \frac{\partial \mathbf{u}}{\partial \x}\right\rrbracket_1 \cdot (\x_i - \boldsymbol{\gamma}_1) +\left\llbracket \frac{\partial \mathbf{u}}{\partial \x}\right\rrbracket_2 \cdot (\x_i - \boldsymbol{\gamma}_2) .
\end{equation}
The distance-weighted jump condition terms correspond to the correction terms $\mathbf{C}_i$ and $\tilde{\mathbf{C}}_i$. Thus, solving for $\mathbf{L}_0(\x_i)$, we arrive at
\begin{equation}
\mathbf{L}_0(\x_i) = \bu(\x_i) + \mathbf{C}_i +\tilde{\mathbf{C}}_i
\end{equation}
It follows that the standard bilinear interpolation exactly reconstructs the value of the linear function at $\boldsymbol{\alpha}_k$.

To demonstrate our two correction method's convergence properties under grid refinement, we present two tests. 
We investigate two parallel plates at various angles submerged in a piecewise linear and piecewise quadratic function. The linear function is
\begin{equation}
f_\mathrm{linear}(\beta) = \begin{cases}
-\beta - 0.01, &\text{if}\,\ \beta < -0.01,\\
\beta + 0.01, &\text{if}\,\ -0.01\leq \beta \leq 0,\\
-\beta + 0.01, &\text{if}\,\ \beta >0,
\end{cases}
\end{equation}
and the quadratic function is
\begin{equation}
f_\mathrm{quadratic}(\beta) = \begin{cases}
\beta^2-\beta - 0.02, &\text{if}\,\ \beta < -0.01,\\
\beta^2 + \beta, &\text{if}\,\ -0.01\leq \beta \leq 0,\\
\beta^2 - \beta, &\text{if}\,\ \beta >0,
\end{cases}
\end{equation}
in which $\beta$ is the local coordinate in the normal direction of the parallel plates.
Figure 5 shows a schematic of the orientations tested.
\begin{figure}[H]
    \centering
    \includegraphics[width=1.0\linewidth]{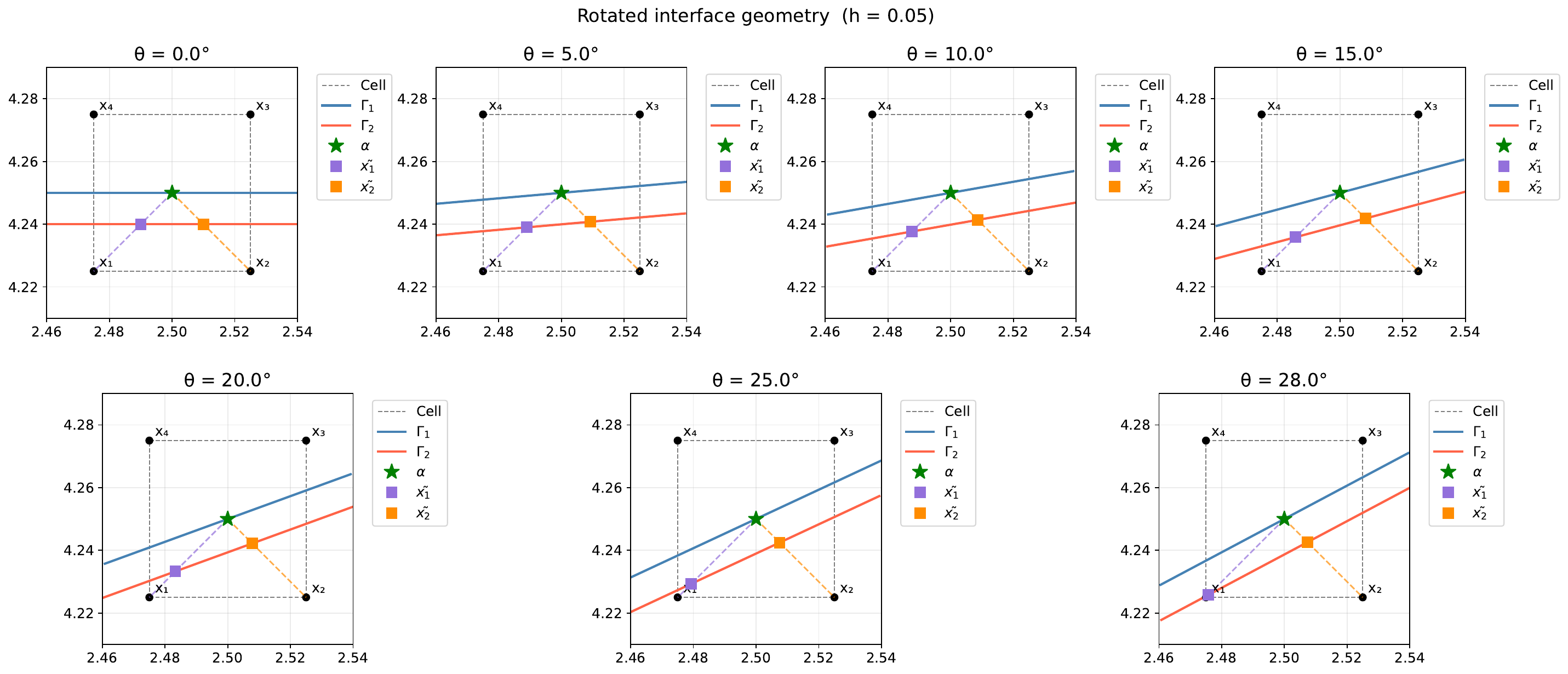}
    \caption{Parallel plates are visualized for various angles of rotation about $\boldsymbol\alpha = (2.5,4.25)$ for rotations of $\theta = 0$, $5$, $10$, $15$, $20$, $25$, and $28$ degrees. The magnitudes of rotation tested here are restricted to these angles because larger angles of rotation would change the number of intersections in a given cell. }
    \label{fig:geometry_angles}
\end{figure}

We analyze the accuracy of the reconstructed velocity at $\boldsymbol\alpha$ using one and two corrections. For $f_\mathrm{linear}$, the two correction method achieves machine precision error, whereas the one correction method exhibits approximately first order convergence. Figure 6 shows this grid convergence study for $h = 0.05$, $0.1$, $0.3$, $0.4$, $0.8$, $1.6$, and $3.2$.
\begin{figure}[H]
    \centering
    \includegraphics[width=1.0\linewidth]{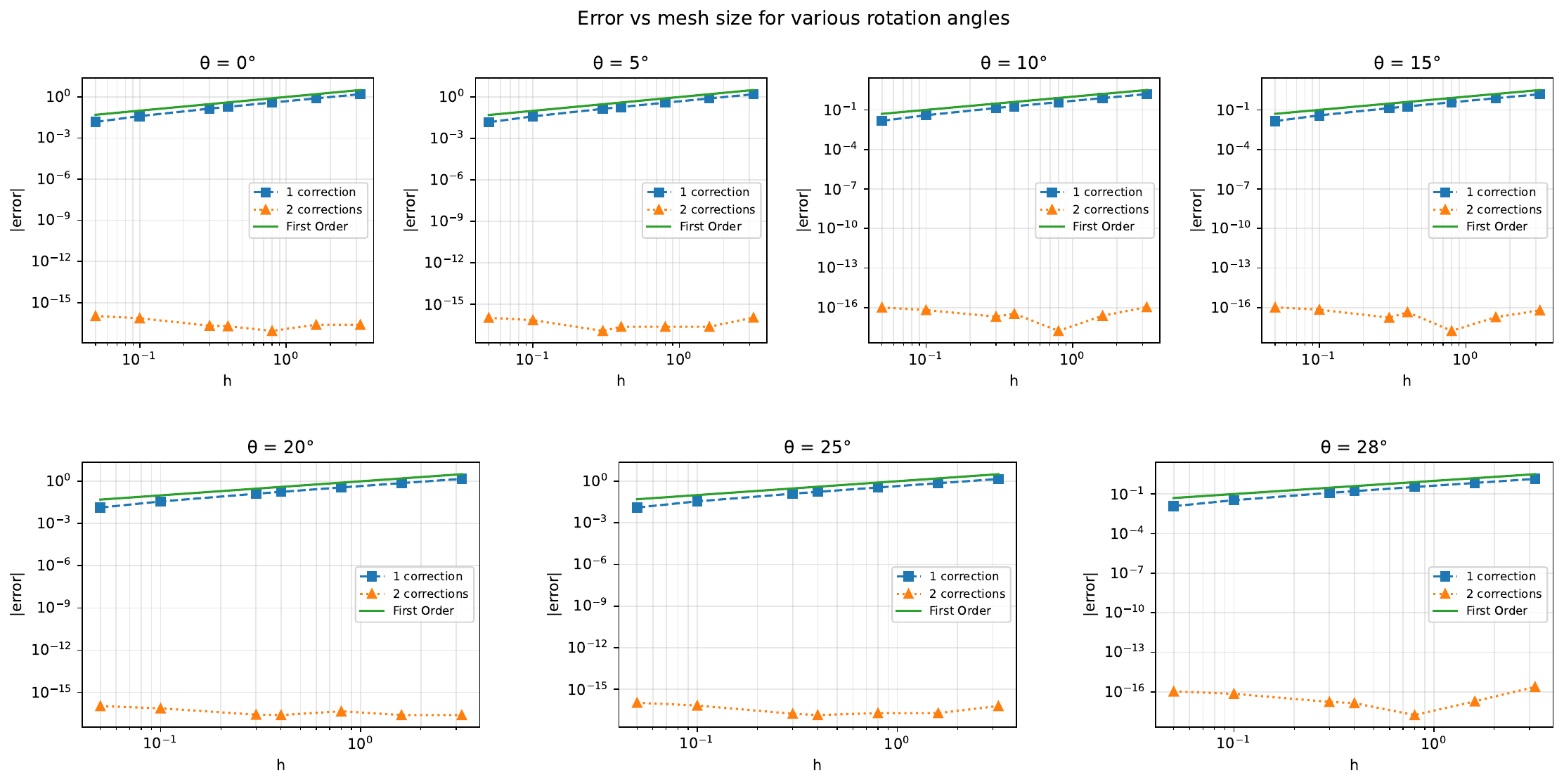}
    \caption{Grid convergence study of $f_\mathrm{linear}$ between two parallel plates at various angles $\theta = 0$, $5$, $10$, $15$, $20$, $25$, and $28$ degrees. Grid spacings $h = 0.05$, $0.1$, $0.3$, $0.4$, $0.8$, $1.6$, and $3.2$ are used. The two correction method is accurate to machine precision and the one correction methods demonstrates approximately linear convergence.}
    \label{fig:lin_f_conv}
\end{figure}

For $f_\mathrm{quadratic}$, the two correction method exhibits second order convergence, and the one correction method demonstrates approximately first order convergence. Figure 7 shows this grid convergence study for $h = 0.05$, $0.1$, $0.3$, $0.4$, $0.8$, $1.6$, and $3.2$.
\begin{figure}[H]
    \centering
    \includegraphics[width=1.0\linewidth]{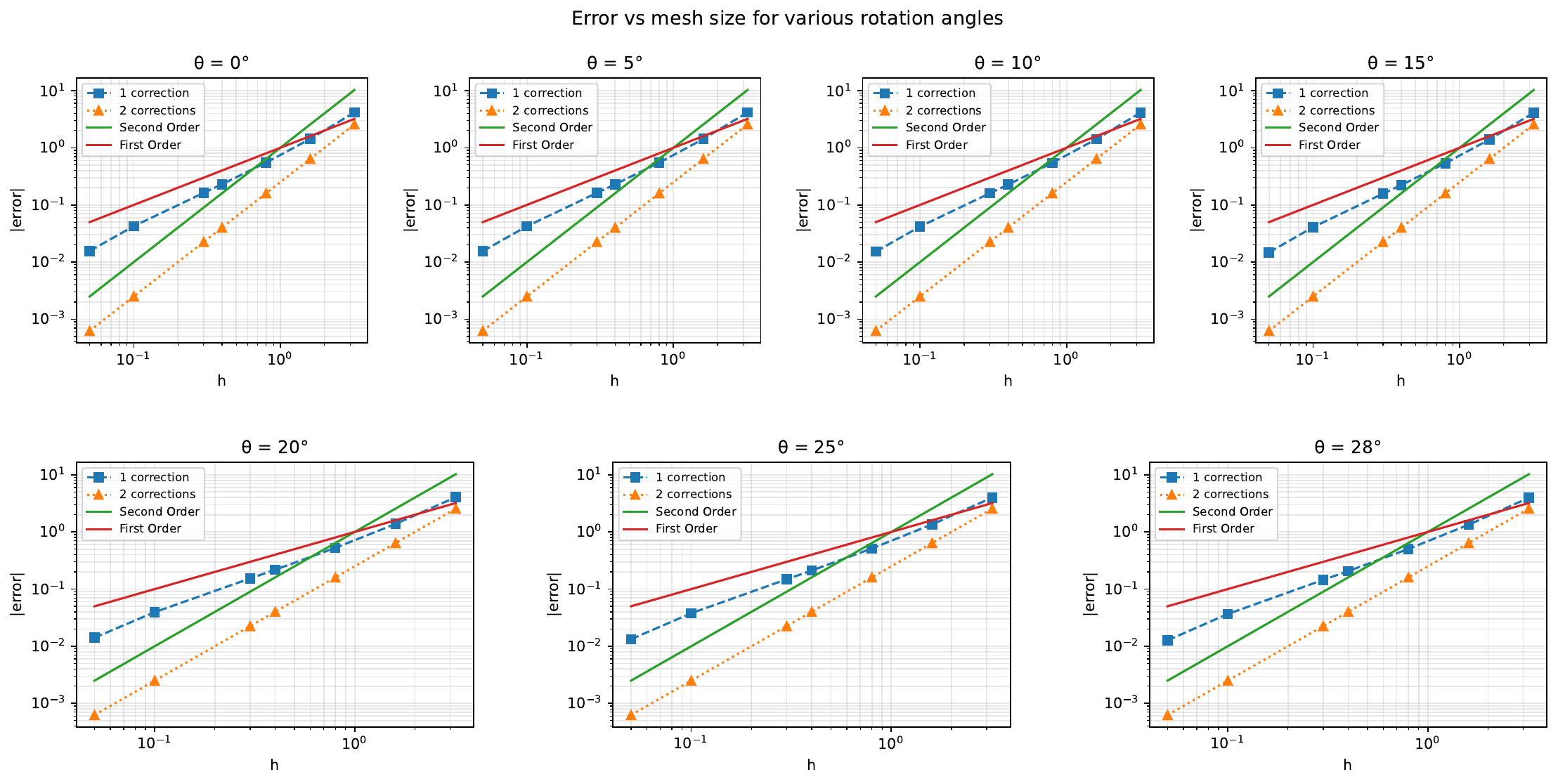}
    \caption{Grid convergence study of $f_\mathrm{quadratic}$ between two parallel plates at various angles $\theta = 0$, $5$, $10$, $15$, $20$, $25$, and $28$. Grid spacings $h = 0.05$, $0.1$, $0.3$, $0.4$, $0.8$, $1.6$, and $3.2$ are used. The two correction method exhibits second order convergence and the one correction methods demonstrates approximately linear convergence.}
    \label{fig:quad_f_conv}
\end{figure}

\subsection{Projected Jump Conditions}
Jump conditions must be evaluated at arbitrary positions along the interface to evaluate the forcing terms associated with stencils that cut the interface. If using the normal vectors determined directly from the $C^0$ interface representation, Eq. (10), pointwise expressions for the jump conditions for pressure, Eq. (3), and shear stress, Eq. (4), are discontinuous between elements. To obtain a continuous representation of these jump conditions along the discretized boundary geometry, previous work by Kolahdouz et al.\cite{kolahdouz_immersed_2020} projected both pressure and shear stress jump conditions onto the same finite element basis used to represent the geometry. For example, the pressure jump condition in the continuous Lagrangian basis is
\begin{equation}
\llbracket  p\rrbracket_h(\mathbf{X}, t)  = \sum_{j = 1}^M \llbracket  p\rrbracket_j(t) \,
  \psi_j (\mathbf{X}),
  \label{projjump}
\end{equation}
in which $\llbracket  p\rrbracket_j(t)$ is the jump condition at a particular Lagrangian nodes at time $t$.
Projecting the pressure jump condition requires $\llbracket  p\rrbracket_h(\mathbf{X}, t)$ to satisfy
\begin{equation}
\int_{\Gamma_0} \llbracket  p\rrbracket_h(\mathbf{X}, t) \, \psi_j(\mathbf{X}) \, {\mathrm d} A = \int_{\Gamma_0} \jmath^{- 1} (\mathbf{X}, t)\, \mathbf{F}_h (\mathbf{X},
  t) \cdot \mathbf{n} (\mathbf{X}, t) \, \psi_j(\mathbf{X}) {\mathrm d} A , \forall j = 1, \ldots, M.
\end{equation}
Models involving this jump condition projection are referred to as CG-IIM.

To mitigate the $O(1)$ errors that arise at sharp corners using continuous representations, Facci et al.~\cite{faccisharp} employed discontinuous Lagrange polynomials. We adopt this formulation here to serve as a benchmark for evaluating our proposed two-correction approach on geometries with sharp features. 
We define a basis $\{ \zeta_j (\mathbf{X}) \}_{j = 1}^K$, where each $\zeta_j$ denotes a piecewise linear Lagrange polynomial that is discontinuous across element boundaries. This allows nodal values to differ between neighboring elements. In contrast to the continuous projection in Eq. (20), we compute the projection into the discontinuous basis by requiring $\llbracket p\rrbracket_h^\text{D}(\mathbf{X}, t)$ to satisfy
\begin{equation}
\int_{\Gamma_0} \llbracket p\rrbracket_h^\text{D}(\mathbf{X}, t)\, \zeta_j(\mathbf{X}) \, {\mathrm d} A = \int_{\Gamma_0} \jmath^{- 1} (\mathbf{X}, t) \mathbf{F}_h (\mathbf{X},
  t) \cdot \mathbf{n} (\mathbf{X}, t) \, \zeta_j(\mathbf{X}){\mathrm d} A , \quad \forall j = 1, \ldots, K.
\end{equation}

The projection of $\llbracket \frac{\partial \mathbf{u} (\mathbf{\boldsymbol{\chi}} (\mathbf{X}, t), t)}{\partial \mathbf{x}_i} \rrbracket$ is handled similarly. In this work, we evaluate these integrals using a seventh order Gaussian quadrature scheme.

\subsection{Finite Difference Approximation}

We use a staggered-grid discretization of the incompressible Navier-Stokes equations. The discretization approximates 
pressure at cell centers, and velocity and forcing terms at the center of cell edges\cite{griffith_volume_2012,griffith_accurate_2009}. Our computations use an isotropic grid, such that $h=\Delta x = \Delta y$. We use second-order accurate finite
difference stencils, and define the discrete divergence of the velocity
$\mathbf{D} \cdot \mathbf{u}$ on cell centers, and the discrete
pressure gradient $\mathbf{G} p$ and discrete Laplacian $L
\mathbf{u} $ on cell edges. 
Finite difference stencils that cut through the interface require additional forcing terms, which are commonly referred to as correction terms, to account for discontinuities in the pressure and velocity gradients, as in previous work\cite{kolahdouz_immersed_2020}.

\subsection{Imposing Jump Conditions in the Navier-Stokes Equations}
In modern IIMs, jump conditions are imposed through correction terms that appear on the right-hand side of the discretized incompressible Navier--Stokes equations.
Formulations that target only the lowest order jump conditions, as considered herein, include corrections only in the discretized momentum equation.
Consequently, these corrections take the form of additional body forces that are concentrated along the immersed boundary.
Because the jump conditions are determined from the configuration of the boundary and the interfacial forces, it is convenient to introduce a force spreading operator, 
$\boldsymbol{\mathcal{S}}[\boldsymbol{\chi}]$, that relates the Lagrangian force
$\mathbf{F}_h (\mathbf{X}, t)  = \sum_{j = 1}^M  \mathbf{F}_j(t) \,\psi_j (\mathbf{X})$
to the Eulerian forces $\mathbf{f}$ on the Cartesian grid by summing all correction terms in the modified finite difference stencils involving the Lagrangian force. Correction terms, which are detailed in previous work \cite{kolahdouz_immersed_2020}, depend on the projected jump conditions, as computed in Section 3.3.

 Here, we discuss cases where a single finite difference stencil is cut by two interfaces or a single interface with a sharp corner. Following the Taylor series expansion approach in Xu et al\cite{xu_systematic_2006}, consider a stencil between two Cartesian grid points $\mathbf{x}_{i,j} \in \Omega^{-}$ and $\mathbf{x}_{i+1,j} \in \Omega^{+}$.
 If this stencil is cut by two interfaces at locations $\mathbf{x}_1$ and $\mathbf{x}_2$, such that $\mathbf{x}_{i,j} < \mathbf{x}_1 < \mathbf{x}_2 < \mathbf{x}_{i+1,j}$ . Then the following approximations for a piecewise differentiable quantity $v$ hold:
\begin{align}
\frac{\partial v}{\partial x}(\mathbf{x}_{i+\frac{1}{2},j}) &= \frac{v_{i+1,j}-v_{i,j}}{h} + \frac{\mathrm{sgn}\{n_1^x\}}{h}\sum_{m=0}^2\frac{(h_1^+)^m}{m!}\jumpinverse{\frac{\partial^{m} v(\mathbf{x}_1)}{\partial x^{m}}} \notag \\
&\quad + \frac{\mathrm{sgn}\{n_2^x\}}{h}\sum_{m=0}^2\frac{(h_2^+)^m}{m!}\jumpinverse{\frac{\partial^{m} v(\mathbf{x}_2)}{\partial x^{m}}} + O(h^2), \label{eqn:taylor_1} \\[1em]
\frac{\partial^2 v}{\partial x^2}(\mathbf{x}_{i,j}) &= \frac{v_{i+1,j}-2v_{i,j}+v_{i-1,j}}{h^2} + \frac{\mathrm{sgn}\{n_1^x\}}{h^2}\sum_{m=0}^3\frac{(h_1^+)^m}{m!}\jumpinverse{\frac{\partial^{m} v(\mathbf{x}_1)}{\partial x^{m}}} \notag \\
&\quad + \frac{\mathrm{sgn}\{n_2^x\}}{h^2}\sum_{m=0}^3\frac{(h_2^+)^m}{m!}\jumpinverse{\frac{\partial^{m} v(\mathbf{x}_2)}{\partial x^{m}}} + O(h^2). \label{eqn:taylor_2}
\end{align}
In these equations, $n_{1}^x$ and $n_{2}^x$ are the x-components of the outward normals at $\mathbf{x}_1$ and $\mathbf{x}_2$, $h_1^+=\mathbf{x}_{i+1,j}-\mathbf{x}_1$, $h_2^+=\mathbf{x}_{i+1,j}-\mathbf{x}_2$, and $h$ is the Cartesian grid size. We substitute  approximations Equation~(\ref{eqn:taylor_1})--(\ref{eqn:taylor_2}) into the discrete finite difference approximations of the pressure gradient $\mathbf{G} p$ and discrete Laplacian $L
\mathbf{u}$ in Equation (2). In this work, we consider only the lowest-order jump conditions, accounting for discontinuities in the pressure and the derivative of the velocity.

\subsection{Time Integration}

Each step begins with known values of $\mathbf{\boldsymbol{\chi}}^n$ and $\mathbf{u}^n$ at time
$t^n$, and $p^{n - \frac{1}{2}}$ at time $t^{n - \frac{1}{2}}$. The goal is to
compute $\mathbf{\boldsymbol{\chi}}^{n + 1}, \mathbf{u}^{n + 1}$, and $p^{n +
\frac{1}{2}}$. First, we approximate the structure location at time $t^{n + 1}$ by
\begin{equation}
  \frac{\mathbf{\boldsymbol{\chi}}^{n + \frac12} - \mathbf{\boldsymbol{\chi}}^n}{\frac{\Delta t}{2}} =
  \mathbf{U}^n (\mathbf{\boldsymbol{\chi}}^n) = \mathcal{I}[\boldsymbol{\chi}^n,\mathbf{F}^n] (\mathbf{u}^n)  .
\end{equation}
Next, we solve for $\mathbf{u}^{n + 1}$ and $p^{n +
\frac{1}{2}}$ in which $\mathbf{f}^{n + \frac{1}{2}} = \boldsymbol{\mathcal{S}}[\boldsymbol{\chi}^{n + \frac{1}{2}} ] \mathbf{F}(\boldsymbol{\chi}^{n + \frac{1}{2}},t^{n + \frac{1}{2}})$:
\begin{eqnarray}
  \rho \left( \frac{\mathbf{u}^{n + 1} - \mathbf{u}^n}{\Delta t} +
  \mathbf{A}^{n+\frac12} \right) & = & - \mathbf{G} p^{n +
  \frac{1}{2}} + \mu L \left( \frac{\mathbf{u}^{n + 1} +
  \mathbf{u}^n}{2} \right) + \mathbf{f}^{n + \frac{1}{2}}, \\
  \mathbf{D} \cdot \mathbf{u}^{n + 1} & = & 0,
\end{eqnarray}
in which the non-linear advection term, $\mathbf{A}^{n + \frac{1}{2}} = \frac{3}{2}\mathbf{A}^{n} - \frac{1}{2}\mathbf{A}^{n-1}$, is handled with the xsPPM7 variant\cite{xsPPM7} of the piecewise
parabolic method.{\cite{colella_piecewise-parabolic_1982}} 
$\mathbf{G}, \mathbf{D} \cdot\mbox{}$, and $L$ are the discrete
gradient, divergence, and Laplacian operators. The system of equations is
iteratively solved via the FGMRES
algorithm with the projection method preconditioner\cite{griffith_accurate_2009}. Last, we update the structure's location, $\mathbf{\boldsymbol{\chi}}^{n + 1}$, with
\begin{equation}
  \frac{\mathbf{\boldsymbol{\chi}}^{n + 1} - \mathbf{\boldsymbol{\chi}}^n}{\Delta t}  = 
  \mathbf{U}^{n + \frac{1}{2}} =  \boldsymbol{\mathcal{I}}[\boldsymbol{\chi}^{n+\frac{1}{2}},\mathbf{F}^{n+\frac{1}{2}}] \left(\frac{\mathbf{u}^n+ \mathbf{u}^{n+1}}{2}\right)  .
\end{equation}

\subsection{Software Implementation}

All computations for were completed through IBAMR \cite{ibamr}, which
utilizes parallel computing libraries and adaptive mesh refinement (AMR).
IBAMR uses other libraries for setting up meshes, fast linear algebra solvers,
and postprocessing, including SAMRAI \cite{samrai}, PETSc \cite{petsc,petsc2,petsc3}, \textit{hypre} \cite{hypre,hypre2}, and libMesh \cite{libmesh,libmesh2}.

\section{Numerical Experiments}
We first use shearing parallel plates as a benchmark case to examine the accuracy of our two correction algorithm on a globally piecewise linear velocity field, for which our interfacial velocity recreation should be exact. 
We also examine benchmark studies in which the velocity field is non-linear in $x$ and $y$, particularly two concentric rotating cylinders. 
This benchmark highlights our method's robust approach to treating two intersections and computing secondary correction terms at a variety of orientations. 
Next, we examine the reconstructed interfacial velocity field on two eccentric rotating cylinders. This study examines geometries in which symmetry is broken along one axis. 
Both the concentric and eccentric cylinders are tested for accuracy for minimal separation distances $\Delta s$ = $\frac{h}{10}$.
In the final set of experiments, we compare the two-correction CG-IIM with one-correction CG-IIM and DG-IIM approaches for geometries involving sharp features. Following tests by Facci et al,\cite{faccisharp} we investigate the method's sensitivity to angular acuteness and examine its handling of both convex and concave corners using anvil and star geometries.
For all experiments, the Lagrangian mesh spacing is set to be one to two times the background Cartesian grid spacing. This spacing is standard in previous work\cite{kolahdouz_immersed_2020}. Because jump conditions are imposed wherever the interface cuts through a stencil, the interface spacing does not cause numerical leaking. Another study \cite{SUN2026114497} has demonstrated that smaller coarsening ratios require additional treatment. 
We impose rigid body motions using a penalty method described in Section 2. Because the penalty parameter is finite, generally there are discrepancies between the prescribed and actual positions of the interface. In numerical tests, we choose $\kappa$ to ensure that the discrepancy in interface configurations $\epsilon_{\mathbf{X}} = \| \mathbf{\xi} (\mathbf{X}, t) -
\mathbf{\boldsymbol{\chi}} (\mathbf{X}, t) \|$ satisfies $\epsilon_{\mathbf{X}} < \frac{h}{10}$. 

\subsection{Shearing Parallel Plates}

This section considers the flow between two shearing parallel plates. The plates are aligned with the $x$-axis, such that the top plate lies at $y= \frac{1}{48}$ and the bottom plate lies at $y= -\frac{1}{48}$. The distance between the plates is $\Delta s = \frac{1}{24}$. 
The two plates extend to the left and right ends of the computational domain, $\Omega = [-1 , 1]^2$, of length $L$. Figure 8 shows a schematic for the model's setup.
The top plate moves to the right at a velocity of $U_\mathrm{upper}=0.003125$, and the bottom plate moves to the left at a velocity of $U_\mathrm{lower}=-0.003125$. We set $\rho = 1$, $\mu = 0.02$, and $\mathrm{Re}=0.006$.
The domain is discretized into $N = 16$, $32$, and $64$ cells in both directions such that the mesh spacing is $h = 0.125$, $0.0625$, and $0.03125$ respectively.
We set the Lagrangian mesh width to be twice as coarse as the background Cartesian grid spacing for computational efficiency. The time step size $\Delta t$ scales with $h$ such that $\Delta t = \frac{h}{100}$. Periodic boundary conditions are used on each side. For this time step size and grid spacing, the Courant-Friedrichs-Lewy (CFL) number is approximately $1.7\cdot10^{-5}$ once the model reaches steady state.
To impose rigid body motion, we set $\kappa = C_{\kappa}(\frac{N}{8})^2$ with $C_{\kappa} = 12.0$. Figure 5 shows a schematic for the shearing plates setup.

\begin{figure}[H]
\center{}

\tikzset{every picture/.style={line width=0.75pt}} 

\begin{tikzpicture}[x=0.75pt,y=0.75pt,yscale=-1,xscale=1]

\draw [color={rgb, 255:red, 74; green, 144; blue, 226 }  ,draw opacity=1 ][line width=2.25]    (76.76,175.08) -- (409.08,175.08) ;
\draw [shift={(414.08,175.08)}, rotate = 180] [fill={rgb, 255:red, 74; green, 144; blue, 226 }  ,fill opacity=1 ][line width=0.08]  [draw opacity=0] (14.29,-6.86) -- (0,0) -- (14.29,6.86) -- cycle    ;
\draw [color={rgb, 255:red, 208; green, 2; blue, 27 }  ,draw opacity=1 ][line width=2.25]    (80.76,189) -- (412.76,189) ;
\draw [shift={(75.76,189)}, rotate = 0] [fill={rgb, 255:red, 208; green, 2; blue, 27 }  ,fill opacity=1 ][line width=0.08]  [draw opacity=0] (14.29,-6.86) -- (0,0) -- (14.29,6.86) -- cycle    ;
\draw  [draw opacity=0][fill={rgb, 255:red, 0; green, 0; blue, 0 }  ,fill opacity=0 ][line width=0.75]  (77,112) -- (413,112) -- (413,252) -- (77,252) -- cycle ; \draw  [color={rgb, 255:red, 155; green, 155; blue, 155 }  ,draw opacity=0.52 ][line width=0.75]  (105,112) -- (105,252)(133,112) -- (133,252)(161,112) -- (161,252)(189,112) -- (189,252)(217,112) -- (217,252)(245,112) -- (245,252)(273,112) -- (273,252)(301,112) -- (301,252)(329,112) -- (329,252)(357,112) -- (357,252)(385,112) -- (385,252) ; \draw  [color={rgb, 255:red, 155; green, 155; blue, 155 }  ,draw opacity=0.52 ][line width=0.75]  (77,140) -- (413,140)(77,168) -- (413,168)(77,196) -- (413,196)(77,224) -- (413,224) ; \draw  [color={rgb, 255:red, 155; green, 155; blue, 155 }  ,draw opacity=0.52 ][line width=0.75]  (77,112) -- (413,112) -- (413,252) -- (77,252) -- cycle ;
\draw  [dash pattern={on 0.84pt off 2.51pt}]  (236.76,112.08) -- (272.76,174.08) ;
\draw  [dash pattern={on 0.84pt off 2.51pt}]  (200.26,189) -- (239.76,251.08) ;
\draw  [dash pattern={on 0.84pt off 2.51pt}]  (200.26,189) -- (272.76,174.08) ;

\draw (148,145.4) node [anchor=north west][inner sep=0.75pt]  [color={rgb, 255:red, 74; green, 144; blue, 226 }  ,opacity=1 ]  {$\mathbf{u} =u_{\text{profile}}$};
\draw (247,199.4) node [anchor=north west][inner sep=0.75pt]  [color={rgb, 255:red, 74; green, 144; blue, 226 }  ,opacity=1 ]  {$\mathbf{\textcolor[rgb]{0.82,0.01,0.11}{u}}\textcolor[rgb]{0.82,0.01,0.11}{=-u}\textcolor[rgb]{0.82,0.01,0.11}{_{\text{profile}}}$};

\end{tikzpicture}
\caption{Shearing parallel plate schematic. Both plates move with equal and opposite velocity.}
\end{figure}

We compare the $L^2$ and $L^\infty$ errors of the Eulerian fluid $x$-velocity, for both of the one and two correction methods at each grid resolution in Figure 9. 

\begin{figure}[H]
    \centering
    \resizebox{240pt}{!}{\includegraphics{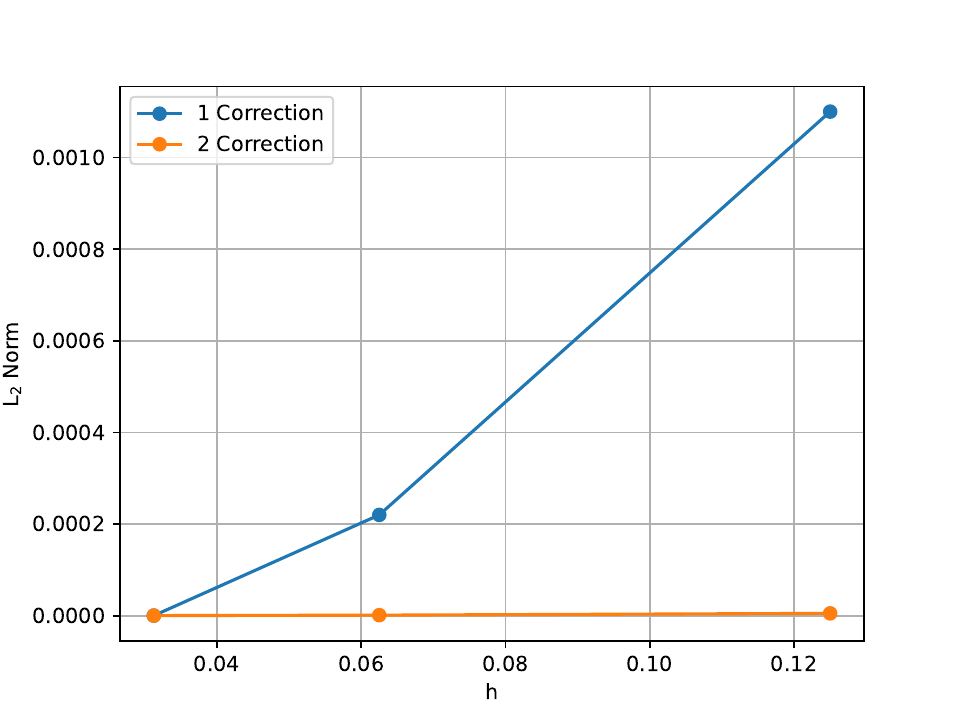}}\resizebox{240pt}{!}{\includegraphics{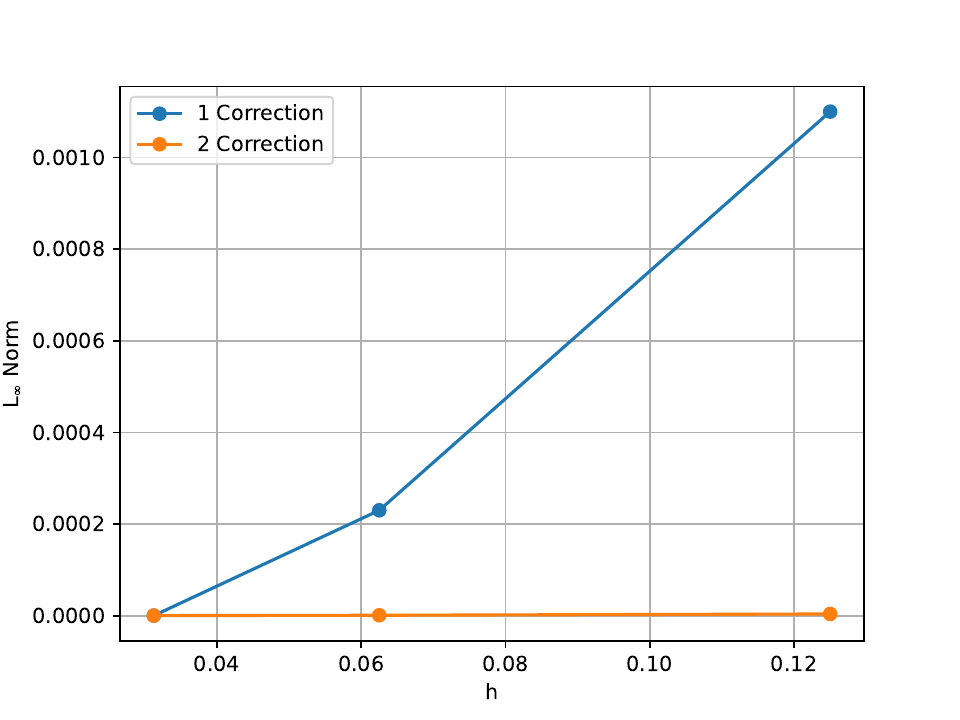}}
    \caption{Convergence of the shearing parallel plates' velocity in the $L^2$ and $L^\infty$ norms. Both methods' $L^2$ and $L^\infty$ errors converge to within machine precision of each other at the finest grid resolution, in which the $\Delta s > h$ and only one interface cuts each stencil. The two correction method exhibits similar error at each grid resolution, approximately $10^{-6}$, due to the finite Stokes solver tolerance.}
\end{figure}

We compare the accuracy of the present one and two correction methods with Fai and Rycroft's \cite{FAI2018319} lubricated IB method. We compute the Eulerian velocity profile  along the line $x = 0$ and compare it to the analytical solution in Figure 10. The analytical solution of the shearing parallel plates is 
\begin{equation}
u(y) = \begin{cases}
    U_\mathrm{lower} + \frac{y + \frac{\Delta s}{2}}{\Delta s}(U_\mathrm{upper} - U_\mathrm{lower}), & |y|\leq\frac{\Delta s}{2},\\
\frac{y + \frac{L}{2}}{\frac{L}{2}-\frac{\Delta s}{2} }(U_\mathrm{lower}), &y<-\frac{\Delta s}{2},\\
U_\mathrm{upper}+\frac{y - \frac{\Delta s}{2}}{\frac{L}{2}-\frac{\Delta s}{2}}, &y>\frac{\Delta s}{2}.
\end{cases}
\end{equation}

For $N = 16$ and $32$ grid cells, $\Delta s < h$. Both correction methods demonstrate convergence under grid refinement. The two correction algorithm is substantially more accuracy at coarser grid resolutions than the one correction and lubricated IB method. When $\Delta s < h$, the two correction method exhibits at least two orders of magnitude greater accuracy. 
\begin{figure}[H]
    \center{}
  \resizebox{220pt}{!}{\includegraphics{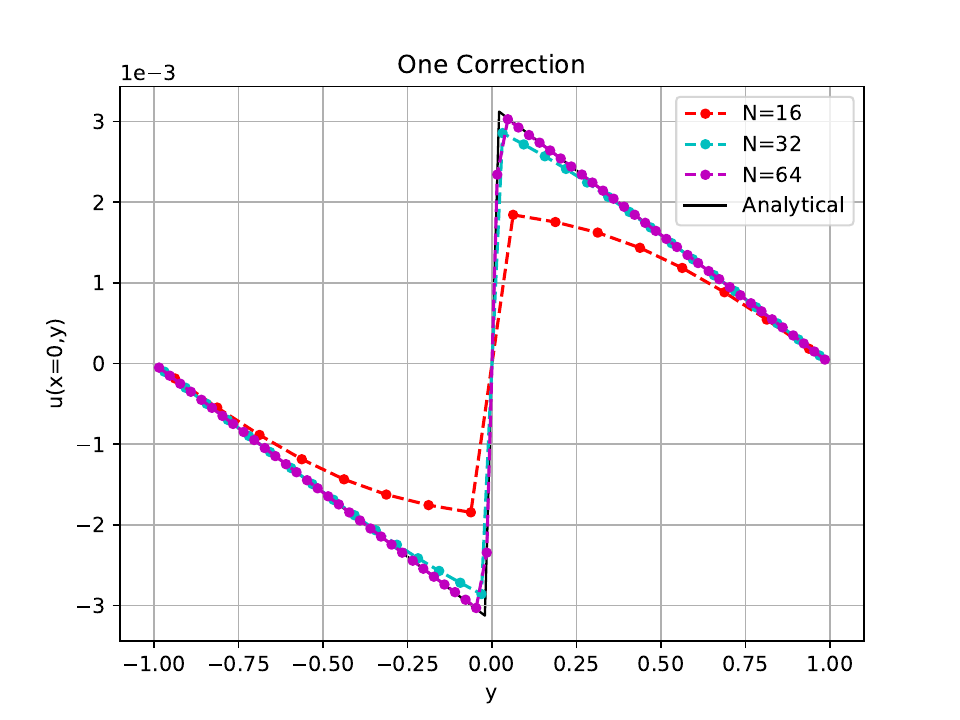}}\resizebox{220pt}{!}{\includegraphics{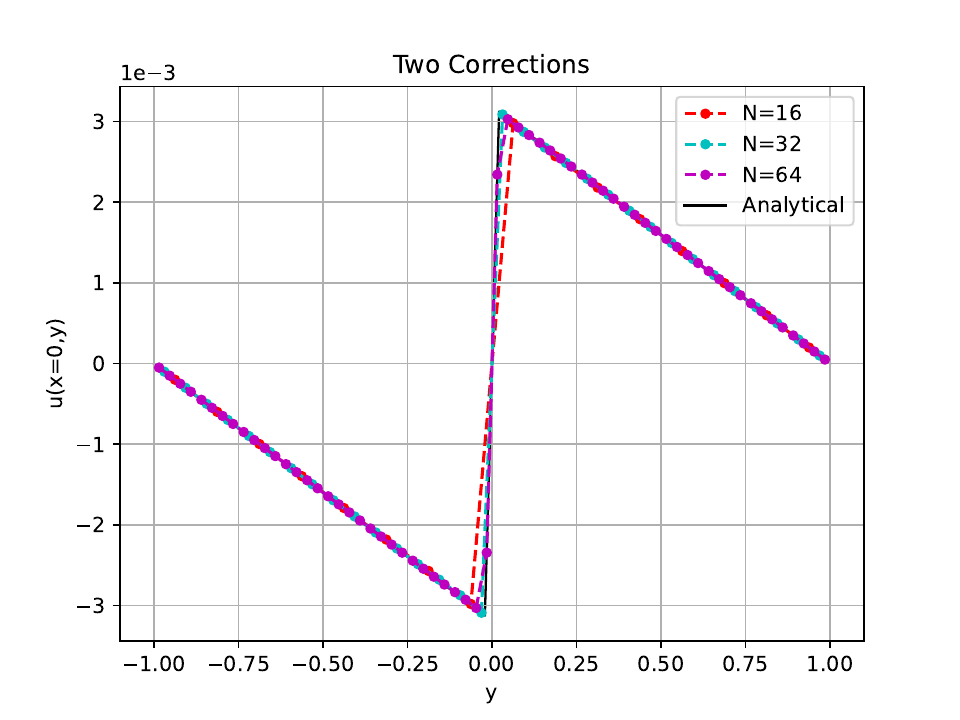}}
  \resizebox{220pt}{!}{\includegraphics{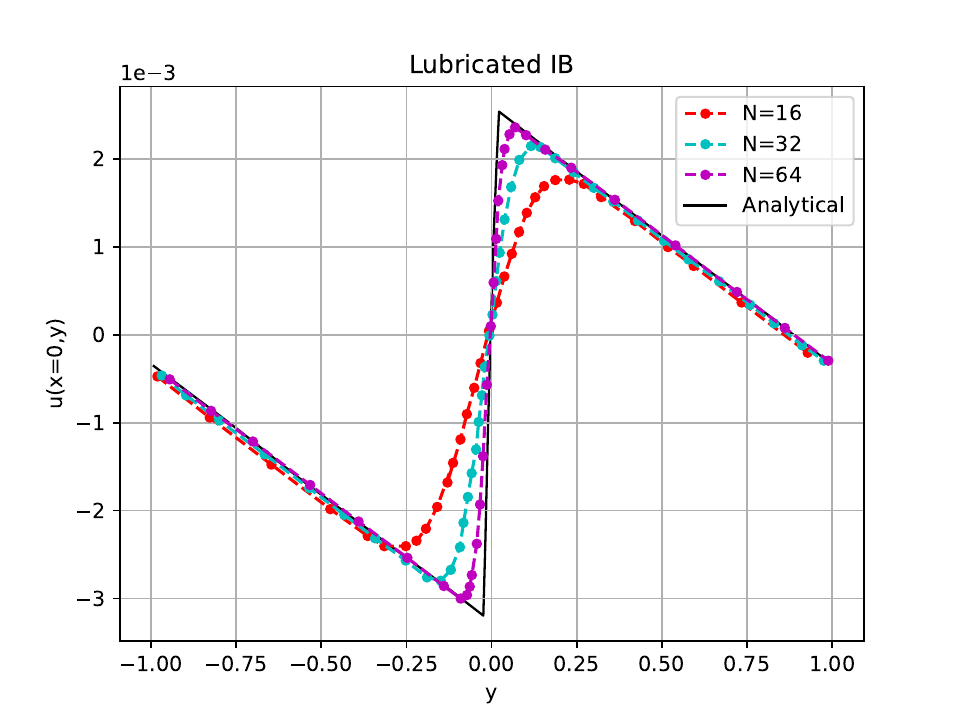}}
  
  \caption{Convergence of the pointwise accuracy of the velocity profile $u(x=0,y)$ across parallel plates. Our present two correction method is compared to the previous one correction method and the results digitized from Fai and Rycroft's lubricated IB method\cite{FAI2018319}. Both correction methods converge under grid refinement, and the two correction algorithm is substantial ly more accurate at coarser Cartesian grid resolutions than both one correction and the lubricated IB method.}
\end{figure}

To assess the accuracy of our method in the limit as the distance between the interfaces, $\Delta s$, approaches zero, we conduct an interface convergence test.
Fixing the Cartesian grid resolution at $N=32$ grid cells in each direction, we set $\Delta s= h$, $\frac{h}{2}$, $\frac{h}{5}$, $\frac{h}{10}$, $\frac{h}{20}$, and $\frac{h}{50}$. 
Figure 11 compares the velocity profile $u(x=0,y)$ using the one and two correction methods.
At all separation distances, the one correction method clearly suffers from decreased accuracy as $\Delta s \rightarrow 0$, whereas the two correction method shows excellent agreement with the analytical velocity profile. 

\begin{figure}[H]
  \centering
  \resizebox{220pt}{!}{\includegraphics{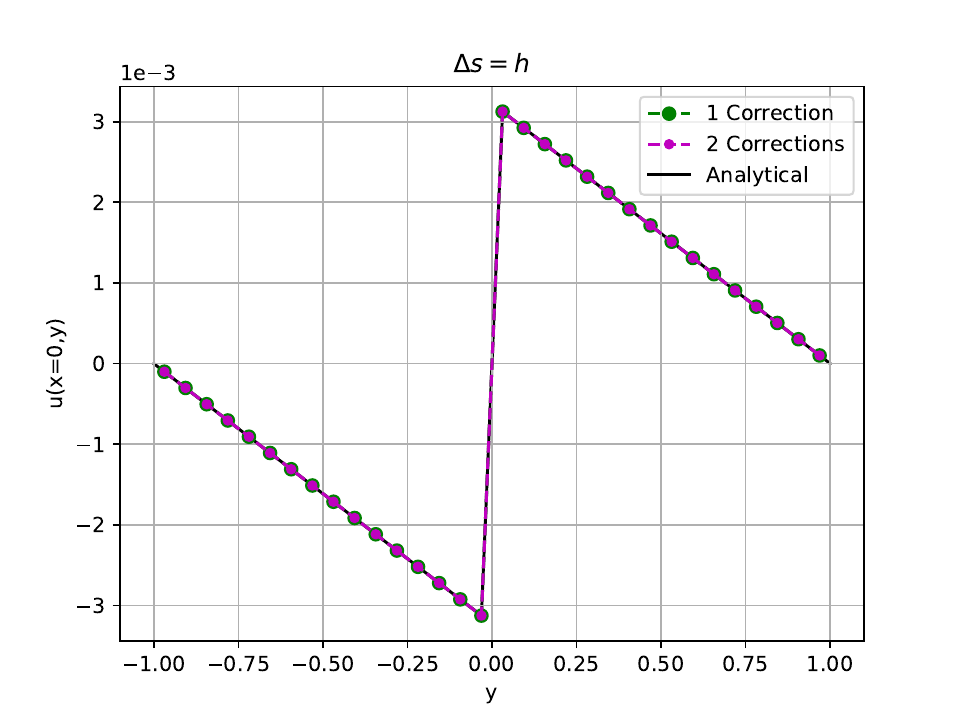}}
  \resizebox{220pt}{!}{\includegraphics{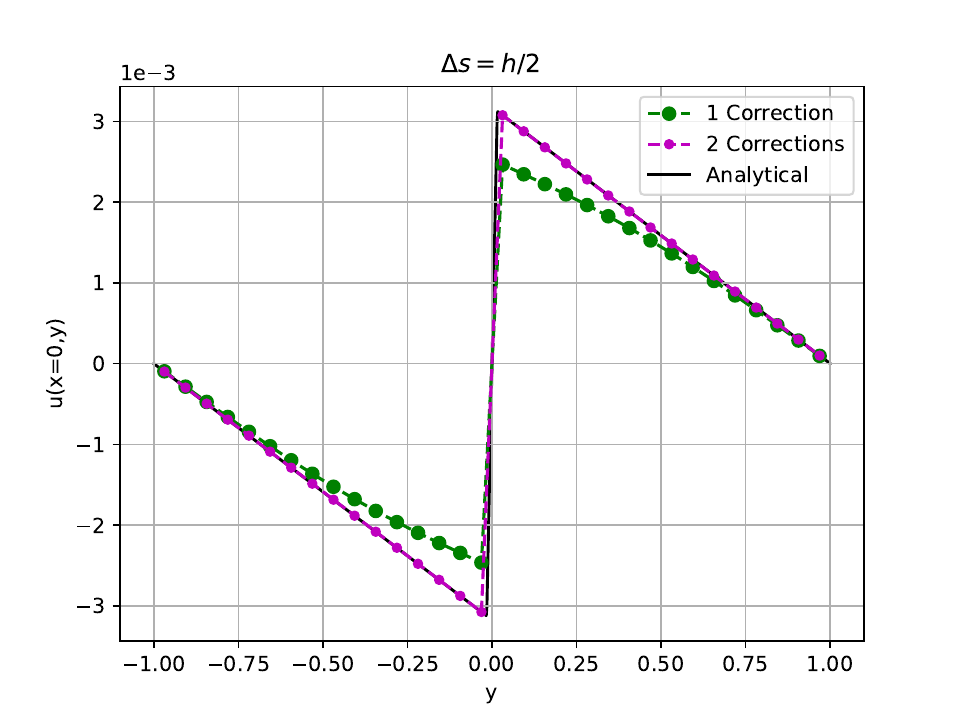}}
  
  \resizebox{220pt}{!}
  {\includegraphics{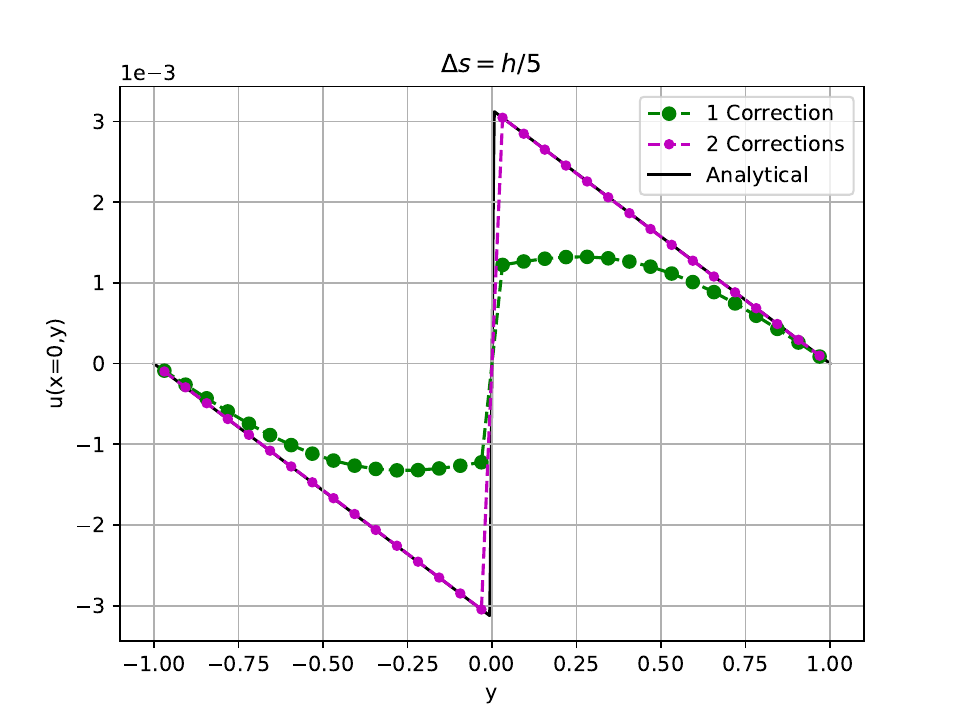}}
  \resizebox{220pt}{!}{\includegraphics{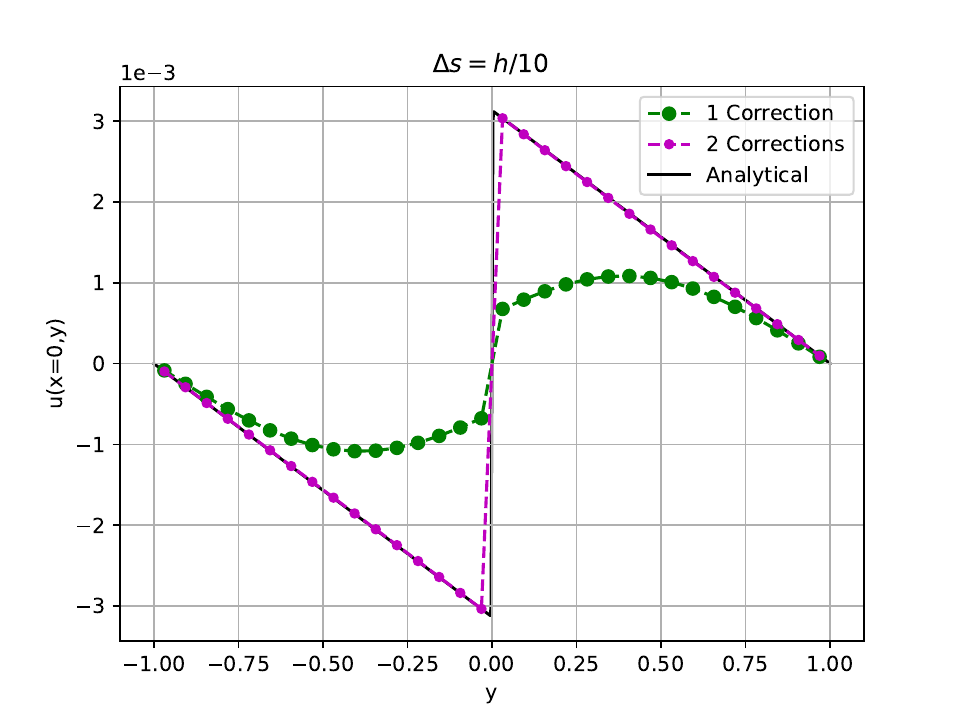}}
  
  \resizebox{220pt}{!}{\includegraphics{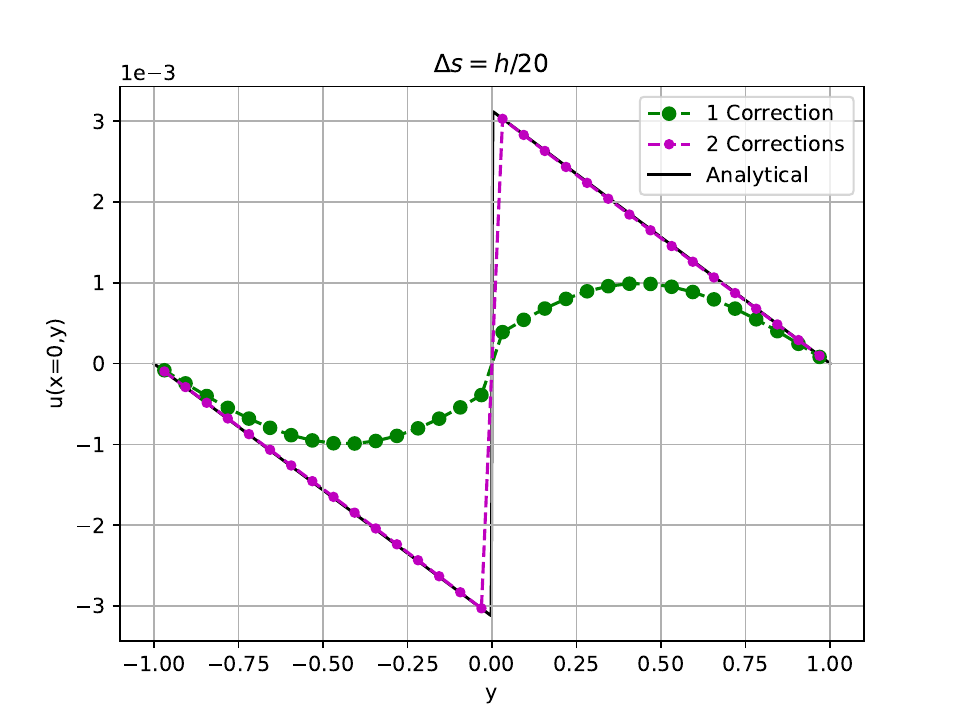}}
  \resizebox{220pt}{!}{\includegraphics{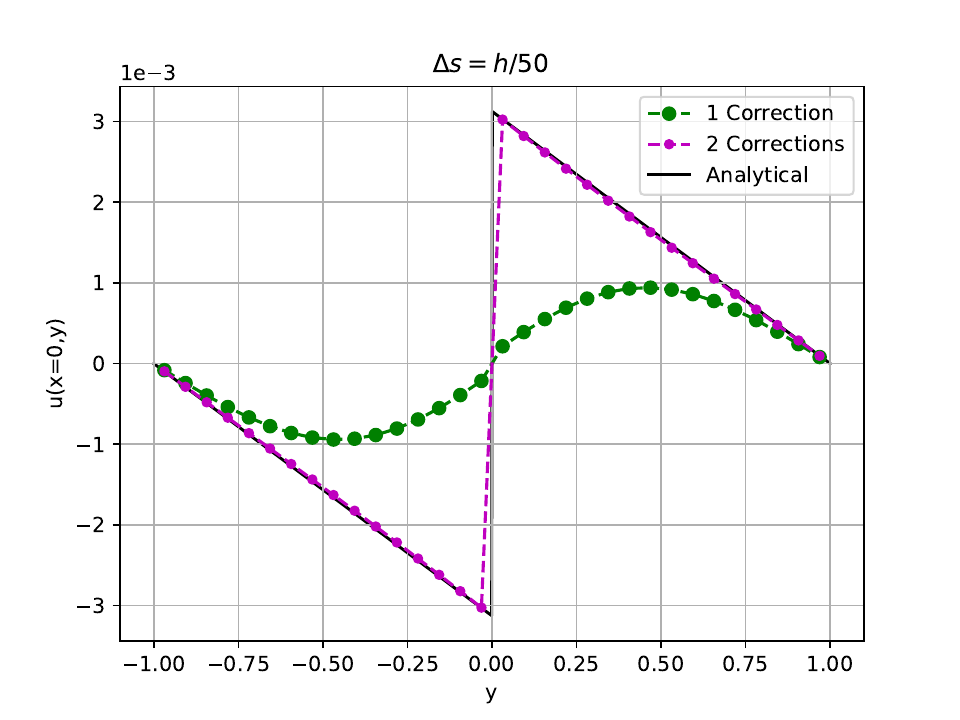}}
  \caption{Comparison of the velocity profile $u(x=0,y)$ for various separations between two shearing parallel plates. When $\Delta s < h$, the two correction method demonstrates substantially more accurate than the one correction method, even when $\Delta s = \frac{h}{50}$.}
\end{figure}

\subsection{Concentric Rotating Cylinders}
This section considers the flow between two concentric rotating cylinders in two spatial dimensions. The cylinders are centered at the origin, with radii $R_\text{1} = 0.75$ and $R_\text{2} =  0.78125$, such that the separation of the interfaces is $\Delta s = 0.03125$. The outer cylinder is fixed in place, and the inner cylinder rotates counter-clockwise at a rotational velocity $\omega = 8.33\cdot10^{-4}$. The fluid domain is $\Omega = [-1,1]^2$.
Figure 12 shows a schematic for the model's setup.
We set $\rho = 1$, $\mu = 0.2$, and $\mathrm{Re}=6.5\cdot 10^{-4}$.
The domain is discretized into $N = 16$, $32$, $64$, and $128$ cells in both directions such that the mesh spacing is $h = 0.125$, $0.0625$, $0.03125$, and $0.015625$ respectively.
We set the Lagrangian mesh width to be equal to the background Cartesian grid spacing. The time step size $\Delta t$ scales with $h$ such that $\Delta t = \frac{h}{100}$. We use zero normal traction and zero tangential velocity boundary conditions on each side of the computational domain. For this time step size and these grid spacings, the CFL number is approximately $10^{-6}$ once the model reaches steady state.
To impose rigid body motion, we set $\kappa = \frac{C_{\kappa}}{(\Delta t)^2}$ with $C_{\kappa} = 2.0\cdot 10^{-4}$.

\begin{figure}[H]
\centering
    \resizebox{180pt}{!}{\includegraphics{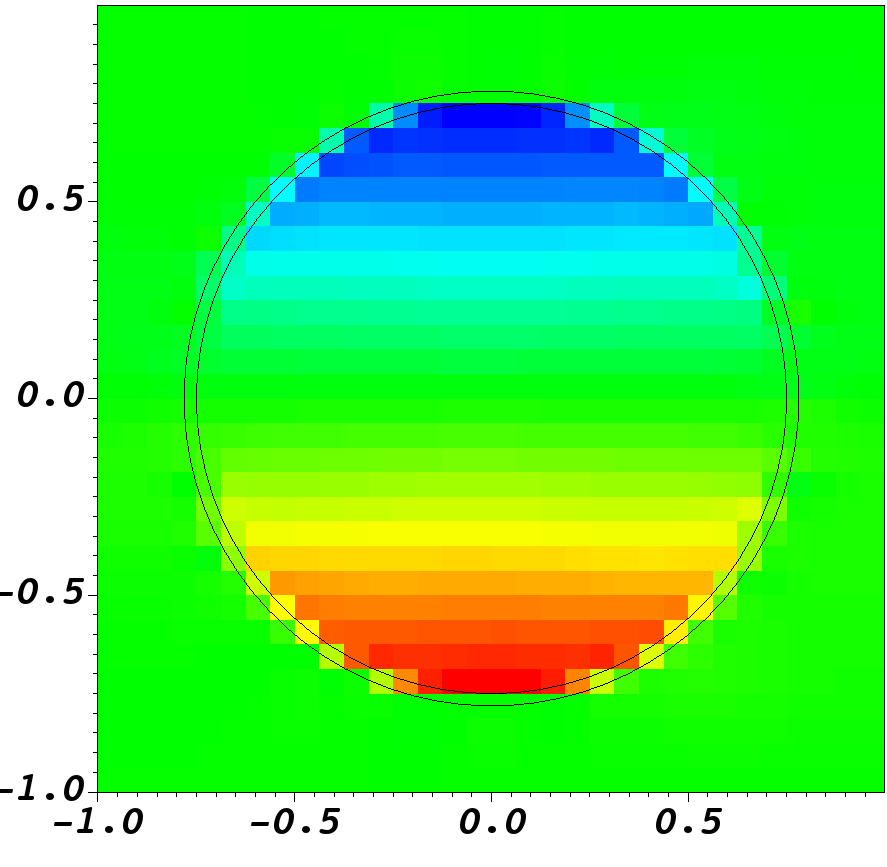}}\resizebox{80pt}{!}{
\tikzset{every picture/.style={line width=0.75pt}} 

\begin{tikzpicture}[x=0.75pt,y=0.75pt,yscale=-1,xscale=1]

\draw (78.97,110.76) node  {\includegraphics[width=50.96pt,height=79.62pt]{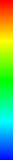}};
\draw [color={rgb, 255:red, 0; green, 0; blue, 0 }  ,draw opacity=1 ]   (112.94,59.2) -- (136.64,59.2) ;
\draw    (112.94,163.84) -- (136.64,163.84) ;
\draw    (112.94,110.76) -- (136.64,110.76) ;
\draw [color={rgb, 255:red, 0; green, 0; blue, 0 }  ,draw opacity=1 ]   (112.94,84.98) -- (136.64,84.98) ;
\draw    (112.94,136.54) -- (136.64,136.54) ;

\draw (70,40) node [anchor=north west][inner sep=0.75pt]  [font=\Large]  {$u_x $};
\draw (147.82,45.42) node [anchor=north west][inner sep=0.75pt]   [align=left] {6.0e-4};
\draw (147.82,72.72) node [anchor=north west][inner sep=0.75pt]   [align=left] {3.0e-4};
\draw (147.82,100.02) node [anchor=north west][inner sep=0.75pt]   [align=left] {6.9e-7};
\draw (147.82,125.8) node [anchor=north west][inner sep=0.75pt]   [align=left] {-3.0e-4};
\draw (147.82,151.58) node [anchor=north west][inner sep=0.75pt]   [align=left] {-6.0e-4};

\end{tikzpicture}}
  \
  \caption{The $x$-component of the velocity field is visualized over the entire fluid domain at $t = 5$ using the two correction method. $N = 32$ grid cells are used in either direction, such that $\Delta s = \frac{h}{2}$.}
\end{figure}

We compare the convergence of $\mathbf{u}$ and the the Lagrangian wall shear stress (WSS) under grid refinement in Figure 13.
At all grid resolutions in which $\Delta s<h$, the two correction method shows increased accuracy compared to the previous one correction method. When $\Delta s \geq h$, however, the interface velocity interpolation stencil intersects the interfaces only once, and the two formulations yield identical results.

\begin{figure}[H]
    \center{}
  \resizebox{220pt}{!}{\includegraphics{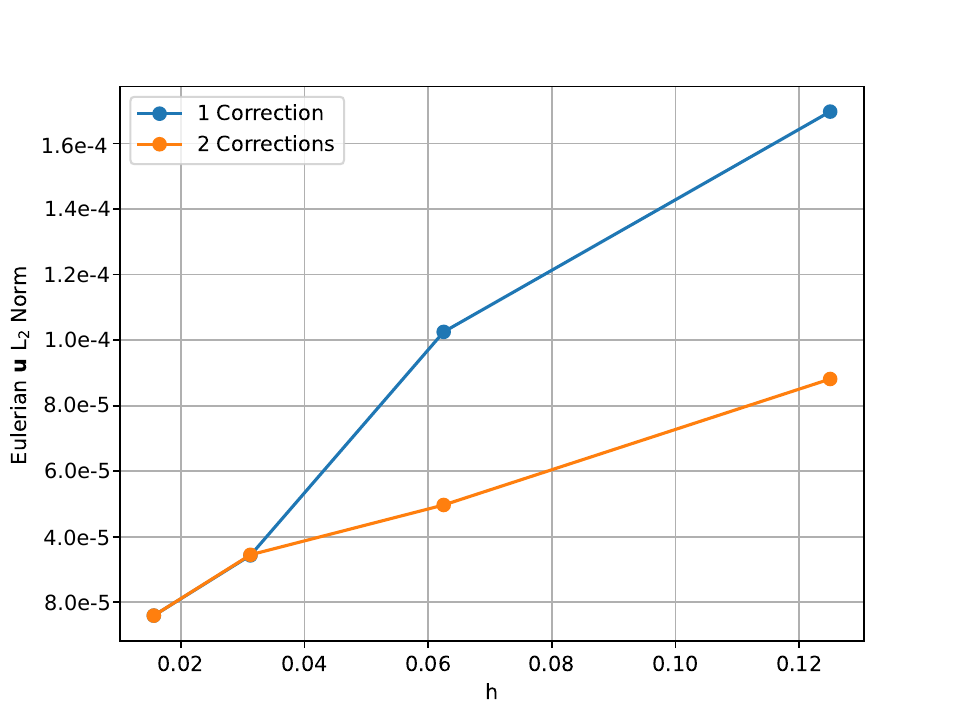}}\resizebox{220pt}{!}{\includegraphics{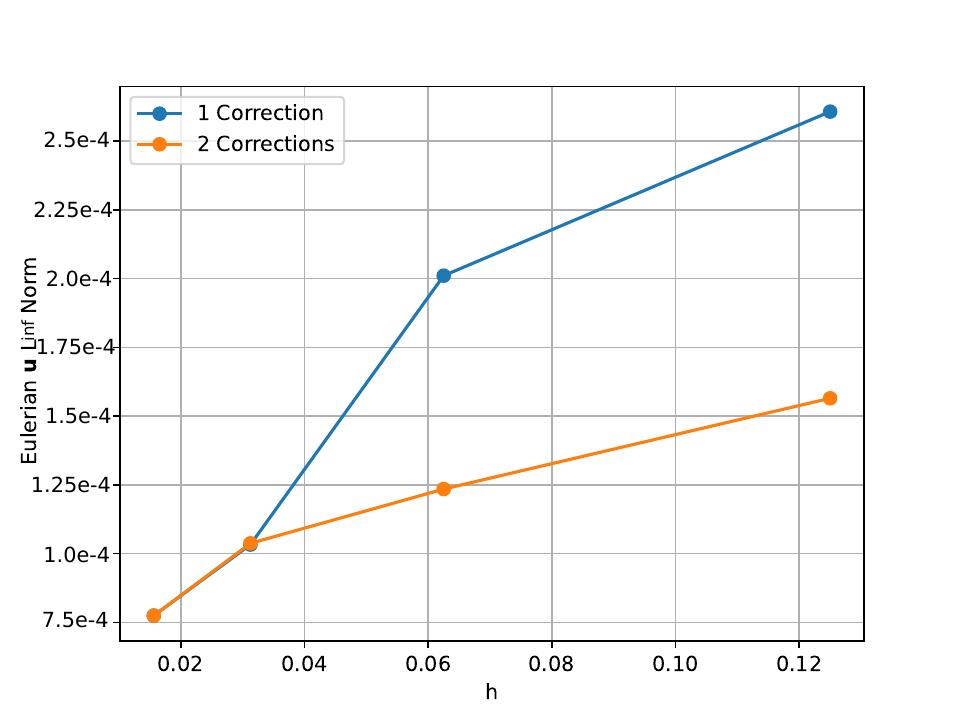}}
  \resizebox{220pt}{!}{\includegraphics{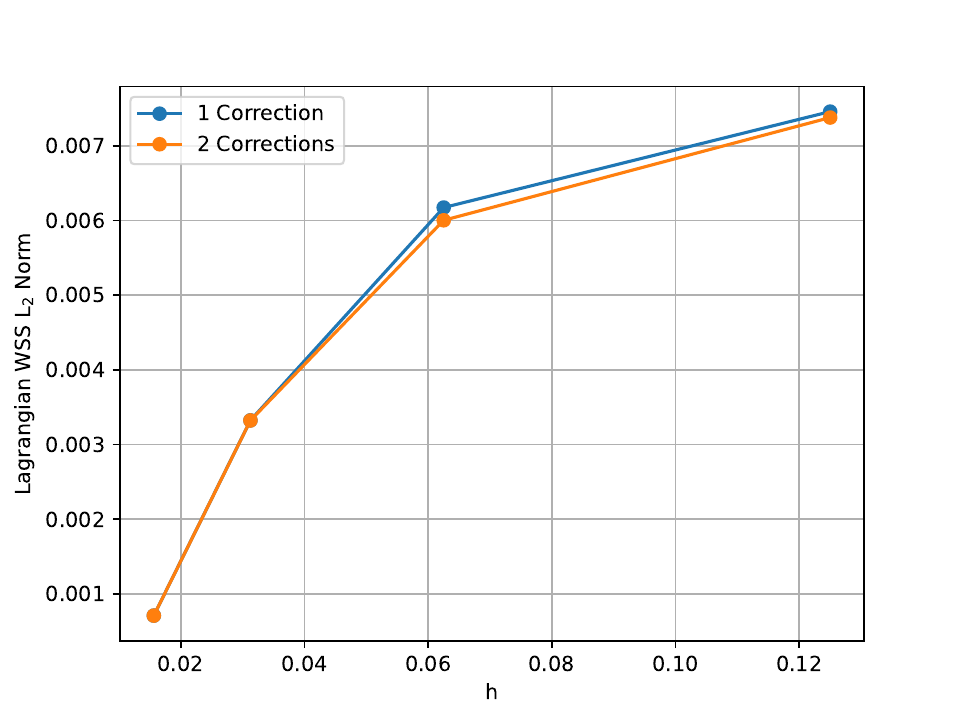}}\resizebox{220pt}{!}{\includegraphics{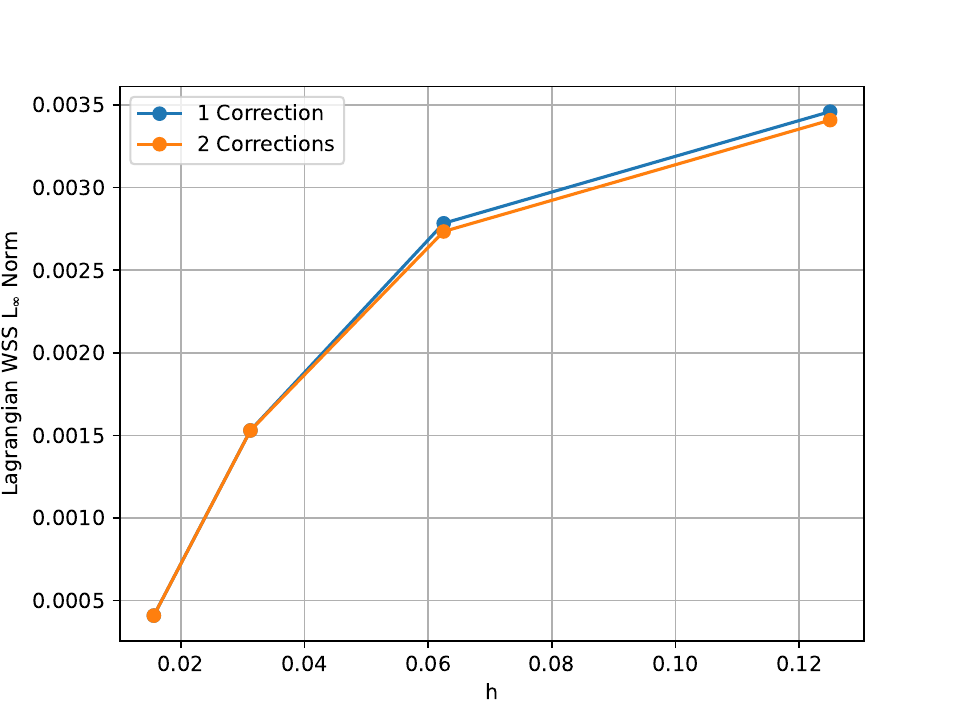}}
  
  \caption{Convergence of $\mathbf{u}$ and Lagrangian wall shear stress for concentric rotating cylinders. Our present two correction method is compared to the previous one correction method. Both methods converge under grid refinement, and the two correction algorithm is more accurate at coarser grid resolutions than the one correction method.}
\end{figure}
The analytical Eulerian velocity is
\begin{equation}
u(x,y) = \begin{cases}
    -\omega_1 y, & 0 \leq r \leq R_1,\\
-y (A+\frac{B}{r^2}), &R_1<r<R_2,\\
0, &R_2\leq r.
\end{cases}
\end{equation}
in which $A = \frac{-\omega R_2^2}{R_2^2-R_1^2}$ and $B = \frac{\omega R_1^2 R_2^2}{R_2^2-R_1^2}$.
We compute the $u(x=0,y)$ profile from $y=-1$ to $y=1$ for both one and two correction methods, and compare their convergence properties in Figure 14. The one correction method demonstrates reduced accuracy near the interfaces in near contact, whereas the two correction method shows excellent agreement with the analytical velocity profile.
\begin{figure}[H]
    \centering
    \resizebox{220pt}{!}{\includegraphics{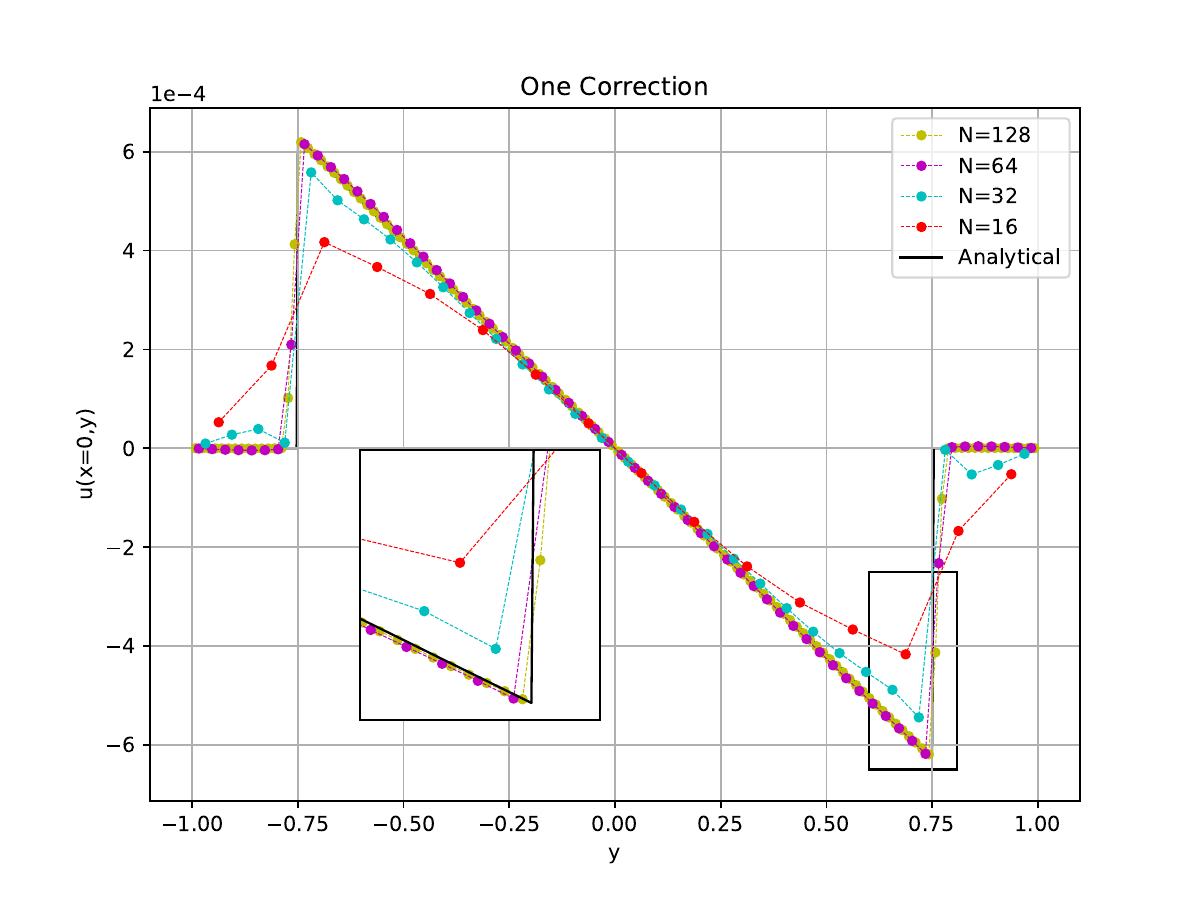}}\resizebox{220pt}{!}{\includegraphics{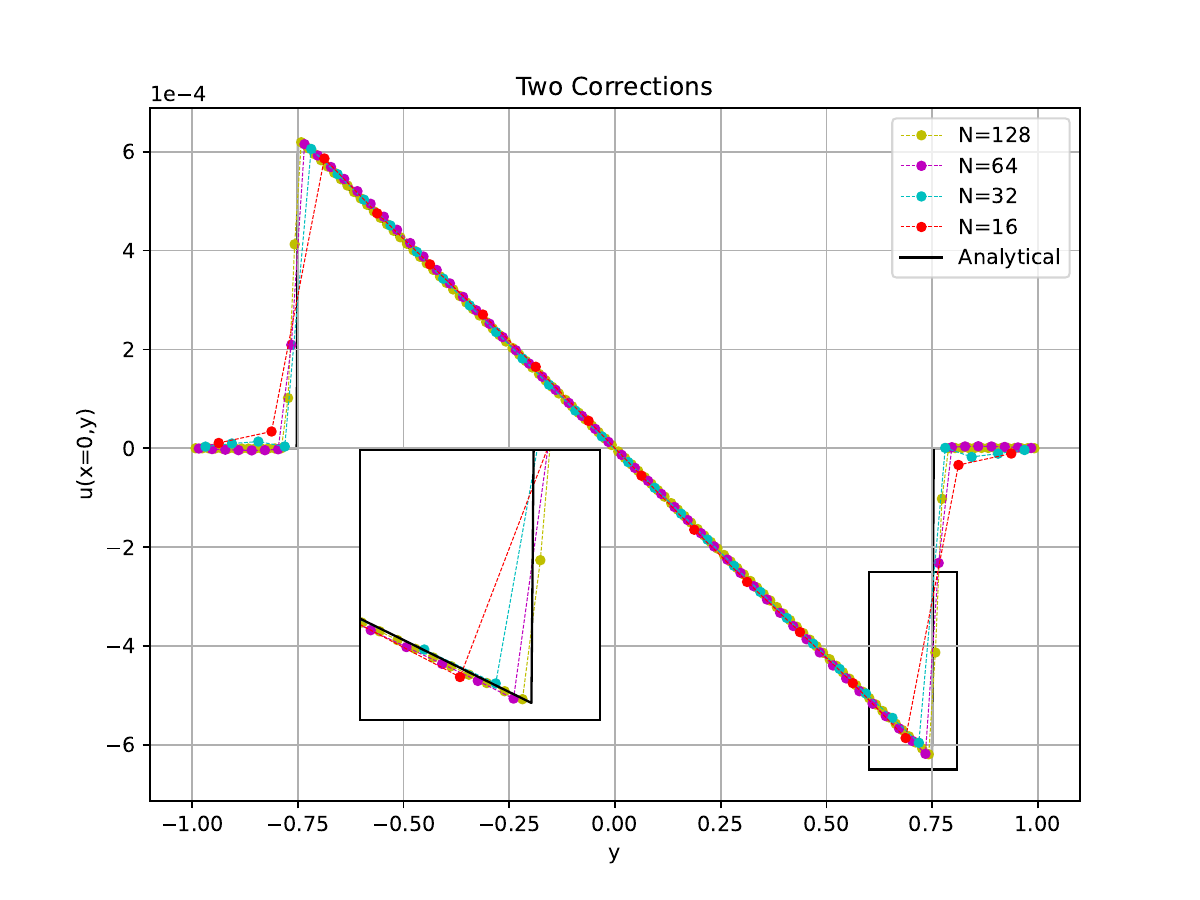}}
    \caption{Convergence of the pointwise accuracy of the velocity profile $u(x=0,y)$ across concentric rotating cylinders. Our present two correction method is compared to the previous one correction method. Both correction methods converge under grid refinement, and the two correction algorithm is significantly more accurate at coarser Cartesian grid resolutions than the one correction method.}
\end{figure}

Last, we analyze the accuracy of the two methods for various separation distances between the cylinders in Figure 15. We fix the Cartesian grid resolution using $N = 64$ grid cells in each direction. 
We vary $\Delta s$ between the two cylinders, particularly $h$, $\frac{h}{2}$, $\frac{h}{5}$, and $\frac{h}{10}$. When $\Delta s < h$, the one correction method clearly suffers from decreased accuracy throughout the computational domain. In contrast, the two correction method maintains substantially better agreement with the analytical profile, even when $\Delta s = \frac{h}{10}$.

\begin{figure}[H]
    \centering
  \resizebox{220pt}{!}{\includegraphics{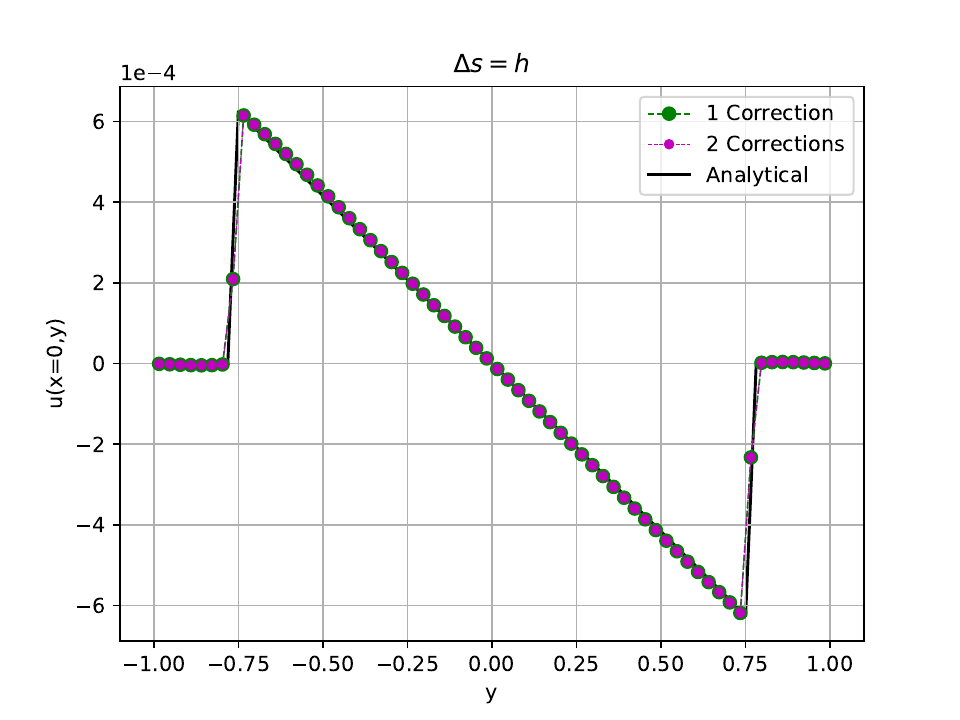}}\resizebox{220pt}{!}{\includegraphics{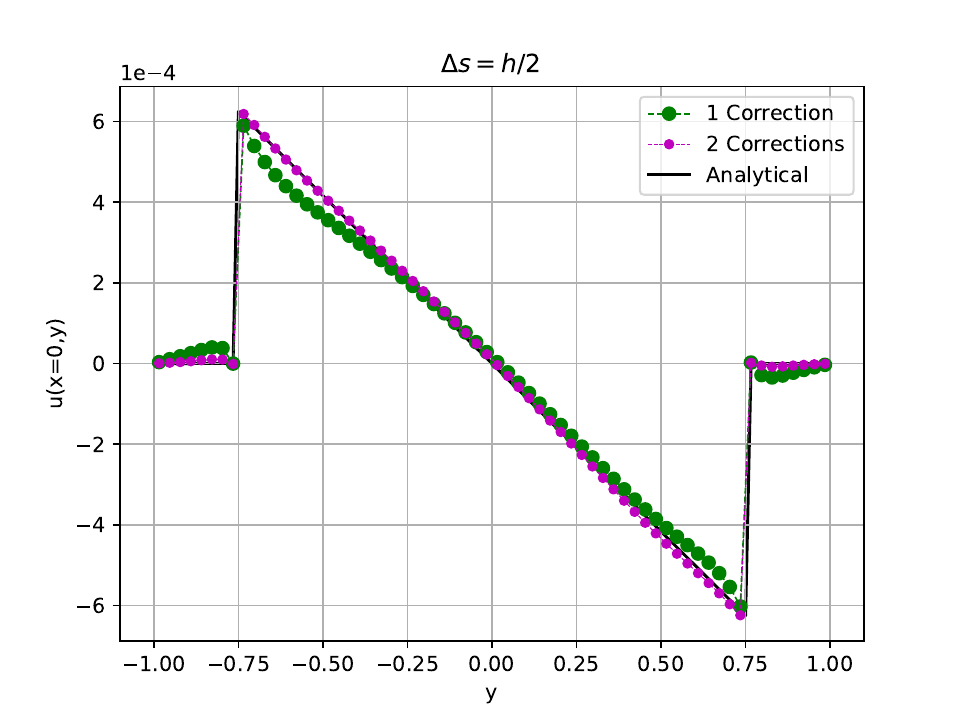}}
  \resizebox{220pt}{!}{\includegraphics{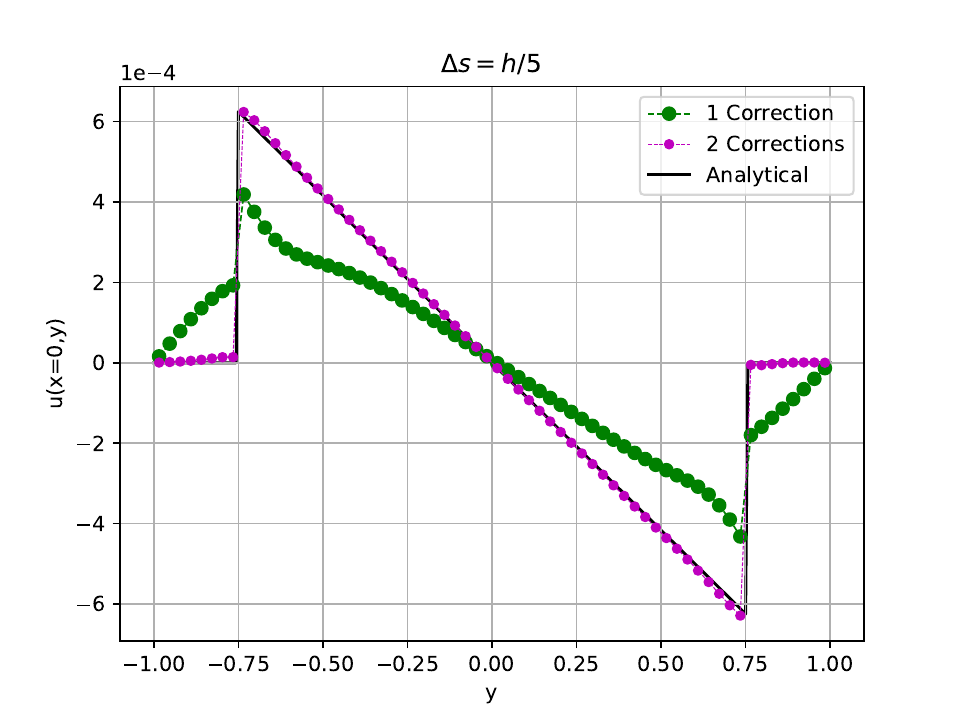}}\resizebox{220pt}{!}{\includegraphics{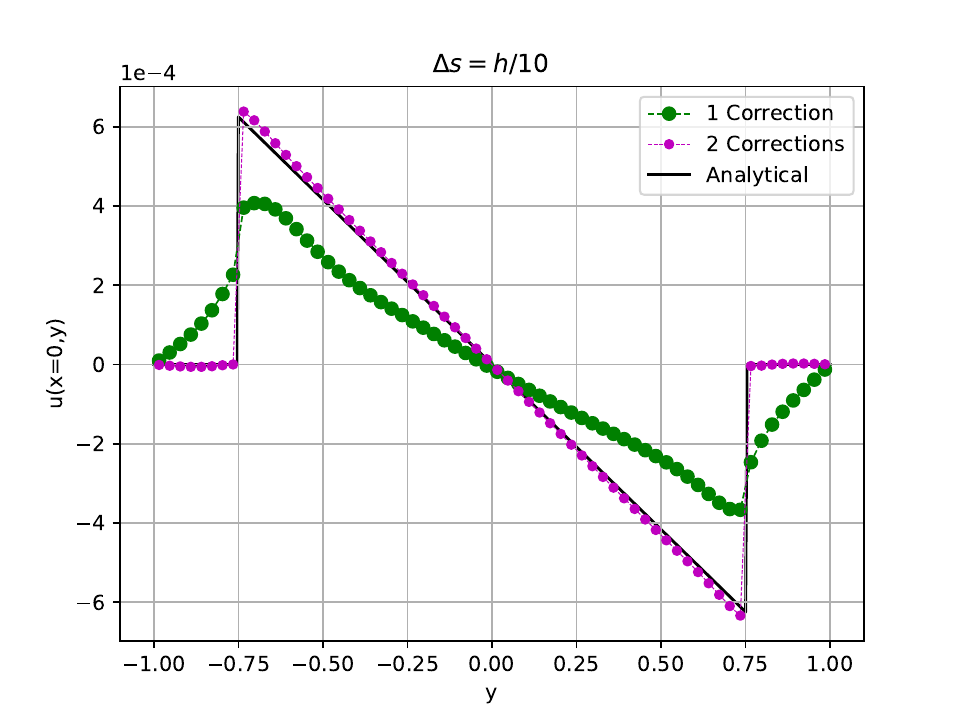}}
  
  \caption{Comparison of the velocity profile $u_x(x=0,y)$ for various separations between two concentric rotating cylinders. When $\Delta s < h$, the two correction method is substantially more accurate than the one correction method.}
\end{figure}

\subsection{Eccentric Rotating Cylinders}
This section considers the flow between two eccentric rotating cylinders in two spatial dimensions. The inner cylinder is centered at the origin, with radii $R_1 = 0.75$, whereas the outer cylinder is centered at $(\tilde{e},0) = (0.03125,0)$ and $R_2 =  0.8125$. The outer cylinder is fixed in place, and the inner cylinder rotates counter-clockwise at a rotational velocity $\omega = 8.33\cdot10^{-4}$. The fluid domain is $\Omega = [-1,1]^2$.
Figure 16 shows a schematic for the model's setup.
We set $\rho = 1$, $\mu = 0.2$, and $\mathrm{Re}=6.5\cdot 10^{-4}$.
The domain is discretized into $N = 16$, $32$, $64$, and $128$ cells in both directions such that the mesh spacing is $h = 0.125$, $0.0625$, $0.03125$, and $0.015625$ respectively.
We set the Lagrangian mesh width to be equal to the background Cartesian grid spacing. The time step size $\Delta t$ scales with $h$ such that $\Delta t = \frac{h}{100}$. We use zero normal traction and zero tangential velocity boundary conditions on each side. For this time step size and these grid spacings, the CFL number is approximately $10^{-6}$ once the model reaches steady state.
To impose rigid body motion, we set $\kappa = \frac{C_{\kappa}}{(\Delta t)^2}$, in which $C_{\kappa} = 2.0\cdot 10^{-4}$.

\begin{figure}[H]
\centering
\resizebox{180pt}{!}{

\tikzset{every picture/.style={line width=0.75pt}} 

\begin{tikzpicture}[x=0.75pt,y=0.75pt,yscale=-1,xscale=1]

\draw (324.41,198.59) node  {\includegraphics[width=254.11pt,height=254.11pt]{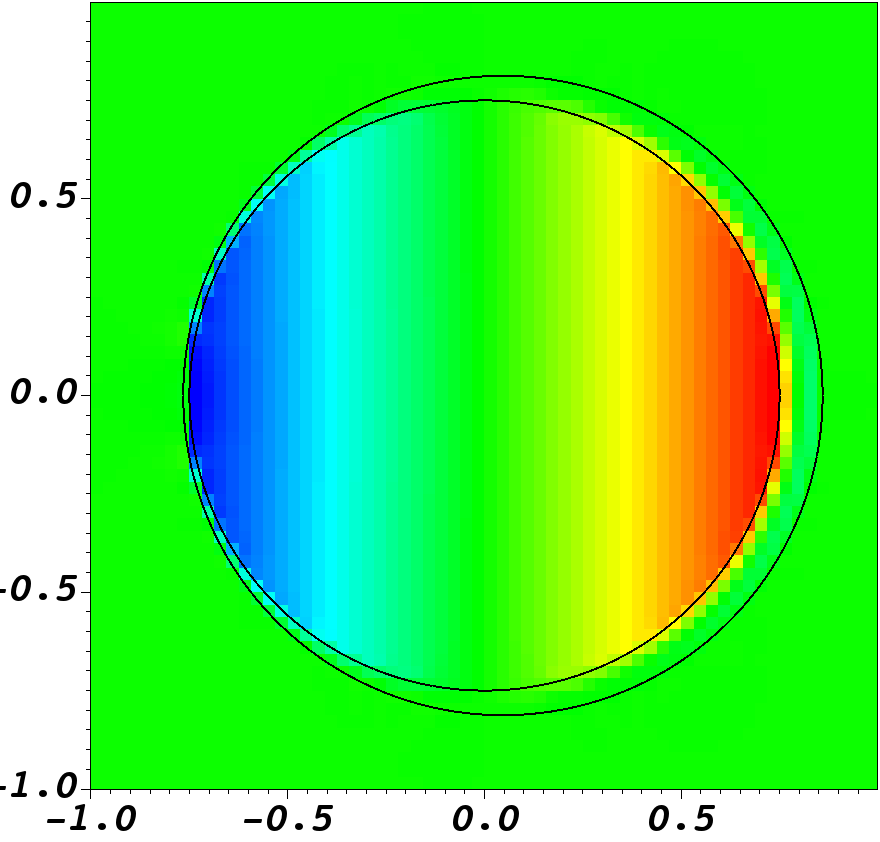}};
\draw (369.41,277.35) node  {\includegraphics[width=86.11pt,height=70.52pt]{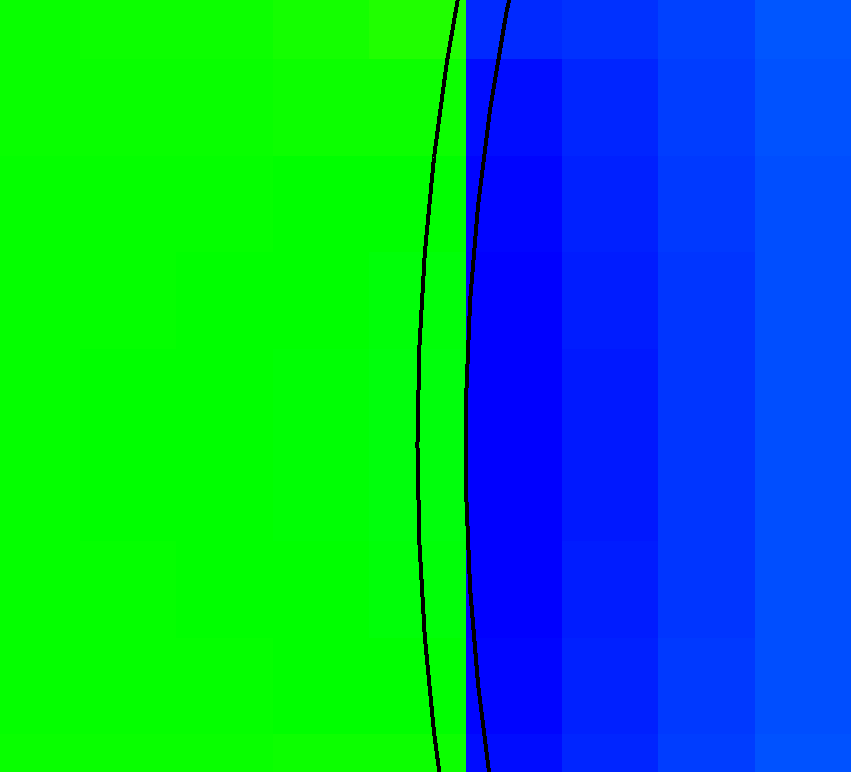}};
\draw   (216.82,172.45) -- (238.82,172.45) -- (238.82,203.45) -- (216.82,203.45) -- cycle ;
\draw   (312,230.34) -- (426.82,230.34) -- (426.82,324.36) -- (312,324.36) -- cycle ;
\draw  [dash pattern={on 0.84pt off 2.51pt}]  (216.82,203.45) -- (312,324.36) ;
\draw  [dash pattern={on 0.84pt off 2.51pt}]  (238.82,172.45) -- (426.82,230.34) ;
\end{tikzpicture}}
    \resizebox{180pt}{!}{
\tikzset{every picture/.style={line width=0.75pt}} 
\begin{tikzpicture}[x=0.75pt,y=0.75pt,yscale=-1,xscale=1]

\draw (324.41,198.59) node  {\includegraphics[width=254.11pt,height=254.11pt]{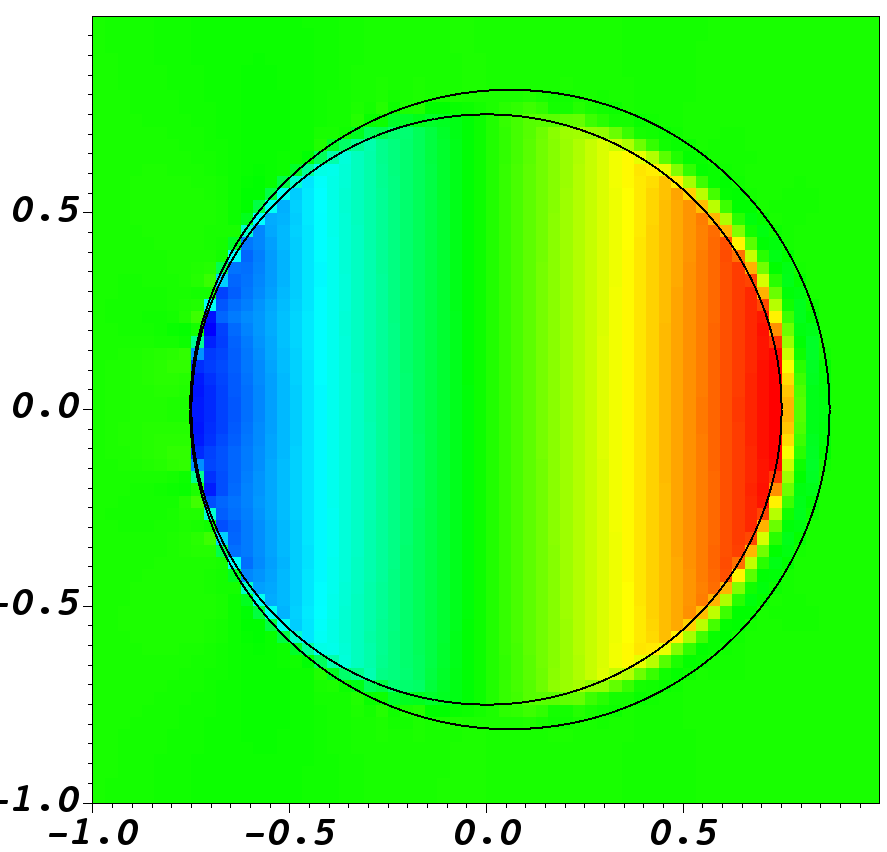}};
\draw (369.41,277.35) node  {\includegraphics[width=86.11pt,height=70.52pt]{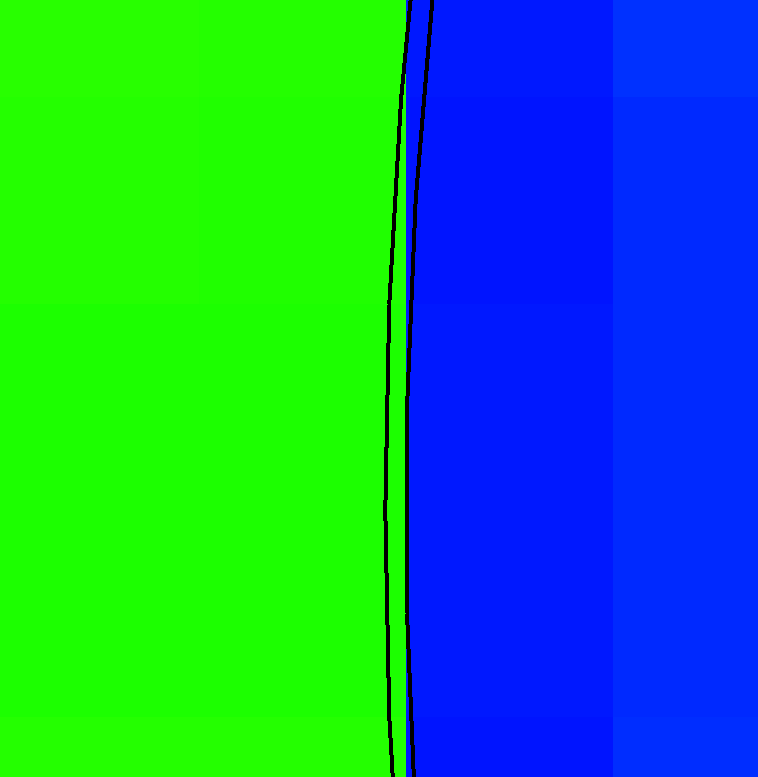}};
\draw   (216.82,182.55) -- (238.82,182.55) -- (238.82,200.64) -- (216.82,200.64) -- cycle ;
\draw   (312,230.34) -- (426.82,230.34) -- (426.82,324.36) -- (312,324.36) -- cycle ;
\draw  [dash pattern={on 0.84pt off 2.51pt}]  (216.82,200.64) -- (312,324.36) ;
\draw  [dash pattern={on 0.84pt off 2.51pt}]  (238.82,182.55) -- (426.82,230.34) ;
\end{tikzpicture}}
\resizebox{80pt}{!}{
\tikzset{every picture/.style={line width=0.75pt}} 

\begin{tikzpicture}[x=0.75pt,y=0.75pt,yscale=-1,xscale=1]

\draw (78.97,110.76) node  {\includegraphics[width=50.96pt,height=79.62pt]{colorbar_displacement.png}};
\draw [color={rgb, 255:red, 0; green, 0; blue, 0 }  ,draw opacity=1 ]   (112.94,59.2) -- (136.64,59.2) ;
\draw    (112.94,163.84) -- (136.64,163.84) ;
\draw    (112.94,110.76) -- (136.64,110.76) ;
\draw [color={rgb, 255:red, 0; green, 0; blue, 0 }  ,draw opacity=1 ]   (112.94,84.98) -- (136.64,84.98) ;
\draw    (112.94,136.54) -- (136.64,136.54) ;

\draw (70,40) node [anchor=north west][inner sep=0.75pt]  [font=\Large]  {$v$};
\draw (147.82,45.42) node [anchor=north west][inner sep=0.75pt]   [align=left] {6.1e-4};
\draw (147.82,72.72) node [anchor=north west][inner sep=0.75pt]   [align=left] {2.9e-4};
\draw (147.82,100.02) node [anchor=north west][inner sep=0.75pt]   [align=left] {2.7e-7};
\draw (147.82,125.8) node [anchor=north west][inner sep=0.75pt]   [align=left] {-3.4e-4};
\draw (147.82,151.58) node [anchor=north west][inner sep=0.75pt]   [align=left] {-6.7e-4};

\end{tikzpicture}}
  \
  \caption{The $y$-component of the velocity field for two eccentric rotating cylinders is visualized over the entire fluid domain at $t = 5$ using the two correction method. The left panel has a minimal $\Delta s$ of $\frac{h}{2}$, and the right panel has a minimal $\Delta s$ of $\frac{h}{10}$. $N = 64$ grid cells are used in either direction.}
\end{figure}

We compare the convergence of $\mathbf{u}$ under grid refinement in both the $L^2$ and $L^\infty$ norms in Figure 17. For grid resolutions in which $\Delta s < h$, the one two correction method demonstrates substantially less error. For grid resolutions in which $\Delta s \geq h$, the interface velocity interpolation stencil intersects the interfaces only once, and the two formulations yield identical results.

\begin{figure}[H]
    \center{}
  \resizebox{220pt}{!}{\includegraphics{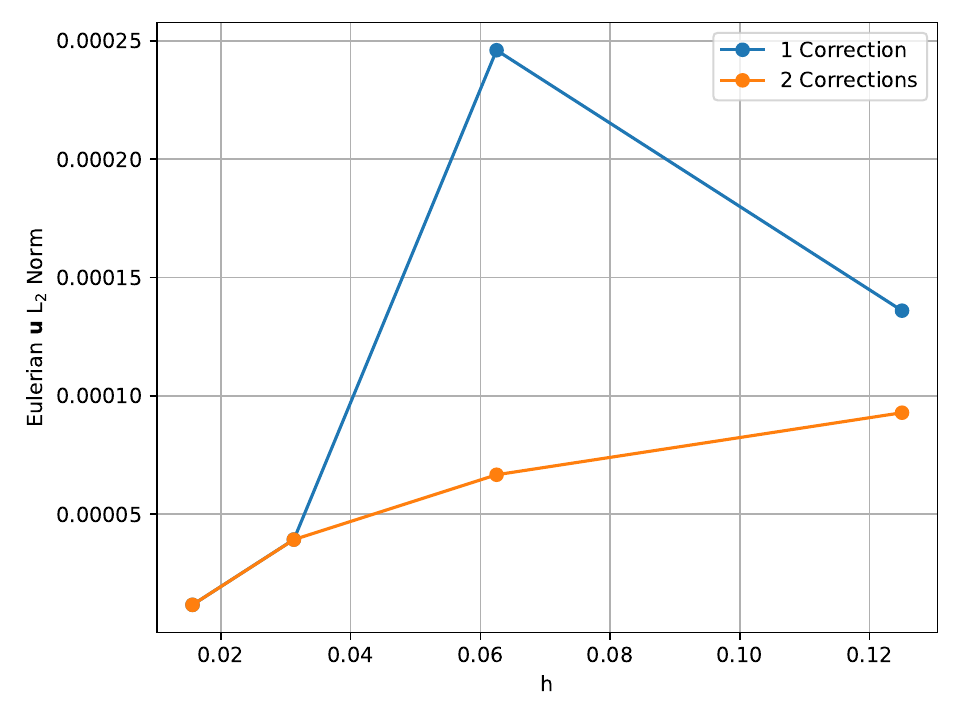}}\resizebox{220pt}{!}{\includegraphics{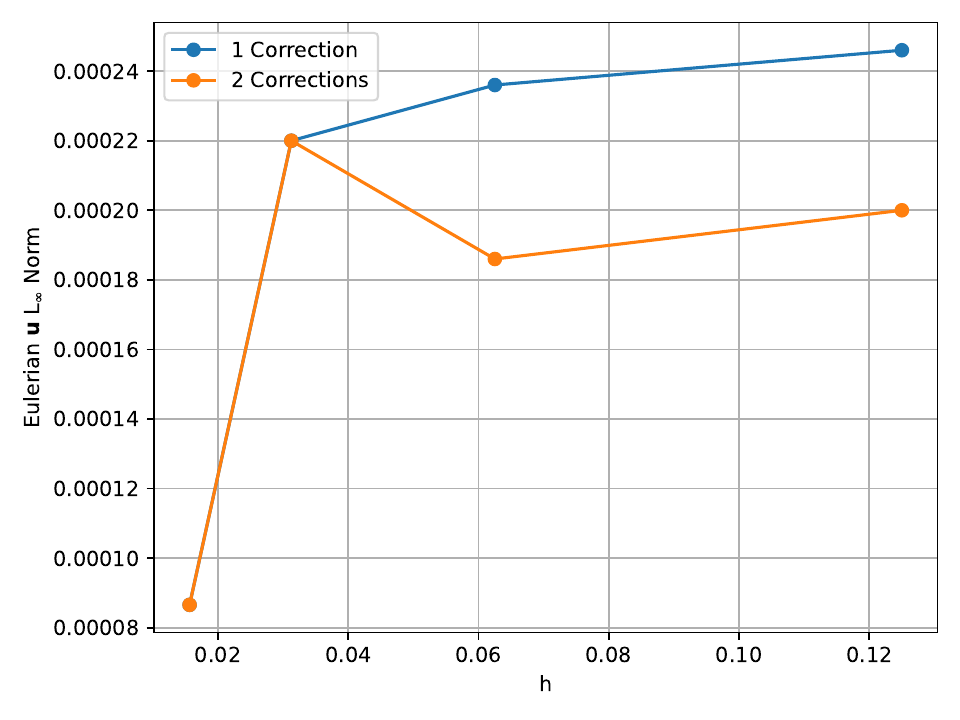}}
  
  \caption{Convergence of $\mathbf{u}$ for eccentric rotating cylinders under grid refinement. Our present two correction method is compared to the previous one correction method. Both methods converge under grid refinement, and the two correction algorithm is more accurate at coarser grid resolutions than the one correction method.}
\end{figure}

The asymptotic analytical Eulerian velocity field may be derived using lubrication theory\cite{lub_ecc}:
\begin{equation}
u(x,y) = \begin{cases}
    -\omega \, y, & 0 \leq \sqrt{x^2+y^2} \leq R_1,\\
-\omega\, y \,(1- \gamma - \frac{3 \epsilon(\gamma-\gamma^2)(2x + 3\epsilon\sqrt{x^2+y^2}+\epsilon^2 x)}{(2+\epsilon^2)(1+\epsilon x)}) &R_1<\sqrt{x^2+y^2} \land \sqrt{(x-\tilde{e})^2+y^2}< R_2,\\
0, & R_2\leq \sqrt{(x-\tilde{e})^2+y^2},
\end{cases}
\end{equation}
in which $\epsilon = \frac{R_2-R_1}{R_1}$ and $\gamma = \frac{x^2+y^2 - R_1\sqrt{x^2+y^2}}{R_2-R_1+\epsilon\,x}$.
We compute $v(x,y=0)$ from $x=-1$ to $x=1$ for both one and two correction methods, and compare their convergence properties in Figure 18. For cases in which $\Delta s < h$, the one correction formulation exhibits localized loss of accuracy near the subgrid interaction, whereas the two correction method maintains substantially improved accuracy. Even for the coarsest resolution with $N=16$ grid cells in each direction, the two correction method captures the sharp velocity transition across the narrow gap.
\begin{figure}[H]
    \centering
    \resizebox{220pt}{!}{\includegraphics{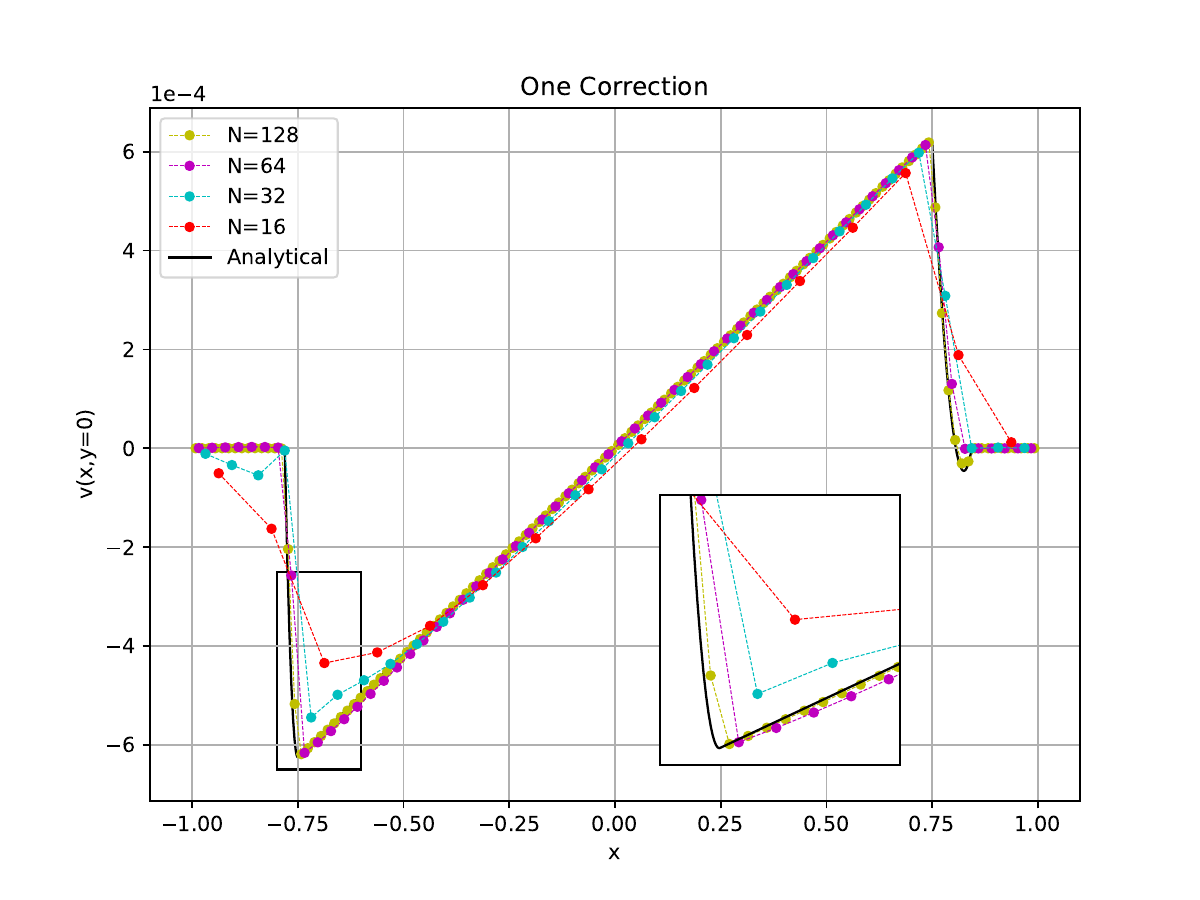}}\resizebox{220pt}{!}{\includegraphics{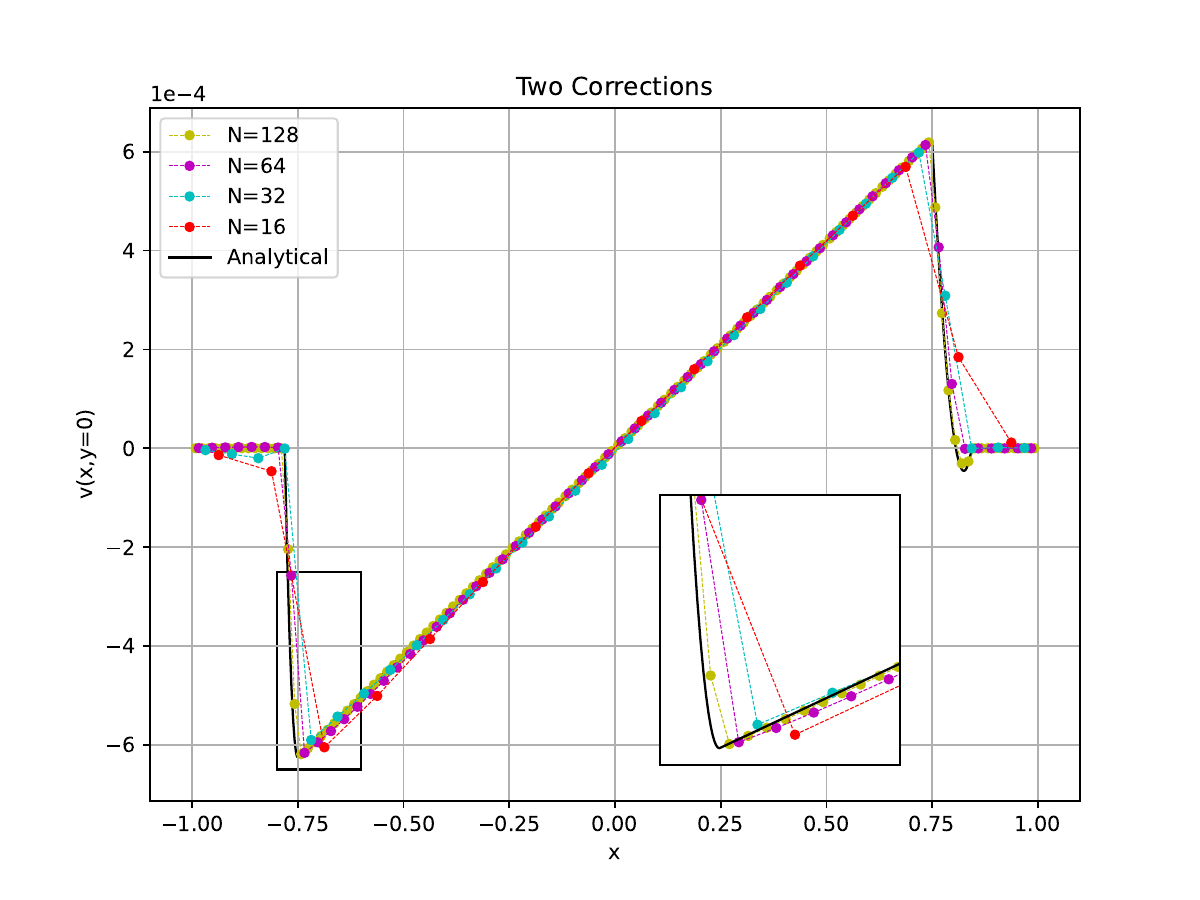}}
    \caption{Convergence of the pointwise accuracy of the velocity profile $v(x,y=0)$ across eccentric rotating cylinders. Both methods converge under grid refinement, and the two correction algorithm is more accurate near the narrow gap at coarser resolutions than the one correction method.}
\end{figure}

Last, we analyze the accuracy of the two methods for various minimal separation distances between the cylinders in Figure 19. We fix the Cartesian grid resolution using $N = 64$ grid cells in each direction. We set the center of the outer cylinder such that the minimal $\Delta s$ is $h$, $\frac{h}{2}$, $\frac{h}{5}$, and $\frac{h}{10}$, and compute $v(x,y=0)$ across the narrow gap between the cylinders. As $\Delta s \rightarrow 0$, the one correction method exhibits a progressive loss of accuracy, whereas the two correction method reliably captures the pronounced velocity transition across the narrow interfacial gap for all tested separations.

\begin{figure}[H]
    \centering
  \resizebox{220pt}{!}{\includegraphics{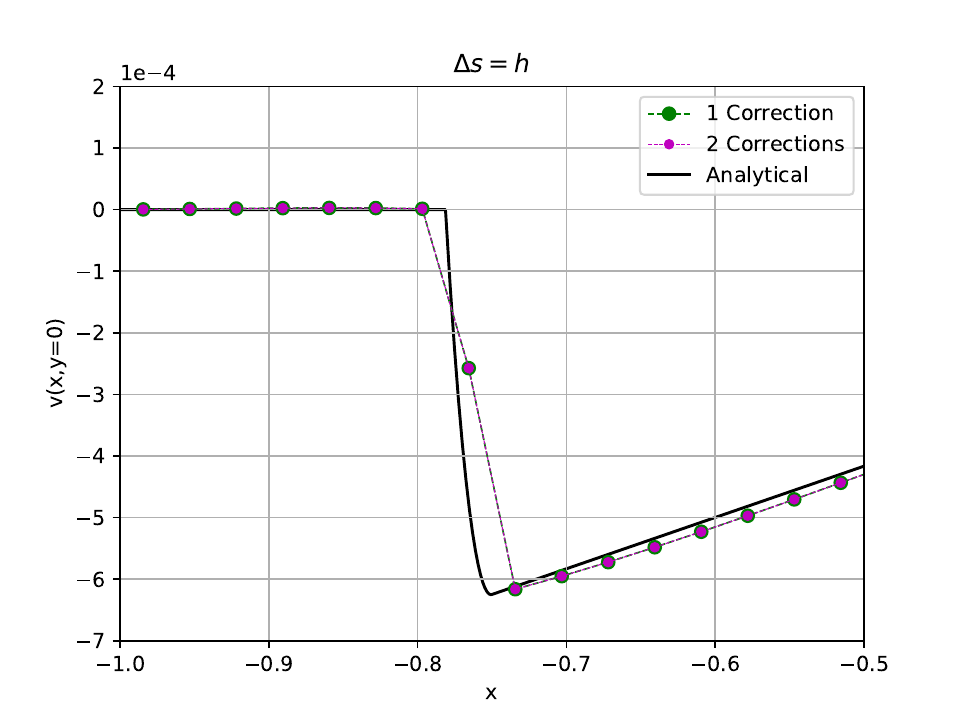}}\resizebox{220pt}{!}{\includegraphics{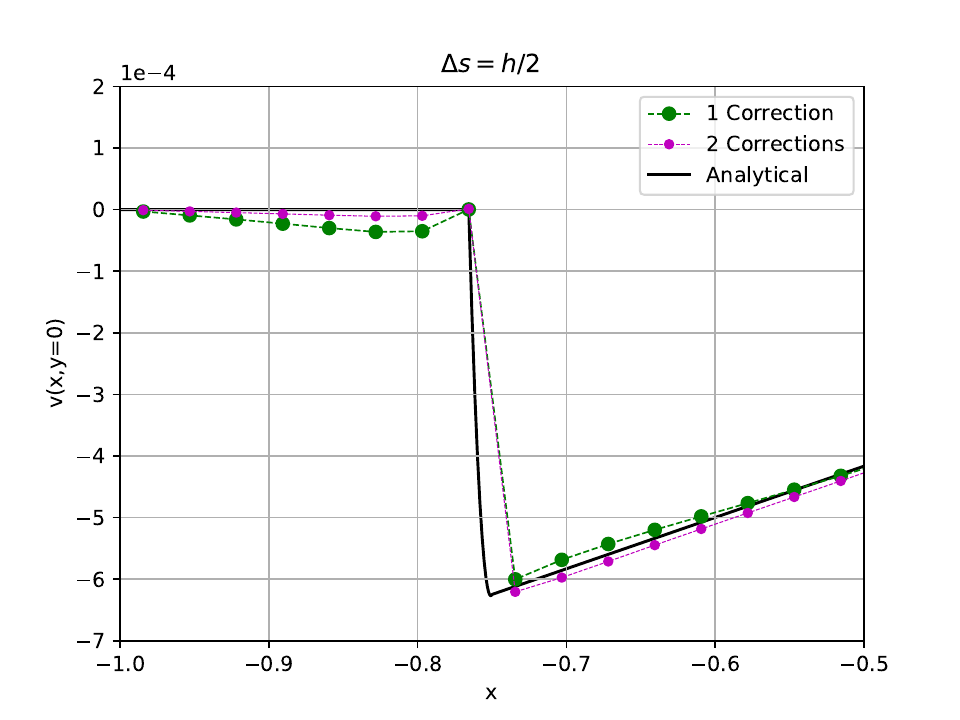}}
  \resizebox{220pt}{!}{\includegraphics{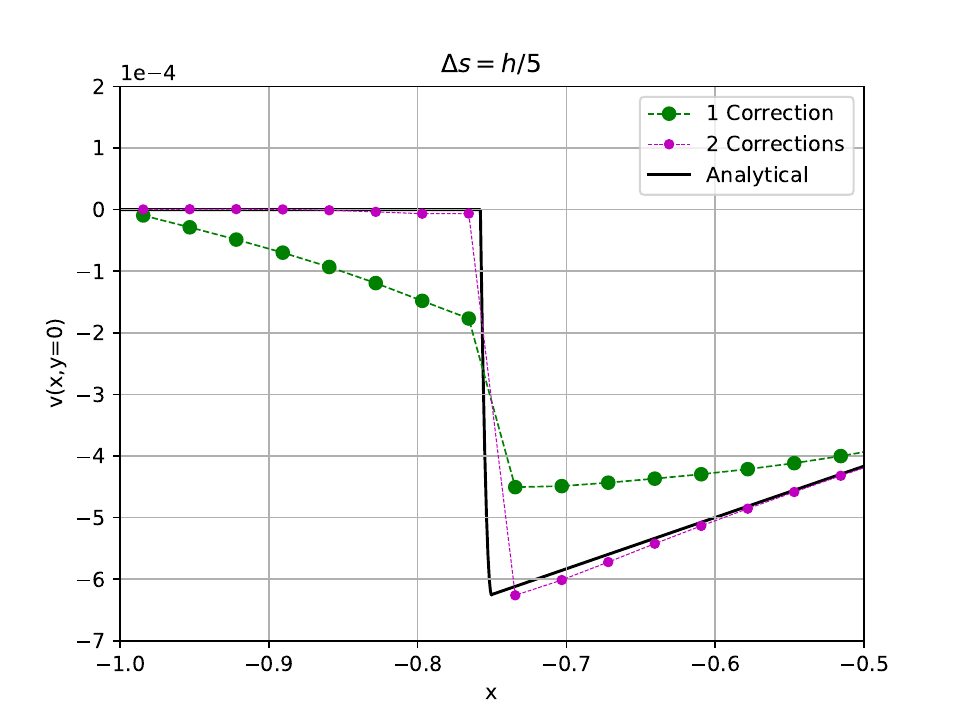}}\resizebox{220pt}{!}{\includegraphics{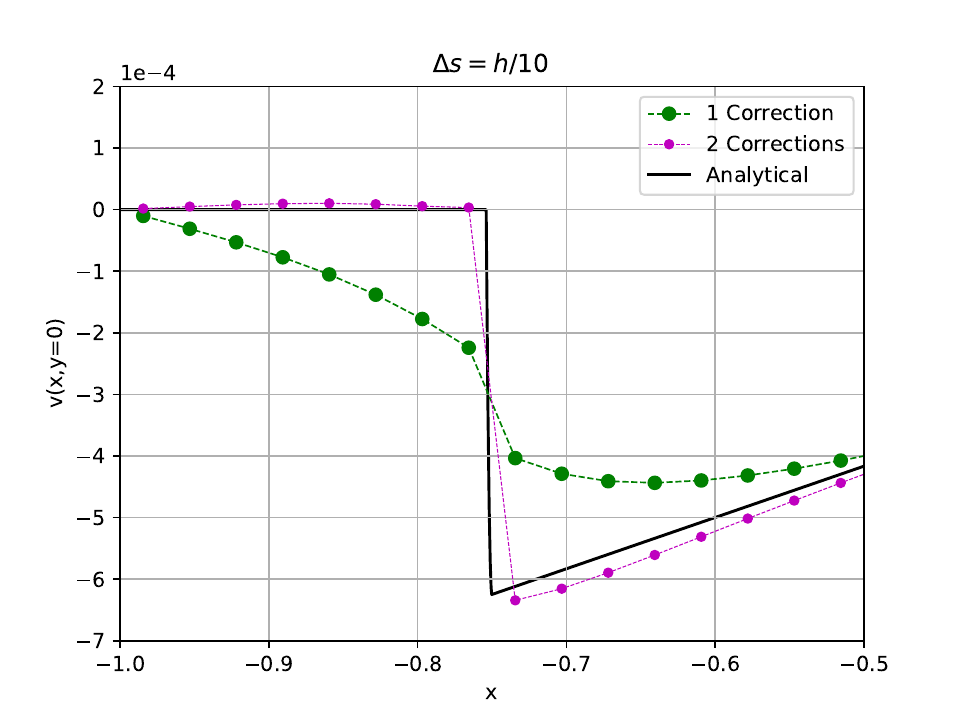}}
  
  \caption{Comparison of the velocity profile $v(x,y=0)$ for various separations between two concentric rotating cylinders. When $\Delta s < h$, the two correction method shows greater accuracy than the one correction method.}
\end{figure}

\subsection{Flow Past Anvil-Shaped Geometry}
This test investigates the flow around an anvil-head shaped interface to compare CG-IIM with two corrections with DG-IIM and CG-IIM with one correction.
 We append a square cylinder of diameter $D=1$ with a rear-facing angle of varying acuteness. Three angles are considered: $\frac
 {\pi}{2}$, $\frac{\pi}{4}$, and $\frac{\pi}{9}$. Vorticity fields around these geometries are visualized in Figure 20. The square is centered at the origin of the computational domain $\Omega = [- 8,8]^2$ with $L=16$. 
We use incoming flow velocity $\mathbf{} u = 1$, and $v = \cos \left( \pi \frac{y}{L} \right) e^{- 2 t}$
to ensure that vortex shedding occurs at a consistent time. We set $\rho = 1$ and $\mu = \frac{1}{{Re}}$, in which $Re$ is the Reynolds number. We use
${Re} = \frac{\rho U D}{\mu}=200$ to ensure that we observe vortex
shedding. 
The effective number of grid cells on the $\ell^{\text{th}}$ level is
${n_{\ell}}  = 2^{\ell - 1} n_\cc$,  with $\ell_{\text{max}}=6$. The Lagrangian mesh is twice as coarse as the background grid. The time step size $\Delta t = 0.002$ . For this time step size and grid spacing, the CFL number is approximately 0.08 once the model reaches periodic steady state. We set $\kappa = 200$ and $\eta = 1$, which are determined using a bisection method. The outflow
boundary uses zero normal and tangential traction, and the top and bottom boundaries use zero tangential traction and $v = 0$.

\begin{figure}[H]
    \centering

    
    \begin{minipage}[b]{0.49\textwidth}
        \centering
        \includegraphics[height=4cm]{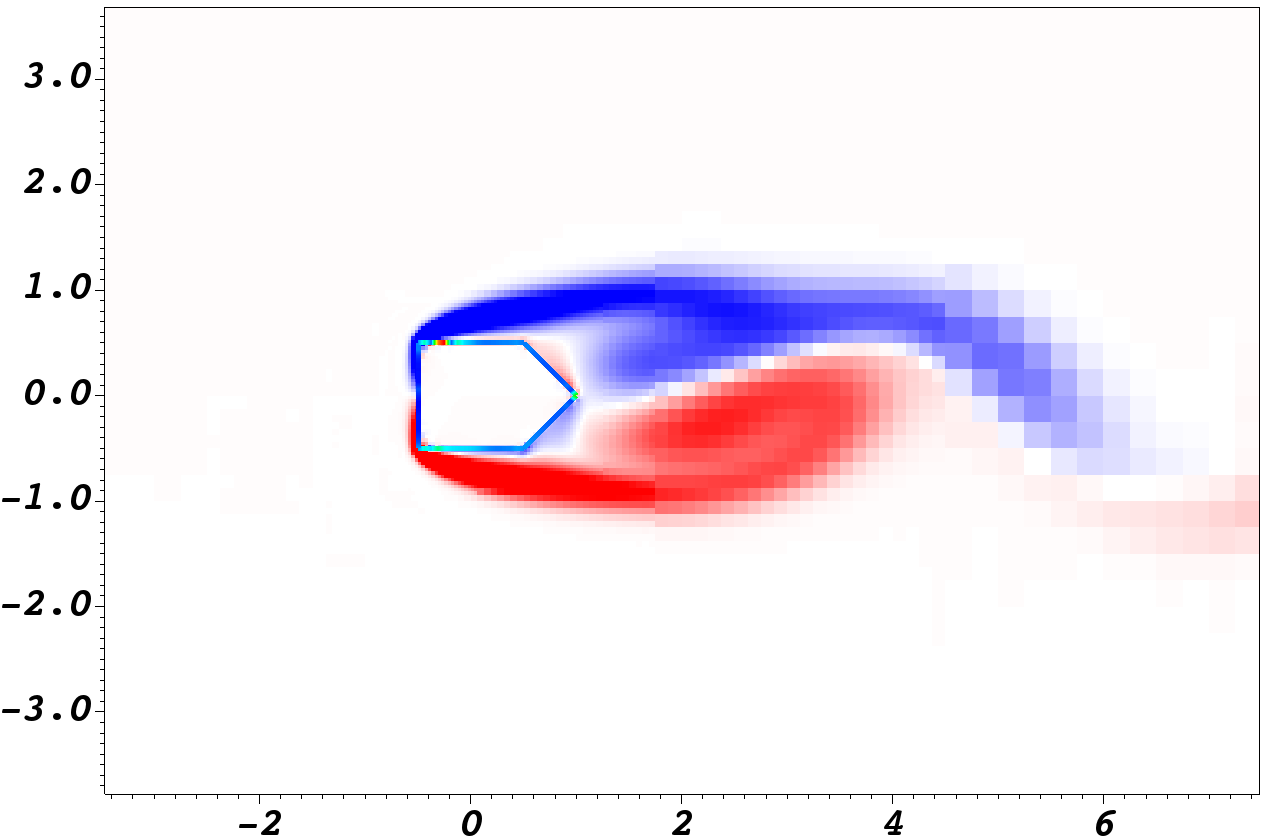}%
        \resizebox{!}{4cm}{
            \tikzset{every picture/.style={line width=0.75pt}} 
            \begin{tikzpicture}[x=0.75pt,y=0.75pt,yscale=-1,xscale=1]
                \draw (56.38,216.48) node {\includegraphics[width=17.07pt,height=238.2pt]{colorbar_displacement.png}};
                \draw [color=black] (67.76,58.68) -- (75.76,58.68);
                \draw (67.76,375.28) -- (76.76,375.28);
                \draw (67.94,215.76) -- (76.76,215.76);
                \draw [color=black] (67.94,137.98) -- (76.76,137.98);
                \draw (67.94,292.54) -- (76.76,292.54);
                \draw (48,24) node [anchor=north west][font=\Large] {$\epsilon_{\mathbf{X}}$};
                \draw (72.82,47.42) node [anchor=north west][align=left] {3.4e-6};
                \draw (73.82,126.72) node [anchor=north west][align=left] {2.5e-6};
                \draw (75.82,204.02) node [anchor=north west][align=left] {1.7e-6};
                \draw (75.82,280.8) node [anchor=north west][align=left] {8.7e-7};
                \draw (77.82,360.58) node [anchor=north west][align=left] {3.1e-8};
            \end{tikzpicture}
        }
    \end{minipage}%
    \hfill 
    \begin{minipage}[b]{0.49\textwidth}
        \centering
        \includegraphics[height=4cm]{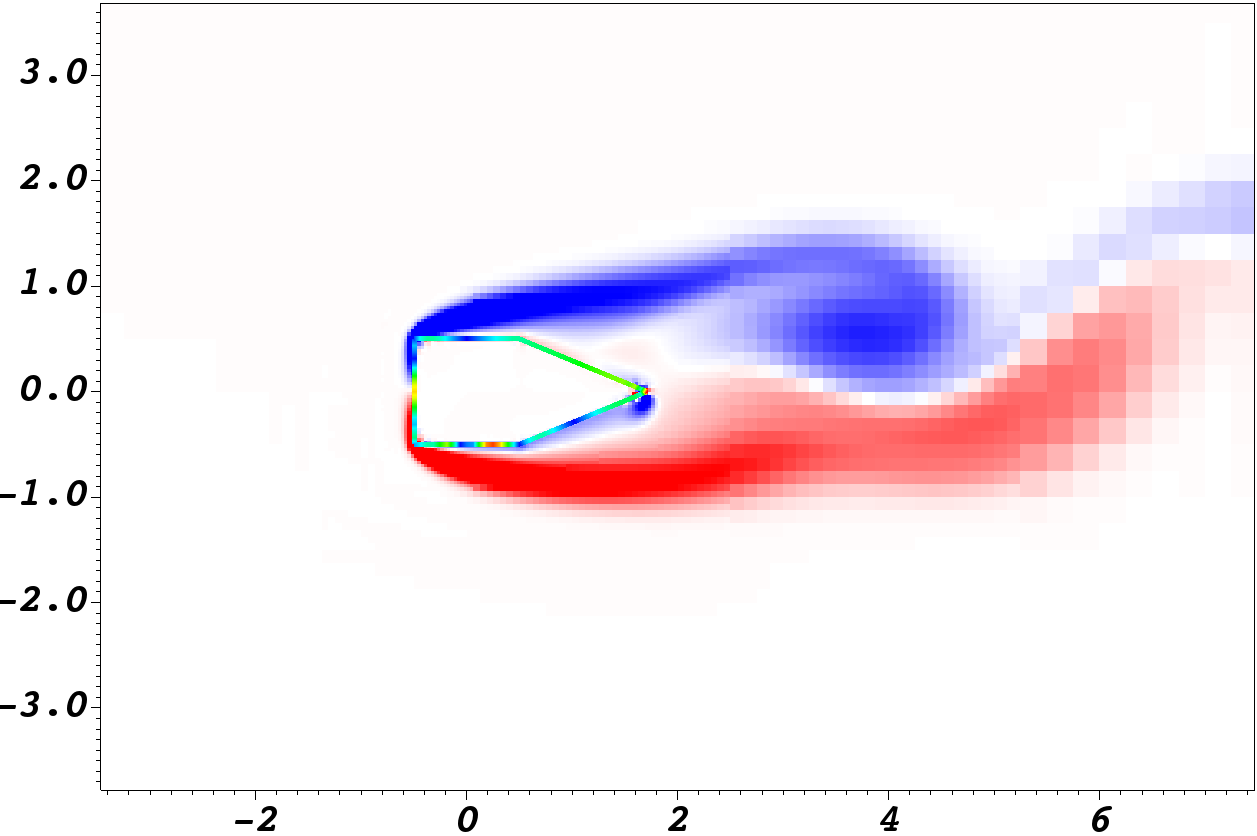}%
        \resizebox{!}{4cm}{
            \tikzset{every picture/.style={line width=0.75pt}} 

\begin{tikzpicture}[x=0.75pt,y=0.75pt,yscale=-1,xscale=1]

\draw (56.38,216.48) node  {\includegraphics[width=17.07pt,height=238.2pt]{colorbar_displacement.png}};
\draw [color={rgb, 255:red, 0; green, 0; blue, 0 }  ,draw opacity=1 ]   (67.76,58.68) -- (75.76,58.68) ;
\draw    (67.76,375.28) -- (76.76,375.28) ;
\draw    (67.94,215.76) -- (76.76,215.76) ;
\draw [color={rgb, 255:red, 0; green, 0; blue, 0 }  ,draw opacity=1 ]   (67.94,137.98) -- (76.76,137.98) ;
\draw    (67.94,292.54) -- (76.76,292.54) ;

\draw (48,24) node [anchor=north west][inner sep=0.75pt]  [font=\Large]  {$\epsilon_{\mathbf{X}}$};
\draw (72.82,47.42) node [anchor=north west][inner sep=0.75pt]   [align=left] {3.2e-6};
\draw (73.82,126.72) node [anchor=north west][inner sep=0.75pt]   [align=left] {2.4e-6};
\draw (75.82,204.02) node [anchor=north west][inner sep=0.75pt]   [align=left] {1.6e-6};
\draw (75.82,280.8) node [anchor=north west][inner sep=0.75pt]   [align=left] {8.1e-7};
\draw (77.82,360.58) node [anchor=north west][inner sep=0.75pt]   [align=left] {1.1e-8};

\end{tikzpicture}
        }
    \end{minipage}

    \vspace{0.5cm} 

    
    \begin{minipage}[b]{0.8\textwidth} 
        \centering
        \resizebox{!}{5cm}{
            \tikzset{every picture/.style={line width=0.75pt}} 
            \begin{tikzpicture}[x=0.75pt,y=0.75pt,yscale=-1,xscale=1]
                \draw (56.38,216.48) node {\includegraphics[width=17.07pt,height=238.2pt]{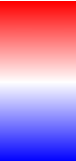}};
                \draw [color=black] (67.76,58.68) -- (75.76,58.68);
                \draw (67.76,375.28) -- (76.76,375.28);
                \draw (67.94,215.76) -- (76.76,215.76);
                \draw [color=black] (67.94,137.98) -- (76.76,137.98);
                \draw (67.94,292.54) -- (76.76,292.54);
                \draw (48,24) node [anchor=north west][font=\Large] {$\omega$};
                \draw (72.82,47.42) node [anchor=north west][align=left] {3.0};
                \draw (73.82,126.72) node [anchor=north west][align=left] {1.5};
                \draw (75.82,204.02) node [anchor=north west][align=left] {0.0};
                \draw (75.82,280.8) node [anchor=north west][align=left] {-1.5};
                \draw (77.82,360.58) node [anchor=north west][align=left] {-3.0};
            \end{tikzpicture}
        }
        \includegraphics[height=5cm]{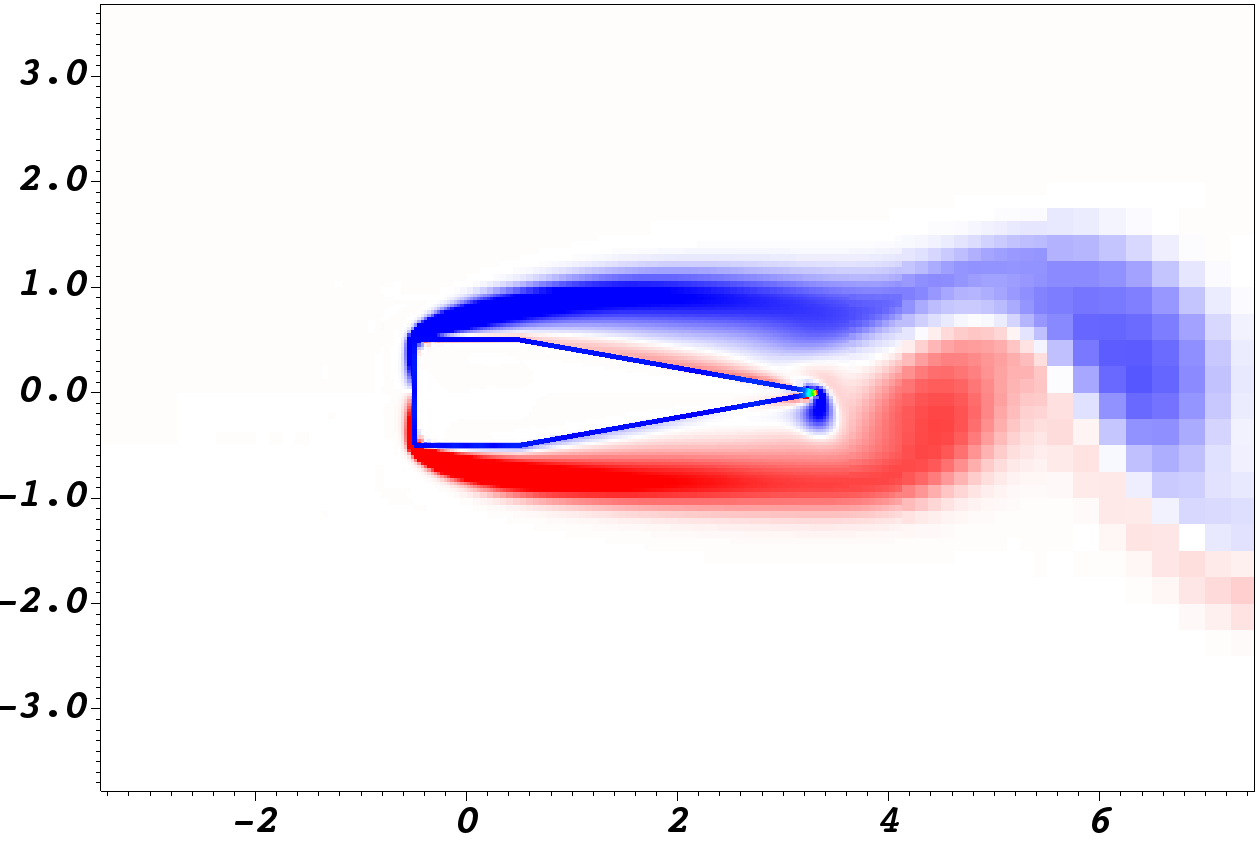}%
        \resizebox{!}{5cm}{
            \tikzset{every picture/.style={line width=0.75pt}} 
            \begin{tikzpicture}[x=0.75pt,y=0.75pt,yscale=-1,xscale=1]
                \draw (56.38,216.48) node {\includegraphics[width=17.07pt,height=238.2pt]{colorbar_displacement.png}};
                \draw [color=black] (67.76,58.68) -- (75.76,58.68);
                \draw (67.76,375.28) -- (76.76,375.28);
                \draw (67.94,215.76) -- (76.76,215.76);
                \draw [color=black] (67.94,137.98) -- (76.76,137.98);
                \draw (67.94,292.54) -- (76.76,292.54);
                \draw (48,24) node [anchor=north west][font=\Large] {$\epsilon_{\mathbf{X}}$};
                \draw (72.82,47.42) node [anchor=north west][align=left] {6.7e-5};
                \draw (73.82,126.72) node [anchor=north west][align=left] {5.0e-5};
                \draw (75.82,204.02) node [anchor=north west][align=left] {3.3e-5};
                \draw (75.82,280.8) node [anchor=north west][align=left] {1.7e-5};
                \draw (77.82,360.58) node [anchor=north west][align=left] {4.2e-9};
            \end{tikzpicture}
        }
    \end{minipage}

    \caption{The vorticity field around anvil-head shaped geometries are visualized for angles $\frac
 {\pi}{2}$, $\frac{\pi}{4}$, and $\frac{\pi}{9}$ using the two correction CG-IIM.  }
\end{figure}

To quantify the accuracy of the proposed method, we monitor the deformation error, $\epsilon_{\mathbf{X}}$, for each of the three rear-facing angles using the three different IIM formulations. 
Figure 21 presents the maximum error observed over the time interval $t\in[0,50]$. The results indicate that the two-correction CG-IIM yields significantly improved accuracy compared to the one-correction approaches. Notably, this formulation also handles sharp geometries more effectively than the DG-IIM approach reported in Facci et al.~\cite{faccisharp}. While the one-correction CG-IIM shows little sensitivity to the angle for sufficiently large (blunter) angles, the two-correction CG-IIM consistently maintains superior accuracy across all three angles tested.

\begin{figure}[H]
\centering
  \begin{minipage}{0.7\textwidth}
    \centering
    \includegraphics[width=1.2\textwidth]{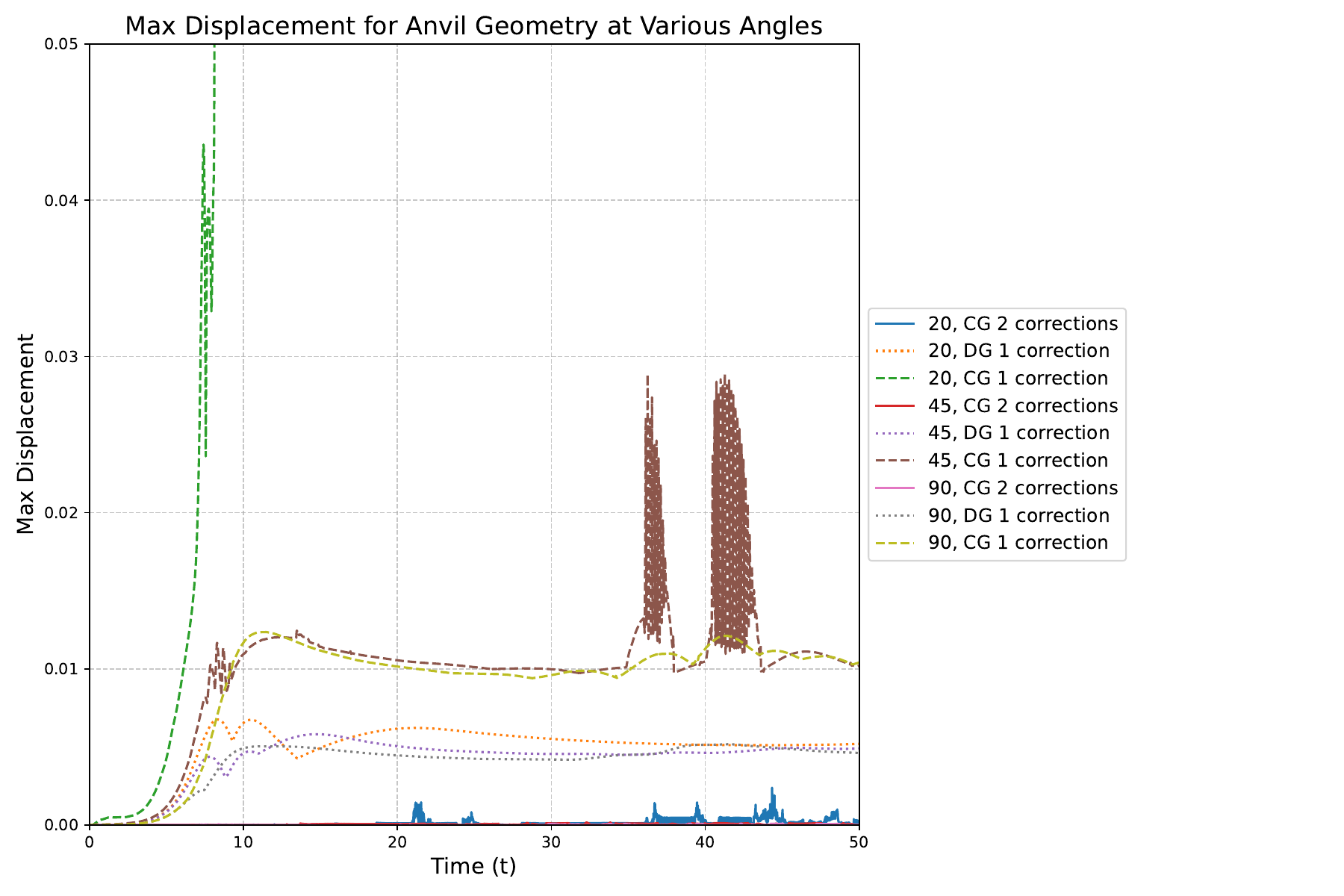}
  \end{minipage}
  
  \begin{minipage}{0.7\textwidth}
    \centering
    \includegraphics[width=1.2\textwidth]
    {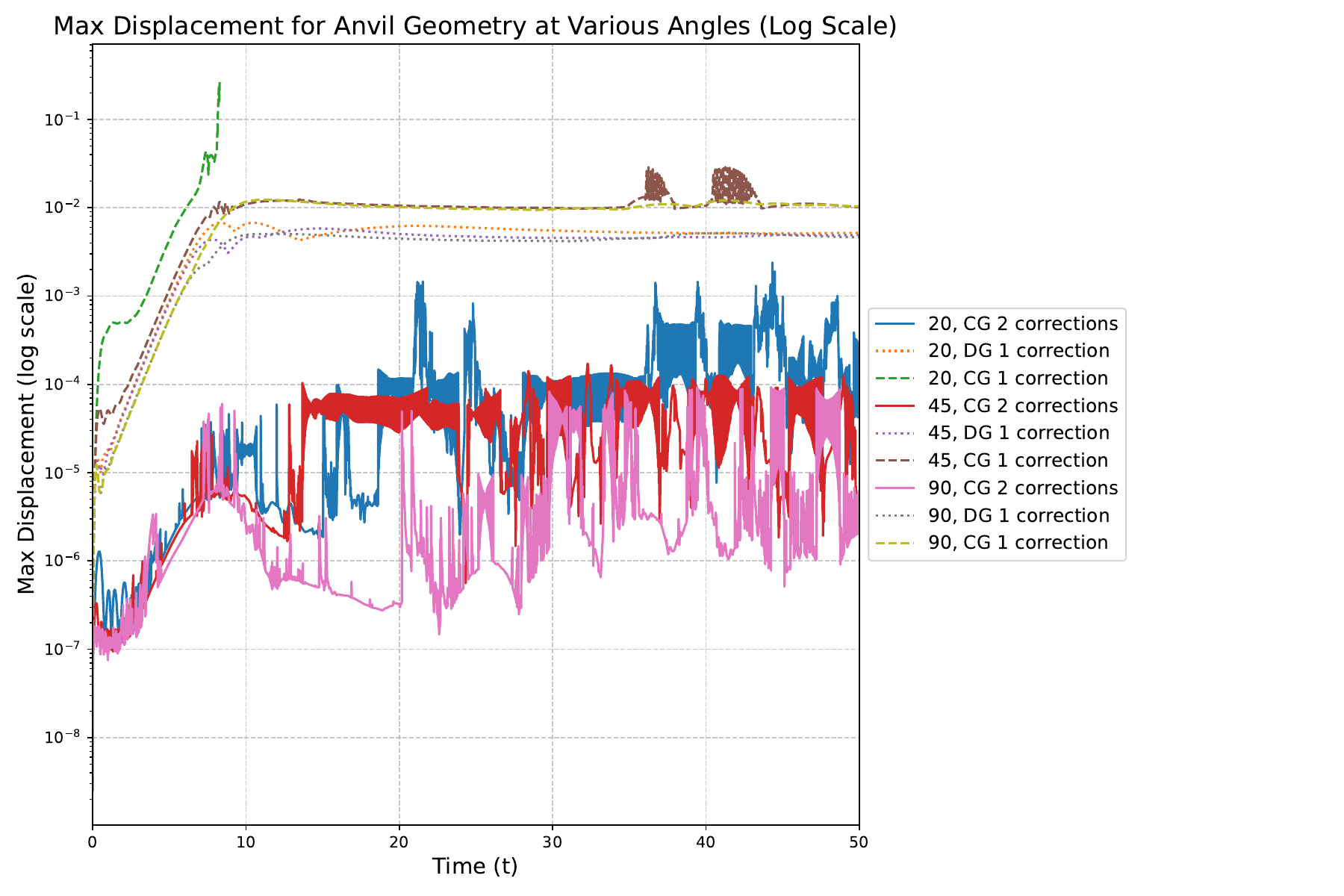}
  \end{minipage}
  \caption{We measure $\mathrm{max}(\epsilon_{\mathbf{X}})$ over the interval $t\in[0,50]$ using the two-correction CG-IIM and the one-correction CG-IIM and DG-IIM. The two-correction CG-IIM demonstrates a substantial improvement in accuracy over both one correction methods at all angles tested.}
\end{figure}

\subsection{Flow Past a Star}
This section considers the flow past a star-shaped interface at various Reynolds numbers. This test serves to compare the accuracy of our two correction CG-IIM with the previous one correction CG-IIM and one correction DG-IIM. 
The star is centered at the origin of the computational domain $\Omega = [- 8, 8]^2 $  with $L=16$. 
We use incoming flow velocity $\mathbf{} u = 1$, and $v = \cos \left( \pi \frac{y}{L} \right) e^{- 2 t}$
to ensure that vortex shedding occurs at a consistent time. We set $\rho = 1$ and $\mu = \frac{1}{\text{Re}}$. We use
$\text{Re} = \frac{\rho U D}{\mu}=150$, $200$, and $300$. 
The effective number of grid cells on the $\ell^{\text{th}}$ level is
${n_{\ell}}  = 2^{\ell - 1} n_\cc$,  with $\ell_{\text{max}}=6$. The Lagrangian mesh's coarsening ratio is 1.6. The time step size $\Delta t$ is fixed at $0.002$. For this time step size and grid spacing, the CFL number is approximately 0.18 once the model reaches periodic steady state. We set $\kappa = 200$ and $\eta = 1$. 
The outflow
boundary uses zero normal and tangential traction, and the top and bottom boundaries use zero tangential traction and $v = 0$.
Figure 22 shows a snapshot of the velocity field at $t=50$ after the onset of vortex shedding.

\begin{figure}[H]
\center{}
  \resizebox{300pt}{!}{\includegraphics{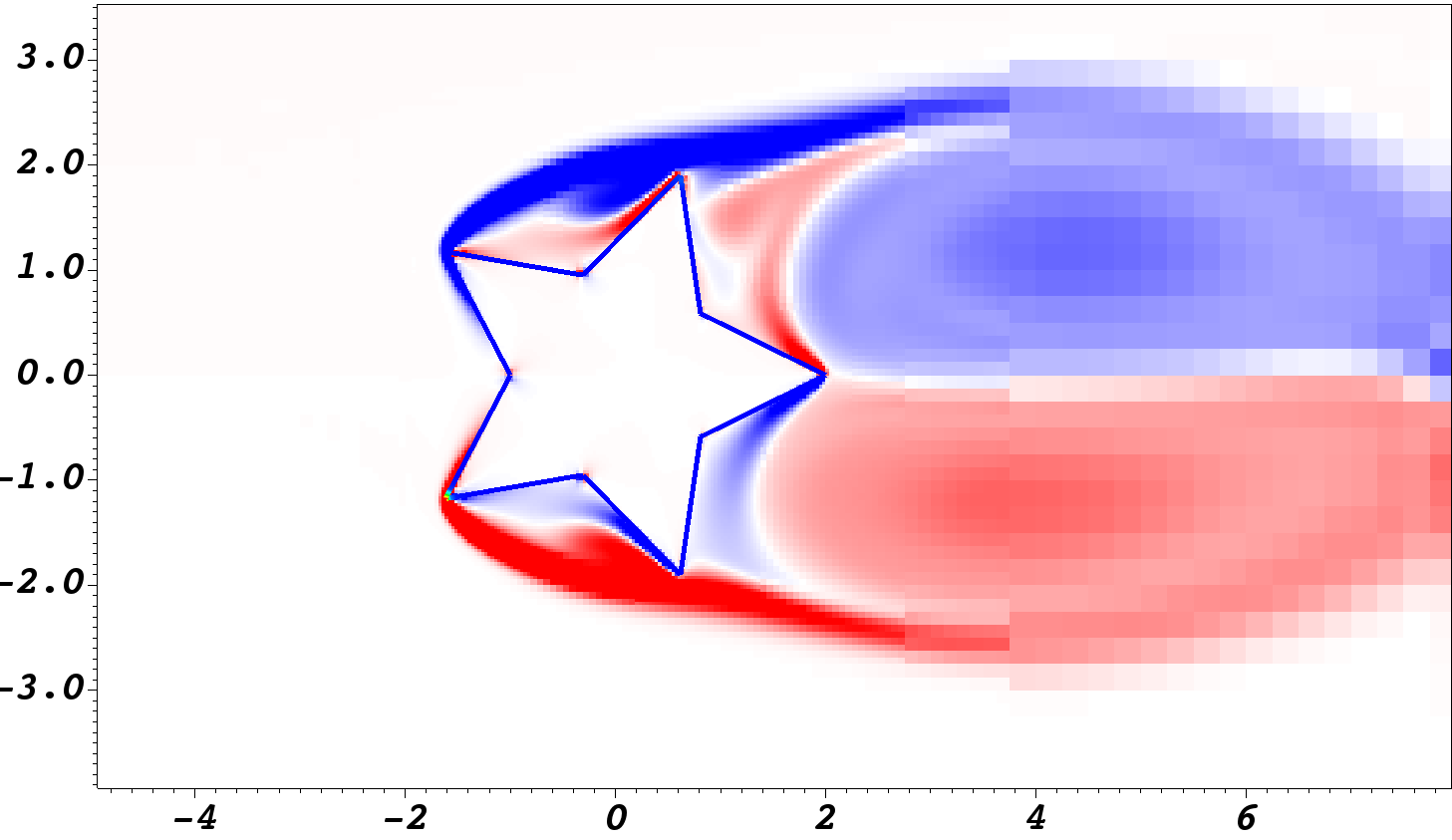}}\resizebox{80pt}{!}{

\tikzset{every picture/.style={line width=0.75pt}} 

\begin{tikzpicture}[x=0.75pt,y=0.75pt,yscale=-1,xscale=1]

\draw (83.97,256.76) node  {\includegraphics[width=50.96pt,height=79.62pt]{colorbar_displacement.png}};
\draw [color={rgb, 255:red, 0; green, 0; blue, 0 }  ,draw opacity=1 ]   (117.94,205.2) -- (141.64,205.2) ;
\draw    (117.94,309.84) -- (141.64,309.84) ;
\draw    (117.94,256.76) -- (141.64,256.76) ;
\draw [color={rgb, 255:red, 0; green, 0; blue, 0 }  ,draw opacity=1 ]   (117.94,230.98) -- (141.64,230.98) ;
\draw    (117.94,282.54) -- (141.64,282.54) ;
\draw [color={rgb, 255:red, 0; green, 0; blue, 0 }  ,draw opacity=1 ]   (119.94,60.2) -- (143.64,60.2) ;
\draw    (119.94,164.84) -- (143.64,164.84) ;
\draw    (119.94,111.76) -- (143.64,111.76) ;
\draw [color={rgb, 255:red, 0; green, 0; blue, 0 }  ,draw opacity=1 ]   (119.94,85.98) -- (143.64,85.98) ;
\draw    (119.94,137.54) -- (143.64,137.54) ;
\draw (82.94,111.58) node  {\includegraphics[width=52.5pt,height=80.07pt]{vorticity_colorbar.png}};

\draw (156.82,155.58) node [anchor=north west][inner sep=0.75pt]   [align=left] {-3.0};
\draw (156.82,129.8) node [anchor=north west][inner sep=0.75pt]   [align=left] {-1.5};
\draw (156.82,104.02) node [anchor=north west][inner sep=0.75pt]   [align=left] {0.0};
\draw (156.82,76.72) node [anchor=north west][inner sep=0.75pt]   [align=left] {1.5};
\draw (156.82,49.42) node [anchor=north west][inner sep=0.75pt]   [align=left] {3.0};
\draw (70,29.4) node [anchor=north west][inner sep=0.75pt]  [font=\Large]  {$\omega$};
\draw (70,180) node [anchor=north west][inner sep=0.75pt]  [font=\Large]  {$\varepsilon_{\mathbf{X}}$};
\draw (154.82,194.42) node [anchor=north west][inner sep=0.75pt]   [align=left] {1.5e-4};
\draw (154.82,221.72) node [anchor=north west][inner sep=0.75pt]   [align=left] {1.1e-4};
\draw (154.82,249.02) node [anchor=north west][inner sep=0.75pt]   [align=left] {7.5e-5};
\draw (154.82,274.8) node [anchor=north west][inner sep=0.75pt]   [align=left] {3.7e-5};
\draw (154.82,300.58) node [anchor=north west][inner sep=0.75pt]   [align=left] {6.5e-9};

\end{tikzpicture}}
  \caption{The vorticity field is visualized near the star-shaped interface at $\mathrm{Re}=300$ using the two-correction CG-IIM at $t=80$.  The largest $\varepsilon_{\mathbf{X}}$ is found on the inflow-facing segments of the mesh.}
\end{figure}

We compare $\varepsilon_\mathbf{X}$ of the two-correction CG-IIM and the one-correction DG-IIM and CG-IIM at $\mathrm{Re}=150$, $200$, and $300$ to assess the methods' impact on accuracy (Figure 23). We observe at least an order of magnitude improvement in accuracy for the two-correction method over both of the one-correction methods. This methood also handles the concave and convex corners more effectively than the DG-IIM approach in Facci et al.~\cite{faccisharp}

\begin{figure}[H]
\centering
  \begin{minipage}{0.7\textwidth}
    \centering
    \includegraphics[width=1.2\textwidth]{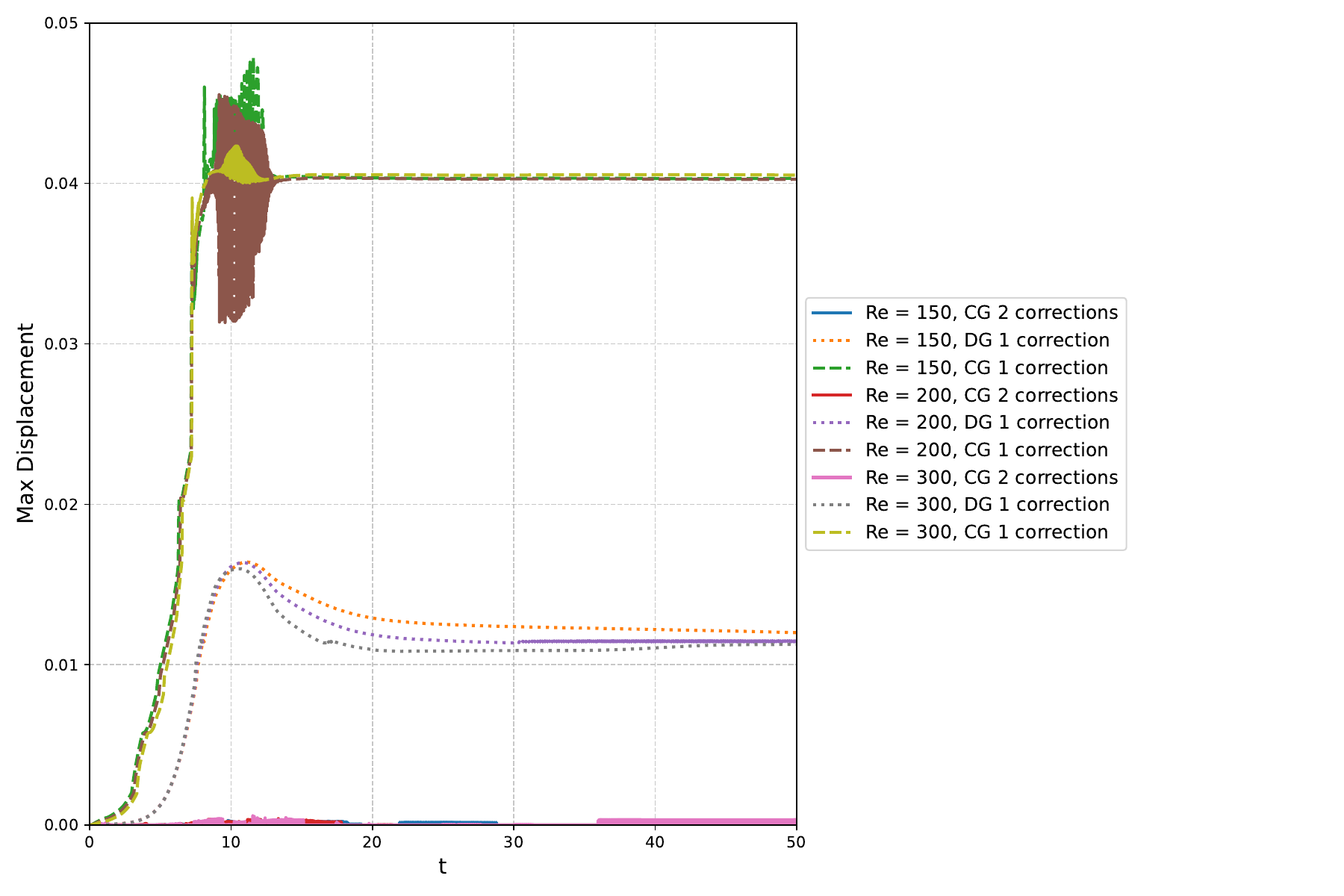}
  \end{minipage}
  
  \begin{minipage}{0.7\textwidth}
    \centering
    \includegraphics[width=1.2\textwidth]
{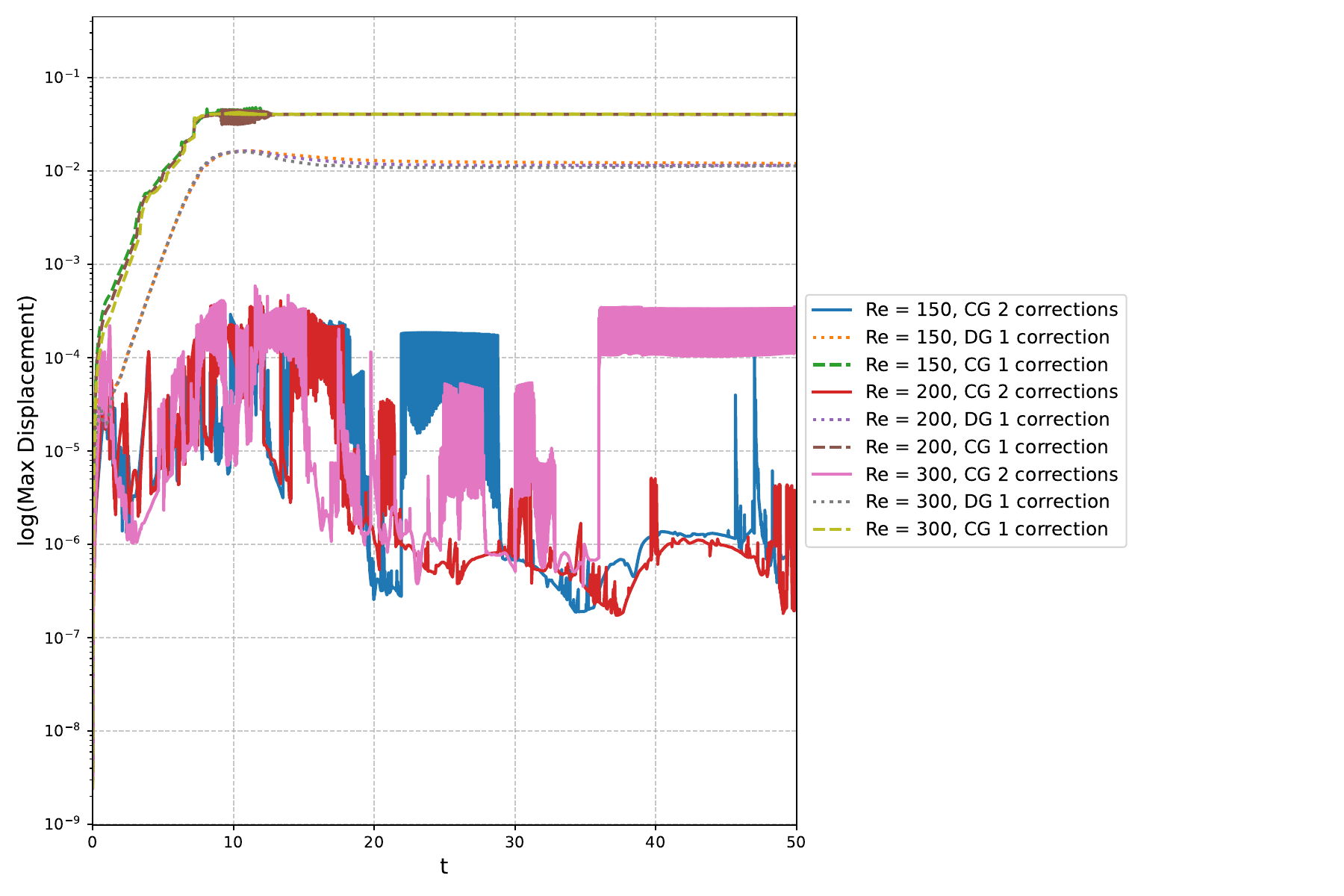}
  \end{minipage}
  \caption{We measure $\mathrm{max}(\epsilon_{\mathbf{X}})$ over the interval $t\in[0,80]$ at $\mathrm{Re} = 150$, $200$, and $300$ using the two-correction CG-IIM and the one-correction CG-IIM and DG-IIM. The left panel uses a linear scale, and the right panel uses a log scale on the vertical axis. The two-correction CG-IIM demonstrates a substantial improvement in accuracy over both one correction methods at all angles tested.}
\end{figure}

\section{Discussion}
By modifying the velocity interpolation and force spreading operators with secondary correction terms, we can effectively model flows between interfaces in near contact and around interfaces with sharp features without requiring excessive grid refinement. 
While our current numerical analyses focus on systems involving two interfaces, the method may be readily extended to more interfaces by collecting additional correction terms at intersections. 
A distinct advantage of this formulation is that it does not rely on prior knowledge of interface orientation, separation distance, or analytic parameterizations, unlike many lubrication-based approaches \cite{FAI2018319,YACOUBI2019108919}. 
Instead, our method reconstructs the interfacial velocity by implicitly assuming that the velocity field within the gap varies linearly. This approach allows for broad compatibility with finite element structural models and complex geometries where analytic gap functions are unavailable.

This linear assumption yields excellent agreement with canonical test problems in tribology. For the piecewise linear velocity field surrounding shearing parallel plates, the method maintains high accuracy for interface separations as small as $\frac{h}{50}$. We observe similar improvements for both concentric and eccentric rotating cylinder models, with substantial gains in the accuracy of both Eulerian and Lagrangian velocity fields compared to standard methods. Furthermore, geometries with sharp features show an increase in accuracy of at least an order of magnitude compared to prior approaches.

To extend this method to three spatial dimensions, addressing numerical stiffness is a crucial. The current explicit time-stepping scheme can be limiting when large penalty parameters are required. Transitioning to an implicit time-stepping scheme would maintain stability for larger time step sizes and accelerate access to high-resolution results, though formulating a fast implicit solver for an IIM in incompressible flows remains an open problem. Some prior works\cite{lee_immersed_2003,LAYTON2009266, FABENDER20241} have designed implicit formulations, but are yet to be adapted for discrete interface representations.

Finally, while the present work utilizes a linear reconstruction of the velocity field in the gap, the framework allows for higher-order extensions. Future work could generalize the interpolation to a quadratic reconstruction. This would capture more complex flow profiles within the gap, further enhancing accuracy and extending the applicability of the method to a broader range of fluid-structure interaction regimes.

\section*{Acknowledgments} 
We gratefully acknowledge research support through NIH Awards R01HL157631, R03HL182166, and U01HL143336, and NSF award OAC 1931516. Computations were performed using
facilities provided by University of North Carolina at Chapel Hill through the Research Computing division of UNC Information Technology Services. We gratefully acknowledge Dr. Ebrahim M. Kolahdouz for providing foundational code that facilitated the setup of our numerical experiments. We are grateful to Dr. Cole Gruninger for his valuable input and helpful discussions concerning the setup of the numerical experiments. We also thank Dr. David R. Wells for developing code infrastructure that facilitated the efficient implementation of our intersection algorithm.

This manuscript is the result of funding in whole or in part by the National Institutes of Health (NIH). It is subject to the NIH Public Access Policy. Through acceptance of this federal funding, NIH has been given a right to make this manuscript publicly available in PubMed Central upon the Official Date of Publication, as defined by NIH.

\section*{Declaration of generative AI and AI-assisted technologies in the manuscript preparation process}

During the preparation of this work the authors used ChatGPT and Gemini in order to improve grammar and readability. After using this tool, the authors reviewed and edited the content as needed and take full responsibility for the content of the published article.

\bibliographystyle{unsrt}

\begin{thebibliography}{10}

\bibitem{griffith_immersed_2012}
B.~E. Griffith.
\newblock Immersed boundary model of aortic heart valve dynamics with physiological driving and loading conditions.
\newblock {\em International Journal for Numerical Methods in Biomedical Engineering}, 28(3):317--345, 2012.
\newblock
\url{https://doi.org/10.1002/cnm.1445}

\bibitem{peskin_flow_1972}
C.~S. Peskin.
\newblock Flow patterns around heart valves: A numerical method.
\newblock {\em Journal of Computational Physics}, 10(2):252--271, 1972.
\newblock
\url{https://doi.org/10.1016/0021-9991(72)90065-4}

\bibitem{zeng_immersed_2023}
Y.~Zeng, Y.~Wang, D.~Yang, and Q.~Chen.
\newblock Immersed boundary methods for simulations of biological flows in swimming and flying bio-locomotion: A review.
\newblock {\em Applied Sciences}, 13(7):4208, 2023.
\newblock
\url{https://doi.org/10.3390/app13074208}

\bibitem{kolahdouz_sharp_2023}
E.~M. Kolahdouz, D.~R. Wells, S.~Rossi, K.~I. Aycock, B.~A. Craven, and B.~E. Griffith.
\newblock A sharp interface {Lagrangian-Eulerian} method for flexible-body fluid-structure interaction.
\newblock {\em Journal of Computational Physics}, 448:112174, 2023.
\newblock
\url{https://doi.org/10.1016/j.jcp.2023.112174}

\bibitem{XU20131810}
D. Xu, E. Kaliviotis, A. Munjiza, E. Avital, C. Ji, and J. Williams.
\newblock Large scale simulation of red blood cell aggregation in shear flows.
\newblock {\em Journal of Biomechanics}, 46(11):1810--1817, 2013.
\newblock
\url{https://doi.org/10.1016/j.jbiomech.2013.05.010}

\bibitem{rbcs}
D. Xu, C. Ji, E. Avital, E. Kaliviotis, A. Munjiza, and J. Williams.
\newblock An investigation on the aggregation and rheodynamics of human red blood cells using high performance computations.
\newblock {\em Scientifica}, 2017:1--10, 04 2017.
\newblock
\url{https://doi.org/10.1155/2017/6524156}

\bibitem{LIU2006905}
F.~Liu, Z.M. Jin, F.~Hirt, C.~Rieker, P.~Roberts, and P.~Grigoris.
\newblock Transient elastohydrodynamic lubrication analysis of metal-on-metal hip implant under simulated walking conditions.
\newblock {\em Journal of Biomechanics}, 39(5):905--914, 2006.
\newblock
\url{https://doi.org/10.1016/j.jbiomech.2005.01.031}

\bibitem{mayo_fast_1984}
A.~Mayo.
\newblock The fast solution of poisson's and the biharmonic equations on irregular regions.
\newblock {\em SIAM Journal on Numerical Analysis}, 21(2):285--299, 1984.
\url{https://doi.org/10.1137/0721021}

\bibitem{leveque_immersed_1994}
R.~J. LeVeque and Z.~Li.
\newblock The immersed interface method for elliptic equations with discontinuous coefficients and singular sources.
\newblock {\em SIAM Journal on Numerical Analysis}, 31(4):1019--1044, 1994.
\url{https://doi.org/10.1137/0731054}

\bibitem{lee_immersed_2003}
L.~Lee and R.~J. LeVeque.
\newblock An immersed interface method for incompressible {Navier--Stokes} equations.
\newblock {\em SIAM Journal on Scientific Computing}, 25(3):832--856, 2003.
\url{https://doi.org/10.1137/S1064827502414060}

\bibitem{li_immersed_2001}
Z.~Li and M.-C. Lai.
\newblock The immersed interface method for the {Navier--Stokes} equations with singular forces.
\newblock {\em Journal of Computational Physics}, 171(2):822--842, 2001.
\url{https://doi.org/10.1006/jcph.2001.6813}

\bibitem{xu_systematic_2006}
S.~Xu and Z.~J. Wang.
\newblock Systematic derivation of jump conditions for the immersed interface method in three-dimensional flow simulation.
\newblock {\em SIAM Journal on Scientific Computing}, 27:1948--1980, 2006.
\url{https://doi.org/10.1137/040604960}

\bibitem{HUA2022111435}
Hua, M., \& Peskin, C. S. (2022). An analysis of the numerical stability of the immersed boundary method. \textit{Journal of Computational Physics}, 467, 111435. \url{https://doi.org/10.1016/j.jcp.2022.111435}



\bibitem{thekkethil_level_2019}
N.~Thekkethil and A.~Sharma.
\newblock Level set function--based immersed interface method and benchmark solutions for fluid flexible-structure interaction.
\newblock {\em International Journal for Numerical Methods in Fluids}, 91(3):134--157, 2019.
\url{https://doi.org/10.1002/fld.4746}

\bibitem{xu_level-set_2020}
J.~Xu, W.~Shi, W.~Hu, and J.~Huang.
\newblock A level-set immersed interface method for simulating the electrohydrodynamics.
\newblock {\em Journal of Computational Physics}, 400:108956, 2020.
\url{https://doi.org/10.1016/j.jcp.2019.108956}

\bibitem{kolahdouz_immersed_2020}
E.~M. Kolahdouz, A.~P.~Singh Bhalla, B.~A. Craven, and B.~E. Griffith.
\newblock An immersed interface method for discrete surfaces.
\newblock {\em Journal of Computational Physics}, 400:108854, 2020.
\url{https://doi.org/10.1016/j.jcp.2019.07.052}

\bibitem{kolahdouz_sharp_2021}
E.~M. Kolahdouz, A.~P.~S. Bhalla, L.~N. Scotten, B.~A. Craven, and B.~E. Griffith.
\newblock A sharp interface {Lagrangian-Eulerian} method for rigid-body fluid-structure interaction.
\newblock {\em Journal of Computational Physics}, 443:110442, 2021.
\url{https://doi.org/10.1016/j.jcp.2021.110442}

\bibitem{Acheson}
D.~J. Acheson.
\newblock {\em Elementary Fluid Dynamics}.
\newblock Oxford University Press, 1990.
\url{https://doi.org/10.1093/oso/9780198596608.001.0001}

\bibitem{FAI2018319}
T.~G. Fai and C.~H. Rycroft.
\newblock Lubricated immersed boundary method in two dimensions.
\newblock {\em Journal of Computational Physics}, 356:319--339, 2018.
\url{https://doi.org/10.1016/j.jcp.2017.11.029}

\bibitem{YACOUBI2019108919}
S.~Xu, A.~E.~Yacoubi, and Z.~J. Wang.
\newblock A new method for computing particle collisions in {Navier-Stokes} flows.
\newblock {\em Journal of Computational Physics}, 399:108919, 2019.
\url{https://doi.org/10.1016/j.jcp.2019.108919}

\bibitem{PESKIN199333}
C.~S. Peskin and B.~F. Printz.
\newblock Improved volume conservation in the computation of flows with immersed elastic boundaries.
\newblock {\em Journal of Computational Physics}, 105(1):33--46, 1993.
\url{https://doi.org/10.1006/jcph.1993.1051}

\bibitem{IMFSI-griffith}
B.~E. Griffith and N.~A. Patankar.
\newblock Immersed methods for fluid–structure interaction.
\newblock {\em Annual Review of Fluid Mechanics}, 52:421--448, 2020.
\url{https://doi.org/10.1146/annurev-fluid-010719-060228}

\bibitem{tan_immersed_2009}
Z.~Tan, D.~Le, K.~Lim, and B.~Khoo.
\newblock An immersed interface method for the incompressible navier--stokes equations with discontinuous viscosity across the interface.
\newblock {\em SIAM Journal on Scientific Computing}, 31:1798--1819, 2009.
\url{https://doi.org/10.1137/080712970}


\bibitem{griffith_volume_2012}
B.~E. Griffith.
\newblock On the volume conservation of the immersed boundary method.
\newblock {\em Communications in Computational Physics}, 12(2):401--432, 2012.
\url{https://doi.org/10.4208/cicp.120111.300911s}

\bibitem{griffith_accurate_2009}
B.~E. Griffith.
\newblock An accurate and efficient method for the incompressible {Navier--Stokes} equations using the projection method as a preconditioner.
\newblock {\em Journal of Computational Physics}, 228(20):7565--7595, 2009.
\url{https://doi.org/10.1016/j.jcp.2009.07.001}

\bibitem{xsPPM7}
W.~J. Rider, J.~A. Greenough, and J.~R. Kamm.
\newblock Accurate monotonicity- and extrema-preserving methods through adaptive nonlinear hybridizations.
\newblock {\em Journal of Computational Physics}, 225(2):1827--1848, 2007.
\url{https://doi.org/10.1016/j.jcp.2007.02.023}

\bibitem{colella_piecewise-parabolic_1982}
P.~Colella and P.~R. Woodward.
\newblock The {Piecewise-Parabolic Method (PPM)} for gas-dynamical simulations.
\newblock {\em Journal of Computational Physics}, 54:174--201, 1984.

\bibitem{ibamr}
{IBAMR}: {An Adaptive and Distributed-Memory Parallel Implementation of the Immersed Boundary Method}.
\newblock \url{https://github.com/IBAMR/IBAMR}.

\bibitem{samrai}
{SAMRAI}: {Structured Adaptive Mesh Refinement Application Infrastructure}.
\newblock \url{http://www.llnl.gov/CASC/SAMRAI}.

\bibitem{petsc}
S.~Balay, S.~Abhyankar, M.~F. Adams, S.~Benson, J.~Brown, P.~Brune, K.~Buschelman, E.~M. Constantinescu, L.~Dalcin, A.~Dener, V.~Eijkhout, J.~Faibussowitsch, W.~D. Gropp, V.~Hapla, T.~Isaac, P.~Jolivet, D.~Karpeev, D.~Kaushik, M.~G. Knepley, F.~Kong, S.~Kruger, D.~A. May, L.~C. McInnes, R.~T. Mills, L.~Mitchell, T.~Munson, J.~E. Roman, K.~Rupp, P.~Sanan, J.~Sarich, B.~F. Smith, S.~Zampini, H.~Zhang, H.~Zhang, and J.~Zhang.
\newblock {PETS}c {W}eb page.
\newblock \url{https://petsc.org/}, 2025.

\bibitem{petsc2}
S.~Balay, S.~Abhyankar, M.~F. Adams, S.~Benson, J.~Brown, P.~Brune, K.~Buschelman, E.~Constantinescu, L.~Dalcin, A.~Dener, V.~Eijkhout, J.~Faibussowitsch, W.~D. Gropp, V.~Hapla, T.~Isaac, P.~Jolivet, D.~Karpeev, D.~Kaushik, M.~G. Knepley, F.~Kong, S.~Kruger, D.~A. May, L.~C. McInnes, R.~T. Mills, L.~Mitchell, T.~Munson, J.~E. Roman, K.~Rupp, P.~Sanan, J.~Sarich, B.~F. Smith, H.~Suh, S.~Zampini, H.~Zhang, H.~Zhang, and J.~Zhang.
\newblock {PETSc/TAO} users manual.
\newblock Technical Report ANL-21/39 - Revision 3.24, Argonne National Laboratory, 2025.
\url{https://doi.org/10.2172/2998643}

\bibitem{petsc3}
S.~Balay, W.~D. Gropp, L.~C. McInnes, and B.~F. Smith.
\newblock Efficient management of parallelism in object-oriented numerical software libraries.
\newblock In E.~Arge, A.~M. Bruaset, and H.~P. Langtangen, editors, {\em Modern Software Tools in Scientific Computing}, pages 163--202. Birkhauser Press, 1997.
\url{https://doi.org/10.1007/978-1-4612-1986-6_8}

\bibitem{hypre}
\textit{hypre}: {High Performance Preconditioners}.
\newblock \url{http://www.llnl.gov/CASC/hypre}.

\bibitem{hypre2}
R.~D. Falgout and U.~M. Yang.
\newblock \textit{hypre}: {A} library of high performance preconditioners.
\newblock In {\em International Conference on Computational Science}, pages 632--641. Springer, 2002.
\url{https://doi.org/10.1007/3-540-47789-6_66}

\bibitem{libmesh}
{libMesh}: a {C++} finite element library.
\newblock \url{http://libmesh.github.io}.

\bibitem{libmesh2}
B.~S. Kirk, J.~W. Peterson, R.~H. Stogner, and G.~F. Carey.
\newblock {libMesh}: A {C++} library for parallel adaptive mesh refinement/coarsening simulations.
\newblock {\em Engineering Computations}, 22(3):237--254, 2006.
\url{https://doi.org/10.1007/s00366-006-0049-3}

\bibitem{lub_ecc}
H.~Christensen and K.~Tonder.
\newblock The hydrodynamic lubrication of rough bearing surfaces of finite width.
\newblock {\em Journal of Lubrication Technology}, 93(3):324--329, 1971.
\url{https://doi.org/10.1115/1.3451579}

\bibitem{faccisharp}
M.~J. Facci, E.~M. Kolahdouz, and B.~E. Griffith.
\newblock An immersed interface method for incompressible flows and geometries with sharp features.
\newblock {\em Journal of Computational Physics}, 537:114119, 2025.
\url{https://doi.org/10.1016/j.jcp.2025.114119}

\bibitem{SUN2026114497}
Q. Sun, E. M. Kolahdouz, and B. E. Griffith,
Improving the robustness of the immersed interface method through regularized velocity reconstruction,
\textit{Journal of Computational Physics}, vol. 546, p. 114497, 2026.
\url{https://doi.org/10.1016/j.jcp.2025.114497}

\bibitem{LAYTON2009266}
A. T. Layton,
Using integral equations and the immersed interface method to solve immersed boundary problems with stiff forces,
\textit{Computers \& Fluids}, vol. 38, no. 2, pp. 266--272, 2009.
\url{https://doi.org/10.1016/j.compfluid.2008.02.003}

\bibitem{FABENDER20241}
C. Fa{\ss}bender, T. B\"urchner, P. Kopp, E. Rank, and S. Kollmannsberger,
``Implicit-explicit time integration for the immersed wave equation,''
\textit{Computers \& Mathematics with Applications}, vol. 163, pp. 1--13, 2024.
\url{https://doi.org/10.1016/j.camwa.2024.02.049}

\end{thebibliography}

\end{document}